\newif\ifpreprint
\preprinttrue % UNCOMMENT THIS LINE TO COMPILE FOR ARXIV
 \ifpreprint % PREPRINT
 \documentclass[twocolumn,10pt,a4paper]{article}
\else % PREPRINT
 \documentclass[amsmath,amssymb,aps,reprint,nofootinbib]{revtex4-2}
 \fi % PREPRINT
\ifpreprint % PREPRINT
\usepackage[cm]{fullpage}
\usepackage{cuted}
\usepackage{amsmath,amssymb,amsthm}
\usepackage[colorlinks,bookmarksopen,bookmarksnumbered,citecolor=red,urlcolor=red]{hyperref}
\usepackage{multicol}
\usepackage[affil-it]{authblk}

\else % PREPRINT
\fi % PREPRINT

\usepackage{graphicx}
\usepackage{color}
\usepackage{multirow}
\usepackage{algorithm,algorithmicx,algpseudocode}
\usepackage{mathtools}
\usepackage{stmaryrd}
\usepackage{comment}
\usepackage{caption}
\usepackage{etoolbox}
\newcommand{\epsstat}{\epsilon_{\text{stat}}}
\newcommand{\epsdisc}{\epsilon_{\text{disc}}}
\newcommand{\figdir}{./}
\makeatletter
\algnewcommand{\StateX}[1]{\Statex \hskip\ALG@thistlm #1}
\makeatother
\renewcommand{\vec}[1]{\boldsymbol{#1}}

\ifpreprint % PREPRINT
\AtBeginEnvironment{algorithmic}{\small}
  \AtBeginEnvironment{tabular}{\small}
  \author[a]{Karl~Jansen}
  \author[b,*]{Eike~Hermann~M\"{u}ller}
  \author[c]{Robert~Scheichl}
   \affil[a]{DESY Zeuthen, Platanenallee 6, 15738 Zeuthen, Germany}
   \affil[b]{Department of Mathematical Sciences, University of Bath, Bath BA2 7AY, Bath, United Kingdom}
  \affil[c]{Institute for Applied Mathematics, Ruprecht-Karls-Universit\"{a}t Heidelberg, Im Neuenheimer Feld 205, 69120 Heidelberg, Germany}
\affil[*]{Email: \texttt{e.mueller@bath.ac.uk}}
\else % PREPRINT
\fi % PREPRINT

\begin{document}
\title{Multilevel Monte Carlo for quantum mechanics on a lattice}
\ifpreprint % PREPRINT
\else % PREPRINT
  \author{Karl~Jansen}
\affiliation{DESY Zeuthen, Platanenallee 6, 15738 Zeuthen, Germany}
\author{Eike~H.~M\"{u}ller}
\email{Email: \texttt{e.mueller@bath.ac.uk}}
\affiliation{Department of Mathematical Sciences, University of Bath, Bath BA2 7AY, United Kingdom}
\author{Robert Scheichl}
\affiliation{Institute for Applied Mathematics, Ruprecht-Karls-Universit\"{a}t Heidelberg, Im Neuenheimer Feld 205, 69120 Heidelberg, Germany}
\fi % PREPRINT

\ifpreprint
  \twocolumn[\begin{@twocolumnfalse}
  \maketitle
\else
\fi
\begin{abstract}
Monte Carlo simulations of quantum field theories on a lattice become increasingly expensive as the continuum limit is approached since the cost per independent sample grows with a high power of the inverse lattice spacing. Simulations on fine lattices suffer from critical slowdown, the rapid growth of autocorrelations in the Markov chain with decreasing lattice spacing $a$. This causes a strong increase in the number of lattice configurations that have to be generated to obtain statistically significant results. In this paper, hierarchical sampling methods to tame this growth in autocorrelations are discussed. Combined with multilevel variance reduction techniques, this significantly reduces the computational cost of simulations for given tolerances $\epsdisc$ on the discretisation error and $\epsstat$ on the statistical error. For an observable with lattice errors of order $\alpha$ and an integrated autocorrelation time that grows like $\tau_{\mathrm{int}}\propto a^{-z}$, multilevel Monte Carlo can reduce the cost from $\mathcal{O}(\epsstat^{-2}\epsdisc^{-(1+z)/\alpha})$ to $\mathcal{O}(\epsstat^{-2}\vert\log \epsdisc \vert^2+\epsdisc^{-1/\alpha})$ or $\mathcal{O}(\epsstat^{-2}+\epsdisc^{-1/\alpha})$. Even higher performance gains are expected for non-perturbative simulations of quantum field theories in $D$-dimensions. The efficiency of the approach is demonstrated on two non-trivial model systems in quantum mechanics, including a topological oscillator that is badly affected by critical slowdown due to freezing of the topological charge. On fine lattices, the new methods are several orders of magnitude faster than standard, single level sampling based on Hybrid Monte Carlo. For high resolutions, multilevel Monte Carlo can be used to accelerate even the cluster algorithm for the topological oscillator. Performance is further improved through perturbative matching. This guarantees efficient coupling of theories on the multilevel lattice hierarchy, which have a natural interpretation in terms of effective theories obtained by renormalisation group transformations.
\end{abstract}
\ifpreprint % PREPRINT
\textbf{keywords}:
\newcommand{\sep}{, }
Multilevel Monte Carlo\sep~Path Integral\sep~Hierarchical Methods\sep~Numerical Algorithms \\
\end{@twocolumnfalse}]
\else % PREPRINT
\maketitle
\fi % PREPRINT
% ==== ACM class ====
% F.2: ANALYSIS OF ALGORITHMS AND PROBLEM COMPLEXITY
% J.2: PHYSICAL SCIENCES AND ENGINEERING

% ==== MSC class ====
% 81-08: Computational methods for problems pertaining to quantum theory
% 81T25: Quantum field theory on lattices
% 65Y20: Complexity and performance of numerical algorithms
% 60J22: Computational methods in Markov chains

%%%%%%%%%%%%%%%%%%%%%%%%%%%%%%%%%%%%%%%%%%%%%%%%%%%%%
\section{Introduction}
%%%%%%%%%%%%%%%%%%%%%%%%%%%%%%%%%%%%%%%%%%%%%%%%%%%%%
The Euclidean path integral formulation of quantum mechanics \cite{Feynman2010} allows the calculation of observable quantities as expectation values with respect to infinite-dimensional and highly peaked probability distributions. After discretising the theory on a lattice with finite spacing $a$, expectation values are computed with Markov Chain Monte Carlo methods (see for example \cite{Creutz1981} for a highly accessible introduction). This approach is elegant and attractive since it can be extended to quantum field theories, where it allows first principles-predictions for strongly interacting theories such as Quantum Chromodynamics (QCD), see e.g.~\cite{Rothe2005,Carleton2006}. Ultimately, however, one is interested in the value of observables in the continuum limit of vanishing lattice spacing $a\rightarrow 0$. Since the cost of the calculation grows with a high power of $a^{-1}$, efficient Monte Carlo sampling techniques are crucial to obtain precise and accurate numerical predictions. Today state-of-the-art techniques \cite{Luscher2010} are routinely used to accelerate the Metropolis-Hastings algorithm \cite{Metropolis1953,Hastings1970} and in particular the Hybrid Monte Carlo (HMC) method \cite{Duane1987} has proved to be highly successful in lattice QCD simulations. However, lattice calculations with HMC methods still become prohibitively expensive as the continuum limit is approached. The reasons for this are twofold:
\begin{enumerate}
\item For quantum mechanical problems the cost $\mathcal{C}_{\mathrm{sample}}$ of generating a single discretised path grows at least in proportion to the number of lattice sites, which increases with $a^{-1}$ if the physical size of the simulation box is kept fixed (for a quantum field theory in $D$ dimensions the growth would be even faster with a cost of $a^{-D}$ per configuration).
  \item As the theory approaches a critical point, subsequent states in the Markov chain are increasingly correlated, which requires the generation of more paths to obtain a given number of \textit{statistically independent} samples.
  \end{enumerate}
  % The second point requires some further explanation. For some local observables which decorrelate across the lattice, the number of statistically independent measurements automatically increases on finer lattices since the observable can be evaluated at a larger number of independent points. However, this is not always the case. If the value of a local quantity of interest is correlated with a fixed physical correlation length or the observable is global and computed by summing over all lattice points, the number of independent measurements has to be increased by producing more lattice configurations. This is the case for the quantities considered in this paper: the squared displacement for the double well potential is correlated across the lattice and the topological susceptibility is computed by summing over all lattice points for the topological oscillator.
Furthermore, the law of large numbers dictates that to reduce the statistical (sampling) error below a given tolerance $\epsstat$, at least $N_{\mathrm{indep}}\propto\epsstat^{-2}$ \textit{independent} samples have to be generated. While in lattice QCD the continuum limit is usually taken by extrapolating simulations at different lattice spacings $a$ and fixed tolerance $\epsstat$ on the statistical error, in the multilevel Monte Carlo literature (see e.g. the classical paper \cite{Giles2008}) it is more common to decrease $\epsstat$ in proportion to the tolerance $\epsdisc$ on the discretisation error as the lattice spacing is reduced. Reducing the combined statistical and discretisation error in this way would make optimal use of computational resources to obtain a result with a given \textit{total} error for a specific fine lattice spacing.
  
The correlation of subsequent samples in the Markov chain is quantified by the integrated autocorrelation time $\tau_{\mathrm{int}}$, which grows particularly rapidly for some quantities, such as the topological susceptibility $\chi_t$ in QCD, where it has been observed that $\tau_{\mathrm{int}}\propto a^{-z}$ with $z=5$ \cite{Schaefer2011}. This is attributed to ``freezing'' of the topological charge, and can lead to observable effects. Those can be both direct, since e.g. the mass of the $\eta'$ meson receives important contributions from the topological susceptibility in a pure Yang-Mills theory \cite{Witten1979,Veneziano1979}, and more indirect due to the coupling of slow modes with large autocorrelation times to other observables. The authors of \cite{Schaefer2011} further report a milder but still significant growth with $z=0.5-1.0$ for a range of other physically relevant observables. While the rapid growth of the integrated autocorrelation time for the topological susceptibility can be addressed by using open boundary conditions in time \cite{Luscher2011,Luscher2013}, this introduces additional complications since it requires lattices with a very large extent in the temporal direction.
  
To estimate the overall growth in cost of a simulation, as the lattice spacing $a$ is reduced, consider a quantum mechanical observable with a discretisation error that is $\mathcal{O}(a^\alpha)$, where values such as $\alpha=1,2$ are typical. To reduce the discretisation error below a tolerance of $\epsdisc$ and the statistical error below $\epsstat$ incurs a cost %that increases as
\begin{equation}
  \begin{aligned}
    \mathcal{C}_{\mathrm{StMC}}(\epsdisc,\epsstat) &= N_{\mathrm{indep}}\times \tau_{\mathrm{int}}\times \mathcal{C}_{\mathrm{sample}}\\
    &= \mathcal{O}(\epsstat^{-2}\epsdisc^{-(1+z)/\alpha}),
  \end{aligned}\label{eqn:cost_StMC}
\end{equation}
with standard Monte Carlo (StMC), since $\epsdisc\propto a^\alpha$. To get an intuitive understanding of this and subsequent complexity estimates it might be instructive to consider the special case $\alpha=2$, $z=0$: since the discretisation error decreases quadratically with $a$, reducing this error by a factor of $4$ can be achieved by halving the lattice spacing, which in turn doubles the cost for generating a single sample if the physical size of the simulation box is kept fixed; in other words, the cost per sample grows in proportion to $\epsdisc^{-1/2}$

In this paper, it is shown how this explosion in computational cost can be significantly reduced with hierarchical sampling~\cite{Christen2005} and Multilevel Monte Carlo (MLMC)~\cite{Heinrich2001,Giles2008}, which has recently been extended to a Markov chain setting~\cite{Dodwell2015,Dodwell2019a}. To generate samples, a hierarchy of $L-1$ coarser lattices with spacings of $2a,4a,\dots,2^{L-1}a$ and corresponding coarse-grained versions of the original theory are constructed. Based on this hierarchy, a recursive implementation of the delayed acceptance method in~\cite{Christen2005} is proposed. Starting on the coarsest level, proposals are successively extended by additional modes and screened with a standard Metropolis-Hastings accept/reject step on increasingly finer lattices. At this point is important to stress that the coarse lattices are only used to accelerate sampling and do not introduce any additional bias because ultimately each new sample is accepted or rejected step with the correct, original action on the finest lattice. Since evaluating the action on the coarse lattices is substantially cheaper, the cost of generating a single fine level sample is not substantially higher than if a single-level sampler was used. In fact, when compared to a method such as HMC, it may be smaller since the cost of generating an HMC trajectory can be shifted to the coarsest level where it is substantially shorter. Since on each level proposals are screened with a Metropolis-Hastings accept/reject step, the method samples from the correct distribution on the original lattice with spacing $a$ and does not introduce any additional bias, cf.~\cite{Christen2005}. Due to the convergence of the lattice theories on subsequent levels of the hierarchy with $a \to 0$, hierarchical sampling eliminates the growth in autocorrelation time, reducing the computational cost 
%in Eq. \eqref{eqn:cost_StMC} 
to
\begin{equation}
  \mathcal{C}_{\mathrm{hierarchical}}(\epsdisc,\epsstat) = \mathcal{O}(\epsstat^{-2}\epsdisc^{-1/\alpha}).
  \label{eqn:cost_hierarchical}
\end{equation}

MLMC is a variance reduction technique, which uses the fact that the expectation value of an observable (or quantity of interest) $Q$ on a lattice with spacing $a$ can be written as a telescoping sum. For this assume that there is some integer $L\in\mathbb{N}$ and a constant $a_0$ such that $a=2^{-L+1}a_0$. Further, let $Q_\ell$ be the observable measured on a lattice with spacing $2^{-\ell}a_0$. Then
\begin{equation}
  \begin{aligned}
    \mathbb{E}[Q] = 
    \mathbb{E}[Q_{L-1}] &= \mathbb{E}[Q_{L-1}-Q_{L-2}] + \mathbb{E}[Q_{{L-2}}] =\\ %-Q_{{L-3}}]\\
    %&\qquad+\;\dots +
    %\mathbb{E}[Q_{1}-Q_{0}] + \mathbb{E}[Q_{0}]\\
    &= \ldots = \sum_{\ell=0}^{L-1} \mathbb{E}[Y_\ell] \approx \sum_{\ell=0}^{L-1}\widehat{Y}_\ell
  \end{aligned}
  \label{eqn:telescoping_sum}
\end{equation}
where
\begin{equation*}
  \begin{aligned}
    Y_\ell&:=\begin{cases} Q_{0} & \text{for $\ell=0$} \\
      Q_{\ell}-Q_{\ell-1} & \text{for $\ell=1,2,\dots,L-1$,}
  \end{cases}\\
  \widehat{Y}_\ell &:=\frac{1}{N_{\ell}} \sum_{j=1}^{N_\ell} Y_{\ell}^{(j)}.
  \end{aligned}
\end{equation*}
Here, the sums in $\widehat{Y}_\ell$ are taken over independent samples, labelled by the superscript ``$(j)$''. The key observation is that, except for the very coarsest level, MLMC estimates \textit{differences} of the observable instead of the quantity of interest itself. Provided that theories on subsequent levels can be coupled efficiently and the variance of the difference $Q_{\ell}-Q_{\ell-1}$ decreases sufficiently rapidly as the lattice spacing $a$ is reduced, significantly lower numbers of samples $N_\ell$ are sufficient on the finer levels of the grid hierarchy. The majority of the cost can be shifted to the coarser levels $\ell\ll L$, where sampling is substantially cheaper. Due to the exactness of the telescoping sum (i.e. the first equality in Eq. \eqref{eqn:telescoping_sum}), MLMC does not introduce any additional bias if the individual MC estimators $\widehat{Y}_\ell$ are unbiased. The algorithms described in the paper allow the construction of estimators $\widehat{Y}_\ell$ which have an arbitrarily small bias. In the numerical results presented below the size of this bias is comparable to the discretisation error on the original, fine level lattice. 
Compared to Eqs.~\eqref{eqn:cost_StMC} and \eqref{eqn:cost_hierarchical}, MLMC further reduces the computational complexity to
\begin{equation}
  \mathcal{C}_{\mathrm{MLMC}}(\epsdisc,\epsstat) = \mathcal{O}(\epsstat^{-2}|\log \epsdisc|^2+\epsdisc^{-1}), \label{eqn:cost_MLMC}
\end{equation}
see below. Similar estimates have been derived in \cite{Giles2008,Dodwell2015,Dodwell2019a} and it has been demonstrated numerically that MLMC leads to a significant reduction in computational complexity and overall runtime for a range of applications, e.g.~in uncertainty quantification (UQ) for sub-surface flow \cite{Cliffe2011,Dodwell2015}, inverse problems \cite{Scheichl2017} or material simulation \cite{Dodwell2019}.

While this paper focuses on the application of these new methods in quantum mechanics, the ultimate goal is to apply them in $D$-dimensional quantum field theories, such as lattice QCD with $D=4$ and $\alpha=2$. For $D>\alpha$, the expected gains are even larger, since the cost to generate a single configuration grows like $a^{-D}$ instead of $a^{-1}$ while the accuracy is still decreasing no faster than $a^{2}$. The predicted improvement in computational performance is summarised in the following diagram, generalising Eqs. \eqref{eqn:cost_StMC}, \eqref{eqn:cost_hierarchical} and \eqref{eqn:cost_MLMC} to $D$ dimensions:
\begin{equation}
  \begin{aligned}
    \mathcal{C}^{\text{(QFT)}}_{\mathrm{StMC}}(\epsdisc,\epsstat) &= \mathcal{O}(\epsstat^{-2}\epsdisc^{-(D+z)/\alpha})\\[1ex]
    &\downarrow\quad\text{(hierarchical sampling)}\\[1ex]
    \mathcal{C}^{\text{(QFT)}}_{\mathrm{hierarchical}}(\epsdisc,\epsstat) &= \mathcal{O}(\epsstat^{-2}\epsdisc^{-D/\alpha})\\[1ex]
    &\downarrow\quad\text{(multilevel Monte Carlo)}\\[1ex]
    \mathcal{C}^{\text{(QFT)}}_{\mathrm{MLMC}}(\epsdisc,\epsstat) &= \mathcal{O}(\epsstat^{-2}\epsdisc^{1-D/\alpha}+\epsdisc^{-D/\alpha}).
    \end{aligned}\label{eqn:cost_QFT}
  \end{equation}
  For example, consider the prediction of the topological susceptibility ($z=5$) in lattice QCD ($D=4$) with improved action ($\alpha=2$). In this case, hierarchical sampling reduces the cost of a Monte Carlo simulation from $\mathcal{O}(\epsstat^{-2}\epsdisc^{-4.5})$ to $\mathcal{O}(\epsstat^{-2}\epsdisc^{-2})$ and MLMC reduces the computational complexity even further to $\mathcal{O}(\epsstat^{-2}\epsdisc^{-1}+\epsdisc^{-2})$.

To discuss this further, consider first the relative advantage of MLMC over standard Monte Carlo in the continuum limit $\epsdisc\rightarrow 0$ for fixed $\epsstat$.  MLMC only requires the generation of a small number of samples on the finest lattice for small $\epsdisc$ (eventually only one for very small $\epsdisc$), whereas the number of configurations that have to be generated with a standard Monte Carlo method is proportional to $\epsstat^{-2}$. As can be seen from the final two lines of Eq. \eqref{eqn:cost_QFT}, MLMC is a factor of $\kappa\epsstat^{-2}$ faster than standard Monte Carlo with hierarchical sampling for $\epsdisc\rightarrow 0$. This argument holds for general $\alpha$ and $D$; the coefficient $\kappa$ depends on the relative cost of generating independent samples on the finest level and the coarser levels. Provided those costs are proportional to the number of unknowns on each level (and the constant of proportionality is independent of $\epsdisc$) we expect $\kappa$ to lie between 1 and 2.

If $\epsstat$ is kept fixed as the continuum limit is taken, eventually the statistical error will dominate the discretisation error. To avoid this, one might consider the case where $\epsdisc=\epsstat=\epsilon/\sqrt{2}$ and the combined root mean square error is reduced below some given tolerance $\epsilon$. This is the common choice in the multilevel Monte Carlo literature (see e.g. \cite{Giles2008}). In that case, the complexity estimates in Eq. \eqref{eqn:cost_QFT} become
\begin{equation}
  \begin{aligned}
    \mathcal{C}^{\text{(QFT)}}_{\mathrm{StMC}}(\epsilon) &= \mathcal{O}(\epsilon^{-2-(D+z)/\alpha}) \\[1ex]
    &\downarrow\quad\text{(hierarchical sampling)}\\[1ex]
    \mathcal{C}^{\text{(QFT)}}_{\mathrm{hierarchical}}(\epsilon) &= \mathcal{O}(\epsilon^{-2-D/\alpha})\\[1ex]
    &\downarrow\quad\text{(multilevel Monte Carlo)}\\[1ex]
    \mathcal{C}^{\text{(QFT)}}_{\mathrm{MLMC}}(\epsilon) &= \mathcal{O}(\epsilon^{-1-D/\alpha})
  \end{aligned}
  \label{eqn:cost_QFT_epsilon}
  \end{equation}

In quantum field theories, coarse grained actions are naturally obtained by integrating out high-frequency modes in a renormalisation group (RG) transformation, which results in an effective theory with less degrees of freedom. In practice, the RG transformation can be carried out either non-perturbatively (e.g. through a block spin transformation) or
through perturbative matching. The latter would, in fact, be sufficient for MLMC as
long as the variance of $Y_\ell$ decays sufficiently rapidly since the coarse levels are only used to accelerate sampling on the original, fine level. For asymptotically free theories, such as lattice QCD, Symanzik-improved actions \cite{Symanzik1983,Luescher1985} can be constructed by systematically adding suitable terms which are proportional to powers of the lattice spacing $a$ and which are multiplied by appropriate, so-called `improvement coefficients'. These coefficients can be tuned non-perturbatively \cite{Luscher1997,Jansen1998}, or they can be computed using perturbation theory for sufficiently small lattice spacing \cite{Symanzik1983,Luescher1985}. In the MLMC approach, the perturbatively calculated improvement coefficients on different levels of the lattice hierarchy are in fact sufficient since the \textit{differences} of these coefficients between subsequent levels are sufficiently small on fine lattices.

As a proof-of-concept, hierarchical sampling and multilevel Monte Carlo are applied to two problems in quantum mechanics ($D=1$): a non-symmetric double-well potential and the topological oscillator studied in \cite{Ammon2016}. The latter case is particularly interesting since it has a topological quantum number, which freezes in the continuum limit ($a\rightarrow 0$). This results in a rapid growth of the autocorrelation time of the topological susceptibility if standard HMC sampling is used. Hierarchical sampling all but eliminates this growth, resulting in a dramatic reduction in runtime. Furthermore, the coarse-grained theories can be improved using a perturbative matching technique for this problem, which further increases the efficiency of the hierarchical approach. As demonstrated in \cite{Ammon2016}, for the topological oscillator the so-called 'cluster algorithm' \cite{Wolff1989} almost entirely eliminates autocorrelations through long-range spin updates. However, this method can be further accelerated with MLMC, leading to a reduction in computational complexity and in absolute runtime for high resolutions. Similar gains are observed for the non-symmetric double-well potential problem with MLMC.

In summary, the main achievements of this work are:
\begin{enumerate}
\item It is described in detail how algorithms for hierarchical sampling and multilevel Monte Carlo acceleration can be applied to the path integral formulation of quantum mechanics.
\item It is shown how hierarchical sampling techniques dramatically reduce autocorrelation times.
\item It is further demonstrated that combining this with MLMC leads to an additional reduction in computational complexity and in the total runtime.
\item It is explained how perturbative matching can further improve performance for the topological oscillator.
\end{enumerate}

It is stressed again that the additional gains due to MLMC  accelerating are expected to be significantly larger in high-dimensional theories, such as lattice QCD (see Eq. \eqref{eqn:cost_QFT}). The present paper therefore aims to lay the foundation for further work on extending the described methods to quantum field theories on a lattice. 
%%%%%%%%%%%%%%%%%%%%%%%%%%%%%%%%%%%%%%%%%%%%%%%%%%%%%
\paragraph*{Structure.}
%%%%%%%%%%%%%%%%%%%%%%%%%%%%%%%%%%%%%%%%%%%%%%%%%%%%%
This paper is organised as follows: after briefly reviewing the literature on related approaches in Section \ref{sec:literature_review}, the application of hierarchical sampling and multilevel Monte Carlo to the path integral formulation of quantum mechanics is discussed in Section \ref{sec:methods}. The quantum mechanical model problems that are used in this work are described in Section \ref{sec:model_systems}, including the construction of coarse-grained actions for those problems. Numerical results for the non-symmetric double-well potential and the topological oscillator are presented in Section \ref{sec:results}, in particular we compare the runtime of all considered algorithms for fixed $\epsstat$. Section \ref{sec:conclusion} contains the conclusion and outlines directions for future work. More technical topics, such as a detailed cost analysis of MLMC and a discussion of how the methods can be extended to higher dimensional problems, are relegated to the appendix where we also show results for $\epsstat=\epsdisc=\epsilon/\sqrt{2}$.
%%%%%%%%%%%%%%%%%%%%%%%%%%%%%%%%%%%%%%%%%%%%%%%%%%%%%
\subsection{Relationship to previous work}\label{sec:literature_review}
%%%%%%%%%%%%%%%%%%%%%%%%%%%%%%%%%%%%%%%%%%%%%%%%%%%%%
While hierarchical sampling techniques have been suggested previously (see e.g. \cite{Goodman1986,Janke1994,Schmidt1983,Faas1986}), the variance reduction techniques from MLMC significantly improve on this. Eqs. \eqref{eqn:cost_hierarchical} and \eqref{eqn:cost_MLMC} show that the additional acceleration will lead to a further dramatic reduction in computational complexity. The presented methods are therefore expected to be superior to the approach in \cite{Endres2015}, which uses a hierarchical method to initialise the simulation, but not for the Monte Carlo sampling. Earlier work in \cite{Schmidt1983,Faas1986} uses renormalisation group techniques to sample close to the critical point of the Ising model where the theories on the coarser levels become self-similar. Similarly, collective cluster-update algorithms \cite{Swendsen1987,Wolff1989,Wang1990} have been applied to models in solid state physics close to phase transitions (see e.g. \cite{Chen1995}). However, the application of all those techniques is limited to spin systems. The approach here applies to general systems and delivers significant additional speedup through multilevel Monte Carlo variance reduction.
%%%%%%%%%%%%%%%%%%%%%%%%%%%%%%%%%%%%%%%%%%%%%%%%%%%%% 
\section{Methods}\label{sec:methods}
%%%%%%%%%%%%%%%%%%%%%%%%%%%%%%%%%%%%%%%%%%%%%%%%%%%%%
%%%%%%%%%%%%%%%%%%%%%%%%%%%%%%%%%%%%%%%%%%%%%%%%%%%%%
\subsection{Path integral formulation of quantum mechanics}
%%%%%%%%%%%%%%%%%%%%%%%%%%%%%%%%%%%%%%%%%%%%%%%%%%%%%
For completeness and to introduce the discretised path integral for non-experts, we recapitulate the key principles here. The path integral formulation of quantum mechanics \cite{Feynman2010} expresses the expectation value of physical observables as the infinite-dimensional sum over all possible configurations or paths $\{x(t)\}$, where $x(t)\in \mathcal{D}\subset \mathbb{R}$ for all times $t\in\mathbb{R}$. In this sum each path $x(t)$ is weighted by a complex amplitude $e^{\frac{i}{\hbar}S(x(t))}$, where $S(x(t))$ is the action, the integral over the Lagrangian $\mathcal{L}$ of the system. This formulation is very elegant since it allows the direct quantisation of any system which can be described by a Lagrangian. In the limit $\hbar\rightarrow 0$ fluctuations around the classical path which minimises the action cancel out, and the Euler-Lagrange equations are recovered. However, for simplicity from now on we will work in atomic units where $\hbar=1$. To make the evaluation of the path integral tractable, two approximations are made: (1) time is restricted to a finite interval $t\in[0,T)$ and (2) the time interval is divided into $d$ intervals of size $a=T/d$, which is known as the lattice spacing. Conditions have to be imposed on the paths at $t=0$ and $t=T$; here we use periodic boundary conditions $x(T)=x(0)$. Each path $x(t)$, which is defined for all times $t\in[0,T)$, is replaced by a vector $\vec{x}=(x_0,x_1,\dots,x_{d-1})\in \Omega =\mathcal{D}^d$. For each $j=0,1,\dots,d-1$ the quantity $x_j$ approximates the position $x(t_j)$ of the particle at the time $t_j=aj$. Those two approximations turn the infinite dimensional integral over all paths into an integral over a finite, but high-dimensional domain $\mathcal{D}^d$. Evaluating the integral in Euclidean time converts it to the canonical ensemble average of a statistical system at a finite temperature. More specifically, the expectation value of an observable (commonly known as ``Quantity of interest'', QoI, in the UQ literature) which assigns a value $Q(\vec{x})$ to each discrete path $\vec{x}$ can be written as the following ratio
\begin{equation}
  \begin{aligned}
    \mathbb{E}[Q] &=\frac{\int_{\mathcal{D}}\;\dots\;\int_{\mathcal{D}} Q(\vec{x})e^{-S(\vec{x})}\;dx_0\;\dots\;dx_{d-1}}{\int_{\mathcal{D}}\;\dots\;\int_{\mathcal{D}} e^{-S(\vec{x})}\;dx_0\;\dots\;dx_{d-1}}\\
    &=\int_{\Omega} \pi^*(\vec{x}) Q(\vec{x})\;d\vec{x}
  \end{aligned}
  \label{eqn:pathintegral}
\end{equation}
with the $d$-dimensional probability density $\pi^*$ given by
\begin{equation}
  \pi^*(\vec{x}) = \mathcal{Z}^{-1}e^{-S(\vec{x})},\quad\text{for all $\vec{x}\in \Omega$},
  \label{eqn:def_pifinelevel}
\end{equation}
with normalisation constant $\mathcal{Z}$.
The action $S(\vec{x})$ is an approximation of the continuum action
\begin{equation*}
S(x(t)) = \int_0^T \mathcal{L}(x(t))\;dt
\end{equation*} 
where $\mathcal{L}$ is the Lagrangian. 

Physically meaningful predictions, which can be compared to experimental measurements, are obtained by extrapolating to the continuum limit $a\rightarrow 0$ and infinite volume $T\rightarrow \infty$. As $d$ is inversely proportional to the lattice spacing, the integrals in Eq. \eqref{eqn:pathintegral} become very high dimensional in the continuum limit. In this paper we do not discuss finite volume errors (due to finite values of $T$). In other words, we take the continuum limit $Q^{\mathrm{exact}}=\lim_{a\rightarrow 0} \mathbb{E}[Q]$ for finite $T$ as the ``true'' value for any observables studied here.
%%%%%%%%%%%%%%%%%%%%%%%%%%%%%%%%%%%%%%%%%%%%%%%%%%%%%
\subsection{Standard Monte Carlo}
%%%%%%%%%%%%%%%%%%%%%%%%%%%%%%%%%%%%%%%%%%%%%%%%%%%%%
Since the distribution $\pi^*$ in Eq. \eqref{eqn:def_pifinelevel} is highly peaked, the expectation value in Eq. \eqref{eqn:pathintegral} is usually computed with importance sampling. For this, the Metropolis-Hastings algorithm \cite{Metropolis1953,Hastings1970} is used to iteratively generate a sequence of samples $\vec{x}^{(0)}$, $\vec{x}^{(1)},\dots,\vec{x}^{(N-1)}\sim\pi^*$. The expectation value can then be approximated as the sample average
\begin{equation}
  \mathbb{E}[Q] \approx \widehat{Q}^{\mathrm{StMC}} := \frac{1}{N}\sum_{j=0}^{N-1} Q(\vec{x}^{(j)}).\label{eqn:StMCsampleaverage}
\end{equation}
A single Metropolis-Hastings step for computing $\vec{x}^{(t+1)}$, given $\vec{x}^{(t)}$, is written down in Alg. \ref{alg:standardMC}.
\begin{algorithm}[H]
\caption{Standard Metropolis-Hastings step.\newline Input: current sample $\vec{x}^{(t)}\sim\pi^*$\newline Output: new sample $\vec{x}^{(t+1)}\sim\pi^*$}
\label{alg:standardMC}
\begin{algorithmic}[1]
  \State{Pick proposal $\vec{y}$ from a probability distribution $q(\cdot\vert \vec{x}^{(t)})$.}
  \State{Compute
    \begin{equation*}
      \begin{aligned}
        \frac{\pi^*(\vec{y})}{\pi^*(\vec{x}^{(t)})}\cdot\frac{q(\vec{x}^{(t)}\vert \vec{y})}{q(\vec{y}\vert \vec{x}^{(t)})} = \exp[-\Delta S]
        \end{aligned}
      \end{equation*}
      with
      \begin{equation*}
        \begin{aligned}
          \Delta S := S(\vec{y}) - S(\vec{x}^{(t)}) + \log q(\vec{y}\vert \vec{x}^{(t)}) - \log q(\vec{x}^{(t)}\vert \vec{y})
          \end{aligned}
        \end{equation*}
      }
      \If{$\Delta S<0$}
      \State{Set $\vec{x}^{(t+1)}\mapsfrom \vec{y}$}
      \Else
      \State{Draw uniformly distributed random number $u\in[0,1)$.}
      \If{$u<\exp[-\Delta S]$}
       \State{Set $\vec{x}^{(t+1)}\mapsfrom \vec{y}$}
       \Else
       \State{Set $\vec{x}^{(t+1)}\mapsfrom \vec{x}^{(t)}$}
      \EndIf
      \EndIf
\end{algorithmic}
\end{algorithm}
The Markov chain $\vec{x}^{(0)},\vec{x}^{(1)},\vec{x}^{(2)},\dots$ is generated by starting from some $\vec{x}^{(0)}$, which is either a given vector or drawn at random. Since this $\vec{x}^{(0)}$ is not drawn from the correct distribution, all subsequent samples $\vec{x}^{(t)}$ are distributed according to some distribution $\pi^{*(t)}$ with $\lim_{t\rightarrow \infty}\pi^{*(t)}=\pi^*$. In practice, the first $n_{\mathrm{burnin}}$ samples are discarded, and throughout this paper we implicitly assume that $n_{\mathrm{burnin}}\gg 1$ is chosen such that for all subsequent samples the error due to the difference between $\pi^{*(t)}$ and $\pi^*$ is much smaller than the discretisation- and sampling- errors.

The law of large numbers states that in the limit of a large number of samples $N\gg 1$ the sample average $\widehat{Q}^{\mathrm{StMC}}$ in Eq. \eqref{eqn:StMCsampleaverage}  is distributed according to a Gaussian $\mathcal{N}(\mu,\sigma)$ with mean $\mu=\mathbb{E}[Q]$ and variance
\begin{equation}
  \sigma^2 = \frac{\tau_{\mathrm{int}} \mathrm{Var}[Q]}{N}.
  \label{eqn:sigma_tauint}
\end{equation}
In this expression $\tau_{\mathrm{int}}$ is the integrated autocorrelation time defined as
\begin{equation}
  \tau_{\mathrm{int}} = 1+2\sum_{s=1}^{\infty} \frac{\mathbb{E}[Q(\vec{x}^{(t_{\text{meas}})})Q(\vec{x}^{(t_{\text{meas}}+s)})]}{\mathbb{E}[Q(\vec{x}^{(t_{\text{meas}})})^2]},
   \label{eqn:tauint_definition}
 \end{equation}
 where $t_{\text{meas}}\gg n_{\text{burnin}}$ is an arbitrary point in time. As can be seen from Eq. \eqref{eqn:sigma_tauint}, the number of samples required to reduce the statistical error below a given tolerance grows with $\tau_{\mathrm{int}}$, and it is therefore important to reduce the correlation between subsequent samples as far as possible. This can be achieved by carefully choosing the proposal $\vec{y}$ in line 1 of Alg. \ref{alg:standardMC}. In lattice QCD with dynamical fermions, Hybrid Monte Carlo \cite{Duane1987} is very popular since it generates global updates. We therefore choose to use this method here, being aware that other algorithms, such as heat bath sampling, might be more efficient for particular applications. We nevertheless believe that HMC is representative, since $\tau_{\mathrm{int}}$ grows with a large power of the inverse lattice spacing as the continuum limit is approached also with other sampling approaches. The only exception are some problem-specific samplers, such as the cluster algorithm \cite{Wolff1989} for the topological oscillator, which we therefore also consider in this work. 
%%%%%%%%%%%%%%%%%%%%%%%%%%%%%%%%%%%%%%%%%%%%%%%%%%%%%
\subsection{Multilevel Monte Carlo}
%%%%%%%%%%%%%%%%%%%%%%%%%%%%%%%%%%%%%%%%%%%%%%%%%%%%%
We now describe hierarchical methods for overcoming the growth in autocorrelations and reducing the variance of the measured observable.
%%%%%%%%%%%%%%%%%%%%%%%%%%%%%%%%%%%%%%%%%%%%%%%%%%%%%
\subsubsection*{Lattice hierarchy}
%%%%%%%%%%%%%%%%%%%%%%%%%%%%%%%%%%%%%%%%%%%%%%%%%%%%%
Recall that a path describes the position of the particle at the discrete points $t_j=aj$ with $j=0,1,2,\dots,d-1$. More formally, define a lattice $\mathcal{T}$ as the set of points
\begin{equation*}
\mathcal{T} = \{t_j=ja, j = 0,1,\dots,d-1\}.
\end{equation*}
Paths $\vec{x}$ on this lattice are objects in the domain $\Omega=\mathcal{D}^d\subset \mathbb{R}^d$. We introduce a hierarchy  of $L$ lattices $\mathcal{T}_{\ell}$ for $\ell=0,1,\dots,L-1$, such that lattice $\mathcal{T}_{\ell}$ has $d_{\ell}=2^{\ell-L+1}d$ points and a lattice spacing of $a_{\ell}=T/d_{\ell}=2^{L-1-\ell} a$, i.e.
\begin{equation*}
  \mathcal{T}_\ell = \{t_j=ja_{\ell}: j = 0,1,\dots,d_{\ell}-1\}.
\end{equation*}
Here $\mathcal{T}_{L-1}=\mathcal{T}$ is the original lattice with $d_{L-1}=d$ points and a spacing of $a_{L-1}=a$. Paths on lattice $\mathcal{T}_{\ell}$ are represented by vectors in the domain $\Omega_{\ell}=\mathcal{D}^{d_{\ell}}\subset \mathbb{R}^{d_{\ell}}$, where obviously $\Omega_{L-1}=\Omega$. 

Note that the lattices are nested, and the points of the lattice $\mathcal{T}_{\ell-1}$ are a subset of the points of $\mathcal{T}_{\ell}$, namely the points with even indices. A path on a particular level $\ell$ stores values at the odd and even lattice points, where the latter are also present on the next-coarser lattice. Formally this can be expressed as
\begin{equation}
\Omega_{\ell} = \Omega_{\ell-1} \oplus \Omega_{\ell-1}\label{eq:Omega_sum}
\end{equation}
such that all $\vec{x}\in\Omega_{\ell}$ can be written as
\begin{equation}
\begin{aligned}
\vec{x} &:= [\tilde{\vec{x}},\vec{x}']\quad\text{with $\tilde{\vec{x}},\vec{x}'\in\Omega_{\ell-1}$ and}\\
x_{j} &= \begin{cases}
x'_{j/2} & \text{for even $j$}\\
\tilde{x}_{(j-1)/2} & \text{for odd $j$}.
\end{cases}
\end{aligned}
\label{eqn:x_decomposition}
\end{equation}

On each lattice we define an action $S_{\ell}:\Omega_{\ell} \rightarrow \mathbb{R}$ such that $S_{L-1}=S$ is the original action. In the simplest case the coarse-level actions are obtained by re-discretising the original action $S$ with the appropriate lattice spacings, but other choices are possible and will be discussed below. On each level the action induces a probability distribution $\pi_{\ell}$ such that
\begin{equation*}
\pi_{\ell}(\vec{x}) = \mathcal{Z}_\ell^{-1}\exp\left[-S_{\ell}(\vec{x})\right]\quad\text{for all $\vec{x}\in\Omega_{\ell}$},
\end{equation*}
where $\mathcal{Z}_\ell^{-1}$ is the normalisation constant. The probability distribution $\pi_{L-1}$ on the finest level is identical to $\pi^*$ defined in Eq. \eqref{eqn:def_pifinelevel}. Further, introduce a conditional probability distribution $\tilde{\pi}_{\ell}(\cdot\vert\vec{x}')$ for the values at the odd points on level $\ell$, given the values at the even points on the same level, namely
\begin{equation}
\tilde{\pi}_{\ell}(\tilde{\vec{x}}\vert \vec{x}') = \tilde{\mathcal{Z}}_\ell(\vec{x}')^{-1} \exp\left[-\tilde{S}_{\ell}([\tilde{\vec{x}}, \vec{x}'])\right].\label{eqn:pi_tilde}
\end{equation}
for all $\tilde{\vec{x}},\vec{x}'\in\Omega_{\ell-1}$. The action $\tilde{S}_{\ell}$ should be some approximation to $S_{\ell}$, such that it is possible to sample from $\tilde{\pi}_{\ell}$ for a given $\vec{x}'$. For the quantum mechanical model problems considered in this work the construction of $\tilde{S}_{\ell}$ is described in Sections \ref{sec:coarse_level_action_doublewell} and \ref{sec:coarse_level_action_rotor}. 

We stress that although in this paper we assume that the lattice can be partitioned into sets of mutually independent even and odd sites, the ideas developed here can be generalised to higher dimensions. This is outlined in Appendix \ref{sec:two_dimensional_lattice}.
%%%%%%%%%%%%%%%%%%%%%%%%%%%%%%%%%%%%%%%%%%%%%%%%%%%%%
\subsubsection*{Hierarchical sampling}
%%%%%%%%%%%%%%%%%%%%%%%%%%%%%%%%%%%%%%%%%%%%%%%%%%%%%
Similar to the delayed-acceptance approach in \cite{Christen2005}, we next introduce a hierarchical algorithm to efficiently construct a Markov chain on a given level $\ell$ using coarser levels: First we define the two-level Metropolis-Hastings step in Alg. \ref{alg:twolevelMC}. Setting $\vec{x}_\ell^{(t)}=[\tilde{\vec{x}}_{\ell}^{(t)},\vec{x}_{\ell-1}^{(t)}]$ this algorithm assumes that on a given level $\ell$ there is a coarse level proposal distribution $q_{\ell-1}(\cdot|\vec{x}_{\ell-1}^{(t)})$ which depends on $\vec{x}_{\ell-1}^{(t)}$. Based on this, it proposes a new fine-level state which is either accepted and returned as the new state $\vec{x}_\ell^{(t+1)}$ or rejected; in the latter case the previous state $\vec{x}_\ell^{(t)}$ is returned as $\vec{x}_\ell^{(t+1)}$. It was shown in \cite{Christen2005} that this defines a correct Metropolis-Hastings algorithm targeting $\pi_\ell$.

\begin{figure}
\begin{center}
\includegraphics[width=\linewidth]{\figdir/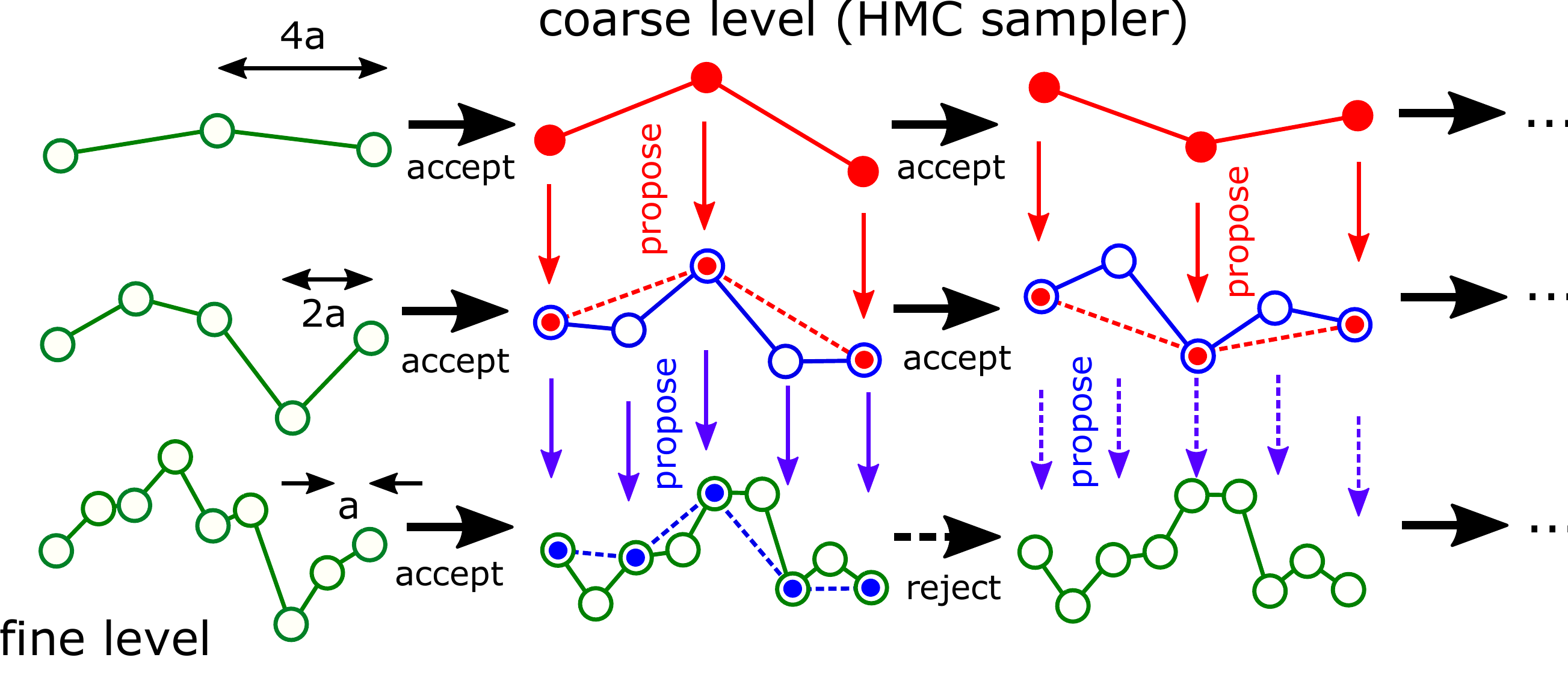}
\end{center}
\caption{Hierarchical sampling,  as described in Alg. \ref{alg:hierarchical_sampler}, for $L=3$ levels.}
\label{fig:hierarchical_schematic}
\end{figure}

Let $q^{(\text{TL})}_{\ell}(\vec{x}_\ell^{(t+1)} | \vec{x}_\ell^{(t)})$ be the transition kernel for the process $\vec{x}_\ell^{(t)}\rightarrow \vec{x}_\ell^{(t+1)}$ implicitly defined by Alg.~\ref{alg:twolevelMC}. 
The key idea is now to use the algorithm recursively by using $q^{(\text{TL})}_{\ell-1}$ as the proposal distribution $q_{\ell-1}$ on level $\ell-1$. The process of picking $\vec{y}_{\ell-1}$ from $q_{\ell-1}(\cdot|\vec{x}_{\ell-1}^{(t)})=q_{\ell-1}^{(\text{TL})}(\cdot|\vec{x}_{\ell-1}^{(t)})$ in the first line of Alg.~\ref{alg:twolevelMC} then corresponds to a recursive call to the same algorithm on the next-coarser level. On the coarsest level ($\ell=0$) $\vec{y}_{0}$ is drawn with the standard Metropolis-Hastings step in Alg.~\ref{alg:standardMC} with corresponding transition kernel $q^{(\text{MH})}_0(\cdot|\vec{x}_0^{(t)})$; here we always assume that the proposal in this Metropolis Hastings step is generated with a symmetric method such as HMC.
\begin{algorithm}[H]
\caption{Two-level Metropolis Hastings step.\newline Input: level $\ell$, current sample $\vec{x}_{\ell}^{(t)}\sim\pi_{\ell}$, proposal distribution $q_{\ell-1}$ \newline Output: new sample $\vec{x}_{\ell}^{(t+1)}\sim\pi_{\ell}$}
\label{alg:twolevelMC}
\begin{algorithmic}[1]
  \State{Let $\vec{x}_\ell^{(t)}=[\tilde{\vec{x}}_{\ell}^{(t)},\vec{x}_{\ell-1}^{(t)}]$ and pick $\vec{y}_{\ell-1}$ from $q_{\ell-1}(\cdot|\vec{x}_{\ell-1}^{(t)})$.}
  \If{$\vec{x}_{\ell-1}^{(t+1)}=\vec{x}_{\ell-1}^{(t)}$ (coarse level proposal rejected)}
  \State{Set $\vec{x}_\ell^{(t+1)}\mapsfrom \vec{x}_\ell^{(t)}$}
  \Else
\State {Pick $\tilde{\vec{y}}_{\ell}$ from $\tilde{\pi}_{\ell}(\cdot\vert\vec{y}_{\ell-1})$ and let $\vec{y}_{\ell}=[\tilde{\vec{y}}_{\ell},\vec{y}_{\ell-1}]$.}
\State {Compute
\begin{equation*}
\begin{aligned}
\frac{\pi_{\ell}(\vec{y}_{\ell})}{\pi_{\ell}(\vec{x}^{(t)}_{\ell})}
\cdot
\frac{\tilde{\pi}_{\ell}(\tilde{\vec{x}}_\ell^{(t)}\vert {\vec{x}}^{(t)}_{\ell-1})}{\tilde{\pi}_{\ell}(\tilde{\vec{y}}_\ell\vert \vec{y}_{\ell-1})}
\cdot
\frac{\pi_{\ell-1}({\vec{x}}^{(t)}_{\ell-1})}{\pi_{\ell-1}(\vec{y}_{\ell-1})}
= \exp[-\Delta S_\ell]
\end{aligned}
\end{equation*}
with
\begin{equation*}
\begin{aligned}
\Delta S_\ell &:= S_{\ell}(\vec{y}_{\ell})-S_{\ell}(\vec{x}^{(t)}_{\ell})\\
&\quad+\;\; \tilde{S}_{\ell}([\tilde{\vec{x}}^{(t)}_{\ell},{\vec{x}}^{(t)}_{\ell-1}]) - \tilde{S}_{\ell}([\tilde{\vec{y}}_{\ell},\vec{y}_{\ell-1}])\\
&\quad+\;\; S_{\ell-1}({\vec{x}}^{(t)}_{\ell-1}) - S_{\ell-1}(\vec{y}_{\ell-1})\\
&\quad+\;\; \log\tilde{\mathcal{Z}}_\ell({\vec{x}}_{\ell-1}^{(t)})-\log\tilde{\mathcal{Z}}_\ell(\vec{y}_{\ell-1}).
\end{aligned}
\end{equation*}
}
      \If{$\Delta S_\ell<0$}
      \State{Set $\vec{x}_\ell^{(t+1)}\mapsfrom \vec{y}_\ell$}
      \Else
      \State{Draw uniformly distributed random $u\in[0,1)$.}
      \If{$u<\exp[-\Delta S_\ell]$}
       \State{Set $\vec{x}_\ell^{(t+1)}\mapsfrom \vec{y}_\ell$}
       \Else
       \State{Set $\vec{x}_\ell^{(t+1)}\mapsfrom \vec{x}_\ell^{(t)}$}
      \EndIf
      \EndIf
      \EndIf
    \end{algorithmic}
\end{algorithm}
More specifically, to construct a sequence of samples $\vec{x}_\ell^{(0)},\vec{x}_\ell^{(1)},\vec{x}_\ell^{(2)},\dots\in\Omega_{\ell}$ distributed according to $\pi_{\ell}$ we use Alg. \ref{alg:hierarchical_sampler}, which is illustrated schematically in Fig. \ref{fig:hierarchical_schematic}.
\begin{algorithm}[H]
\caption{Hierarchical delayed-acceptance sampler (recursive implementation).\newline Input: level $\ell$, current sample $\vec{x}_\ell^{(t)}\sim\pi_\ell$\newline Output: new sample $\vec{x}^{(t+1)}_\ell\sim\pi_\ell$}
\label{alg:hierarchical_sampler}
\begin{algorithmic}[1]
  \State{Generate $\vec{x}_{\ell}^{(t+1)}$ using Alg.~\ref{alg:twolevelMC} with level $\ell$, current sample $\vec{x}_{\ell}^{(t)}\sim\pi_{\ell}$ and proposal distribution
    \begin{equation*}
     q_{\ell-1}(\cdot|\vec{x}_{\ell-1}^{(t)})=
     \begin{cases}
       q^{(\text{MH})}_{0}(\cdot|\vec{x}_{0}^{(t)}) & \text{for $\ell=1$}\\
       q^{(\text{TL})}_{\ell-1}(\cdot|\vec{x}_{\ell-1}^{(t)}) & \text{for $\ell=2,3,\dots,L-1$}
     \end{cases}
  \end{equation*}
}
\end{algorithmic}
\end{algorithm}
Note that $\vec{x}_{\ell}^{(t+1)}=\vec{x}_{\ell}^{(t)}$ unless the proposals on all levels $0,1,\dots,\ell$ get accepted. At first sight this seems to imply that the overall acceptance probability of Alg.~\ref{alg:hierarchical_sampler} drops as the number of levels increases, and subsequent samples are highly correlated. However, this turns out not to be the case if the theories on subsequent level converge with $\ell \to \infty$: in this case the proposal from the two-level step in Alg.~\ref{alg:twolevelMC} is almost certainly accepted on finer levels. Our numerical experiments confirm this observation.

In practice, it is more convenient to implement Alg.~\ref{alg:hierarchical_sampler} iteratively, starting from the coarsest level. As discussed in Appendix \ref{sec:MLMC_cost_analysis}, the cost of executing Alg.~\ref{alg:hierarchical_sampler} on level $\ell$ can be bounded by a constant times the number of unknowns $d_\ell$ on this particular level. Observe also that setting $\ell=L-1$ in Alg.~\ref{alg:hierarchical_sampler} allows drawing a new sample $\vec{x}^{(t+1)}\sim\pi^*$ from the original fine level probability distribution defined in Eq.~\eqref{eqn:def_pifinelevel}.
\paragraph*{Relationship to the literature.}
The two-level %Metropolis-Hastings 
step in Alg.~\ref{alg:twolevelMC} is closely related to similar algorithms in \cite{Christen2005,Dodwell2015}. If the coarse level sample is drawn with an arbitrary Metropolis-Hastings kernel $q_{\ell-1}(\cdot|\vec{x}_{\ell-1}^{(t)})=q_{\ell-1}^{\text{(MH)}}(\cdot|\vec{x}_{\ell-1}^{(t)})$, then Alg.~\ref{alg:twolevelMC} above is a variant of the delayed-acceptance method in \cite[Alg.~1]{Christen2005} with proposal distribution
$q(\vec{y}_\ell|\vec{x}_\ell^{(t)})=\tilde{\pi}_{\ell}(\tilde{\vec{y}}_\ell|\vec{y}_{\ell-1})q^{(\text{MH})}_{\ell-1}(\vec{y}_{\ell-1}|\vec{x}_{\ell-1}^{(t)})$ and approximation $f_x^*(\vec{y}_\ell)=\tilde{\pi}_{\ell}(\tilde{\vec{y}}_\ell|\vec{y}_{\ell-1})\pi_{\ell-1}(\vec{y}_{\ell-1})$, recalling the notation $\vec{x}^{(t)}_\ell = [\tilde{\vec{x}}^{(t)}_{\ell},\vec{x}^{(t)}_{\ell-1}]$, $\vec{y}_\ell = [\tilde{\vec{y}}_{\ell},\vec{y}_{\ell-1}]$.

On the other hand, if the coarse level sample is drawn from the exact coarse level distribution, i.e. if $q(\cdot|\vec{x}_{\ell-1}^{(t)})=\pi_{\ell-1}(\cdot)$, Alg.~\ref{alg:twolevelMC} is identical to \cite[Alg.~2]{Dodwell2015}.
%%%%%%%%%%%%%%%%%%%%%%%%%%%%%%%%%%%%%%%%%%%%%%%%%%%%%
\subsubsection*{Multilevel Monte Carlo algorithm}
%%%%%%%%%%%%%%%%%%%%%%%%%%%%%%%%%%%%%%%%%%%%%%%%%%%%%
As discussed in the introduction, the multilevel Monte Carlo algorithm computes the quantity of interest $Q_0$ on the coarsest level and adds corrections to this by computing the \textit{difference} $Y_\ell$ of the observable on subsequent levels $\ell=1,2,\dots,L-1$ according to the telescoping sum in Eq. \eqref{eqn:telescoping_sum}. Since those differences $Y_\ell$ have a smaller variance, this allows shifting the cost to the coarser levels where samples can be generated cheaply. The original MLMC algorithm described in \cite{Giles2008} assumes that it is possible to draw independent identically distributed (i.i.d.) samples from a distribution on each level. For the Markov chain Monte Carlo setting considered here this is not possible since subsequent samples in the chain are correlated and, as discussed in \cite{Dodwell2015}, this introduces an additional bias. This bias can be reduced by constructing sequences $\vec{z}_{\ell}^{(0)}, \vec{z}_{\ell}^{(1)}, \vec{z}_{\ell}^{(2)}, \dots$ of samples for each level $\ell=0,\dots,L-1$ with Alg. \ref{alg:hierarchical_sampler} and sampling those sequences with sufficiently large sub-sampling rates $t_\ell$. The typical rule in statistics is to use twice the integrated autocorrelation time $\tau_{\text{int},\ell}$ to achieve (sufficient) independence. In our numerical experiments, we set $t_\ell=\lceil 2\tau_{\text{int},\ell}\rceil$ and observe that the additional bias due to computing the coarse level samples which are only approximately independent is comparable to the discretisation error.

\begin{figure}
\begin{center}
\includegraphics[width=0.5\linewidth]{\figdir/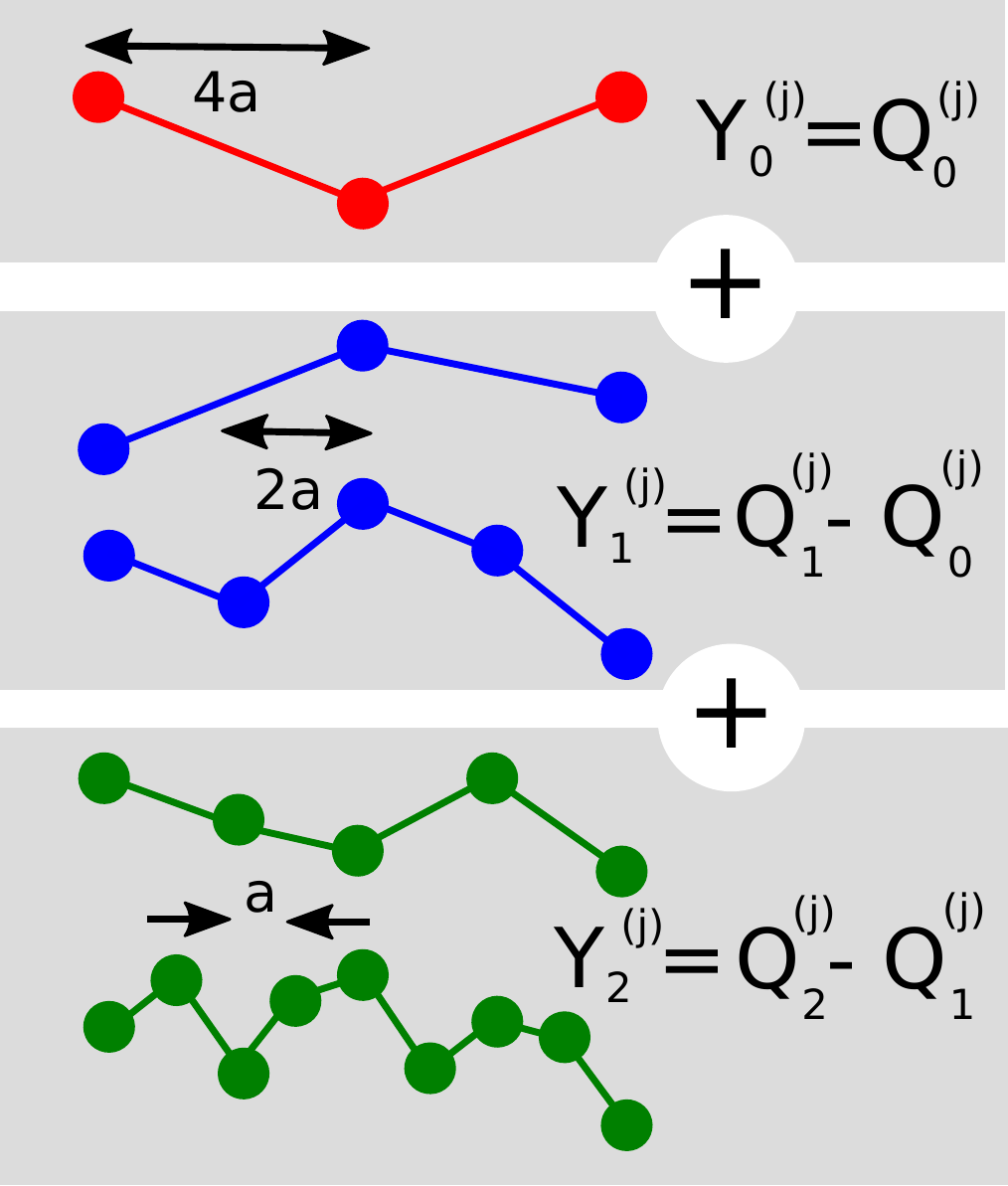}
\end{center}
\caption{Schematic visualisation of Multilevel Monte Carlo,  as described in Alg. \ref{alg:MLMC}, for $L=3$ levels.}
\label{fig:MLMC_schematic}
\end{figure}

The multilevel Monte Carlo algorithm which we use in this work is presented in Alg.~\ref{alg:MLMC} and visualised in Fig.~\ref{fig:MLMC_schematic}. It is similar to the multilevel algorithm in \cite{Dodwell2015}, but with the recursive independent sampler in \cite[Alg.~3]{Dodwell2015} replaced by the (suitably sub-sampled) hierarchical delayed-acceptance sampler in our Alg.~\ref{alg:hierarchical_sampler} above. Multilevel Monte Carlo computes 
\begin{equation}
\widehat{Q}^{\mathrm{MLMC}}_{L,\{N_\ell^{\mathrm{eff}}\}}=\sum_{\ell=0}^{L-1}\widehat{Y}_{\ell,N_\ell^{\mathrm{eff}}}\quad\text{with}\quad
\widehat{Y}_{\ell,N_\ell^{\mathrm{eff}}}=\frac{1}{N_\ell^{\mathrm{eff}}}\sum_{j=1}^{N_{\ell}^{\mathrm{eff}}} Y_\ell^{(j)},
\label{eqn:Yhatell_definition}
\end{equation}
which as unbiased estimator for the expectation $\mathbb{E}[Q]$ in Eq.~\eqref{eqn:telescoping_sum}.
On each level $\ell$, the number of samples is chosen to be
\begin{equation}
N_\ell^{\mathrm{eff}} = \max\left\{\,1\;,\;\epsstat^{-2} \left(\sum_{\ell=0}^{L-1}\sqrt{V_\ell\mathcal{C}_\ell^{\mathrm{eff}}}\right)\!\sqrt{\frac{V_\ell}{\mathcal{C}_\ell^{\mathrm{eff}}}}\,\right\},
\label{eqn:N_ell_eff}
\end{equation}
where $\mathcal{C}_\ell^{\mathrm{eff}}$ is the effective cost of generating an independent sample (taking into account autocorrelations) and $V_\ell=\mathrm{Var}[Y_\ell]$ is the variance of the quantity $Y_\ell$ on level $\ell$, which converges to zero as $\ell \to \infty$.
\begin{algorithm}[H]
\caption{Multilevel Monte Carlo.\newline Input: Number of levels $L$, number of samples per level $N_\ell^{\mathrm{eff}}$ and sub-sampling rates $t_\ell$ for $\ell=0,\dots,L-1$\newline Output: MLMC estimate for QoI.}
\label{alg:MLMC}
\begin{algorithmic}[1]
\For{level $\ell=0,\dots,L-1$}
\For{$j=1,\dots,N_{\ell}^{\textrm{eff}}$}
\If{$\ell=0$}
\State{Create a new sample $\vec{x}_{0}^{(t+t_0)}$ from $\vec{x}_{0}^{(t)}$ with a}
\StateX{standard Metropolis-Hastings method.}
\State{Compute $Y_{0}^{(j)}=Q_0(\vec{x}_{0}^{(t+t_0)})$}
\Else
\State{Create a new sample $\vec{x}_{\ell}^{(t+1)}$  from $\vec{x}_\ell^{(t)}$ with}
\StateX{Alg.~\ref{alg:twolevelMC} and $q_{\ell-1}(\cdot|\vec{x}_\ell^{(t)})=\pi_{\ell-1}$\,;}\vspace{1ex}
\StateX{In practice, use $t_{\ell-1}$ steps of Alg.~\ref{alg:hierarchical_sampler} to compute}
\StateX{an approximately independent sample $\vec{z}_{\ell-1}^{(t+t_{\ell-1})}$}
\StateX{on level $\ell-1$.}
\vspace{1ex}
\State{Compute $Y_{\ell}^{(j)}=Q_\ell(\vec{x}_{\ell}^{(t+1)})-Q_{\ell-1}(\vec{z}_{\ell-1}^{(t+t_{\ell-1})})$.}
\EndIf
\EndFor
\EndFor
\State{Compute the MLMC estimator defined in Eq. \eqref{eqn:Yhatell_definition}.}
\end{algorithmic}
\end{algorithm}
We now discuss how the cost of Alg. \ref{alg:MLMC} increases as the tolerance on the total error is tightened, assuming for simplicity i.i.d. samples on all levels. Let the exact value of the observable in the continuum limit be $Q^{\mathrm{exact}}$. The total mean square error of the multilevel Monte Carlo estimator defined in Eq.~\eqref{eqn:Yhatell_definition} can be expanded
\begin{equation}
\begin{aligned}
&\mathbb{E}\Big[\Big(\widehat{Q}^{\text{MLMC}}_{L,\{N_\ell^{\mathrm{eff}}\}}-Q^{\mathrm{exact}}\Big)^2\Big]\\
%&= \mathbb{E}\Big[\Big(\big(\widehat{Q}^{\text{MLMC}}_{L,\{N_\ell^{\mathrm{eff}}\}}-\mathbb{E}\big[\widehat{Q}^{\mathrm{MLMC}}_{L,\{N_\ell^{\mathrm{eff}}\}}\big]\big)\\
%&\qquad+\;\;\big(
%\mathbb{E}\big[\widehat{Q}^{\mathrm{MLMC}}_{L,\{N_\ell^{\mathrm{eff}}\}}\big]
%-Q^{\mathrm{exact}}\big)\Big)^2\Big]\\
&= \mathrm{Var}\Big[\widehat{Q}^{\mathrm{MLMC}}_{L,\{N_\ell^{\mathrm{eff}}\}}\Big] + 
\left(\mathbb{E}\Big[\widehat{Q}^{\mathrm{MLMC}}_{L,\{N_\ell^{\mathrm{eff}}\}}\Big]-Q^{\mathrm{exact}}\right)^2\,,
\end{aligned}
\label{eqn:total_error}
\end{equation}
where the first term in the final line of Eq.~\eqref{eqn:total_error} is the squared statistical error, whereas the second term is the squared discretisation error. An easy calculation shows that choosing $N_\ell^{\mathrm{eff}}$ as in Eq.~\eqref{eqn:N_ell_eff} guarantees that 
\begin{equation*}
\mathrm{Var}\left[\widehat{Q}^{\mathrm{MLMC}}_{L,\{N_\ell^{\mathrm{eff}}\}}\right] = \sum_{\ell=0}^{L-1} \frac{V_\ell}{N_{\ell}^{\mathrm{eff}}} \le \epsstat^2.
\end{equation*}

To analyse the complexity we assume that
\begin{enumerate}
    \item[(i)] the discretisation error is of order $\mathcal{O}(a_\ell^\alpha)$, 
    \item[(ii)] $V_\ell$ converges with order $\mathcal{O}(a_\ell^\beta)$ for some $\beta>0$,
    \item[(iii)] the integrated autocorrelation times of $Y_\ell$, and thus also the sub-sampling rates $t_\ell$, can be bounded by a constant independent of $\ell$ such that the cost $\mathcal{C}_\ell^{\mathrm{eff}}$ of generating an independent sample does not grow faster than the number of unknowns $d_\ell$ for all $\ell$.
\end{enumerate}
As shown in more detail in Appendix \ref{sec:MLMC_cost_analysis}, it is then possible to choose the number of levels $L$ such that the discretisation error in Eq.~\eqref{eqn:total_error} does not exceed $\epsdisc$. As a consequence, the cost $\mathcal{C}_{\mathrm{MLMC}}(\epsdisc,\epsstat)$ of computing the MLMC estimator in Eq.~\eqref{eqn:Yhatell_definition} with a statistical error less than $\epsstat$ and a discretisation error less than $\epsdisc$ has the following computational complexity:
\begin{equation}
\mathcal{C}_{\mathrm{MLMC}} = 
\begin{cases}
\mathcal{O}\left(\epsstat^{-2}+\epsdisc^{-1/\alpha}\right) & \text{for $\beta>1$,}\\
\mathcal{O}\left(\epsstat^{-2}\vert\log \epsdisc\vert^2+\epsdisc^{-1/\alpha}\right) & \text{for $\beta=1$,}\\
\mathcal{O}\left(\epsstat^{-2}\epsdisc^{-\frac{1-\beta}{\alpha}}+\epsdisc^{-1/\alpha}\right) & \text{for $\beta<1$.}
\end{cases}
\label{eqn:MLMC_complexity}
\end{equation}
For the choice $\epsdisc=\epsstat=\epsilon/\sqrt{2}$, the total mean square error in Eq.~\eqref{eqn:total_error} does not exceed $\epsilon^2$ and Eq.~\eqref{eqn:MLMC_complexity} becomes
\begin{equation}
\mathcal{C}_{\mathrm{MLMC}}(\epsilon) = \begin{cases}
\mathcal{O}\left(\epsilon^{-2}\right) & \text{for $\beta>1$}\\
\mathcal{O}\left(\epsilon^{-2}\vert\log \epsilon\vert^2\right) & \text{for $\beta=1$}\\
\mathcal{O}\left(\epsilon^{-2-\frac{1-\beta}{\alpha}}\right) & \text{for $\beta<1$}
\end{cases},\label{eqn:MLMC_complexity_epsilon}
\end{equation}
which is a special case of the well-known estimate in \cite{Giles2008}.

However, the samples created by Alg.~\ref{alg:MLMC} on each of the levels $\ell$ are generated with a Markov chain and thus only asymptotically distributed according to $\pi_\ell$. As discussed in \cite{Dodwell2015}, the complexity analysis can be modified to address this issue, leading to an additional factor $|\log \epsdisc|$ in Eqs.~\eqref{eqn:MLMC_complexity} and \eqref{eqn:MLMC_complexity_epsilon}. This seems to be not visible in the numerical results below or in \cite{Dodwell2015} (at least pre-asymptotically).
%%%%%%%%%%%%%%%%%%%%%%%%%%%%%%%%%%%%%%%%%%%%%%%%%%%%%
\subsection{Memory requirements}
%%%%%%%%%%%%%%%%%%%%%%%%%%%%%%%%%%%%%%%%%%%%%%%%%%%%%
Although the one dimensional quantum mechanical problems considered here do not require significant storage, the memory requirements of the algorithms introduced in this paper need to be considered in addition to their runtimes. This is particularly important for simulations of higher dimensional quantum field theories on modern many-core architectures where the memory per compute core is limited.

As discussed in detail in Appendix \ref{sec:memory_requirements}, on a given level the hierarchical sampler in Alg.~\ref{alg:hierarchical_sampler} requires less memory than a standard Metropolis Hastings method with a HMC proposal distribution. The memory footprint of the multilevel Monte Carlo method in Alg.~\ref{alg:MLMC} is less than three times that of a HMC based Metropolis Hastings algorithm.
%%%%%%%%%%%%%%%%%%%%%%%%%%%%%%%%%%%%%%%%%%%%%%%%%%%%%
\section{Quantum mechanical model systems}\label{sec:model_systems}
%%%%%%%%%%%%%%%%%%%%%%%%%%%%%%%%%%%%%%%%%%%%%%%%%%%%%
To demonstrate the performance of the methods discussed in the previous section we consider two non-trivial quantum mechanical problems.
%%%%%%%%%%%%%%%%%%%%%%%%%%%%%%%%%%%%%%%%%%%%%%%%%%%%%
\subsection{Non-symmetric double-well potential}
\begin{figure}
\begin{center}
\includegraphics[width=\linewidth]{\figdir/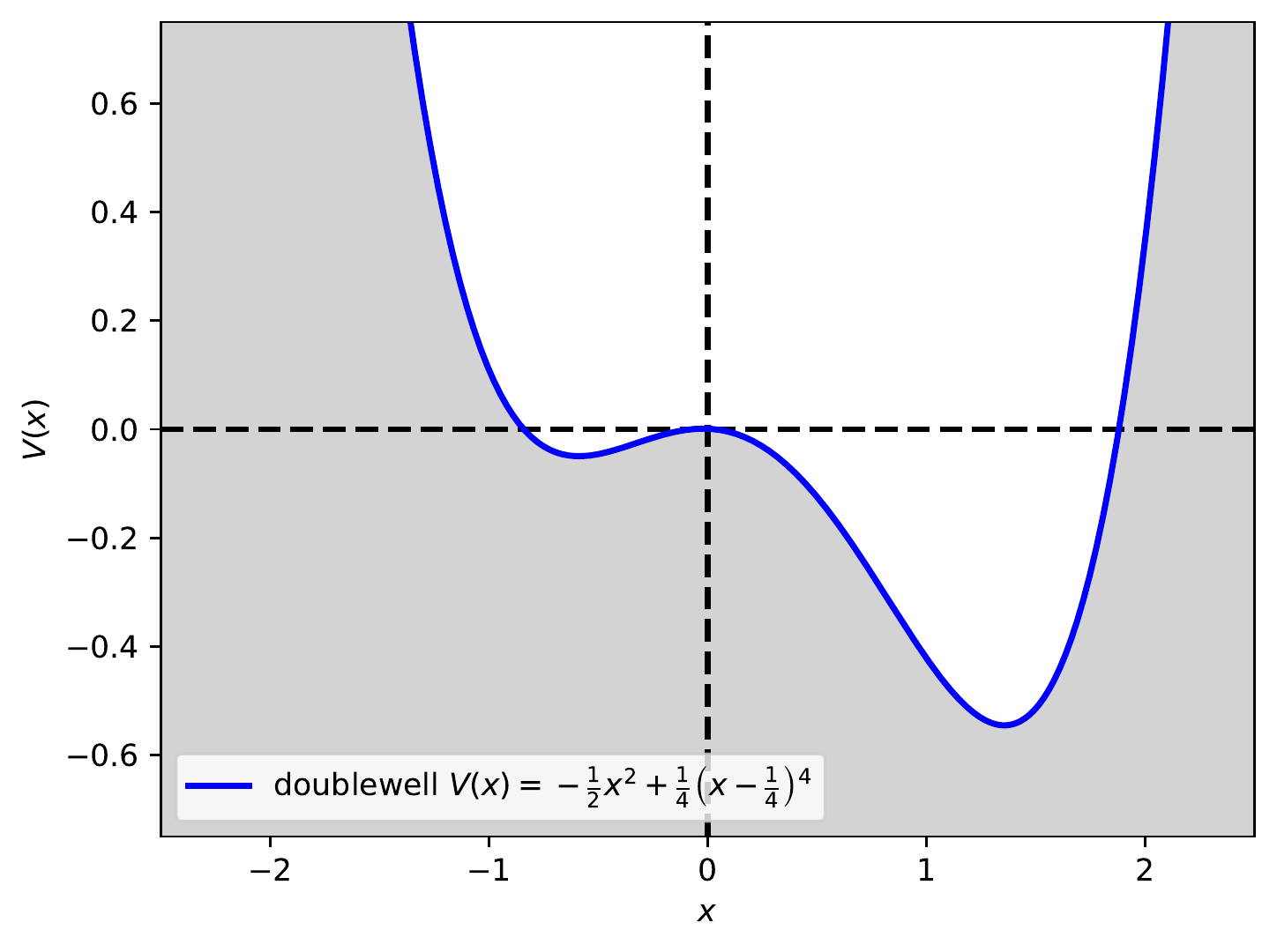}
\end{center}
\caption{Double well potential used for numerical experiments}
\label{fig:potential_doublewell}
\end{figure}

%%%%%%%%%%%%%%%%%%%%%%%%%%%%%%%%%%%%%%%%%%%%%%%%%%%%%
The first system describes a particle with mass $m_0$ moving subject to a non-symmetric double-well potential $V(x)=\frac{m_0\mu^2}{2}x^2+\frac{\lambda}{4}(x-\eta)^4$. Fig. \ref{fig:potential_doublewell} shows this potential for the choice of parameters that were used in our numerical experiments, namely $m_0=1$, $\mu^2=-1$, $\lambda=1$, $\eta=\frac{1}{4}$.
In the Euclidean time formulation of the path integral the corresponding Lagrangian is
\begin{equation}
\mathcal{L}(x(t)) = \frac{m_0}{2}\left(\frac{dx}{dt}\right)^2 + \frac{m_0\mu^2}{2}x^2+\frac{\lambda}{4}(x-\eta)^4\label{eqn:doublewell_lagrangian}
\end{equation}
where $x(t)\in\mathbb{R}$. For a given path $\vec{x}=(x_0,x_1,\dots,x_{d-1})\in\mathbb{R}^d$ the discretised lattice action is
\begin{equation}
  \begin{aligned}
    S(\vec{x}) &= a \sum_{j=0}^{d-1}\Bigg\{\frac{m_0}{2}\left(\frac{x_j-x_{j-1}}{a}\right)^2\\
    &\qquad\qquad+\;\frac{m_0\mu^2}{2}x_i^2+\frac{\lambda}{4}(x_i-\eta)^4\Bigg\}.
    \end{aligned}
\label{eqn:doublewell_action}
\end{equation}
The observable we consider is the average squared displacement
\begin{equation}
  Q(\vec{x})=\frac{1}{d}\sum_{j=0}^{d-1}x_j^2.\label{eqn:QoI_doublewell}
\end{equation}
Note that since points on the lattice are correlated with a correlation length which is constant in physical units, the variance of this observable does not go to zero in the continuum limit. In other words, the sampling error is not automatically reduced on finer lattices.
%%%%%%%%%%%%%%%%%%%%%%%%%%%%%%%%%%%%%%%%%%%%%%%%%%%%%
\subsubsection{Coarse level action}\label{sec:coarse_level_action_doublewell}
%%%%%%%%%%%%%%%%%%%%%%%%%%%%%%%%%%%%%%%%%%%%%%%%%%%%%
Coarse grained versions $S_{\ell}$ of the action in Eq. \eqref{eqn:doublewell_action} are obtained by re-discretising the Lagrangian in Eq. \eqref{eqn:doublewell_lagrangian} on the lattice $\mathcal{T}_\ell$ with \mbox{$d_\ell=2^{\ell-L+1}d$} points and lattice spacing $a_\ell=2^{L-1-\ell}a$ on level $\ell$ to obtain
\begin{equation*}
  \begin{aligned}
    S_\ell(\vec{x}) &= a_\ell \sum_{j=0}^{d_\ell-1}\Bigg\{\frac{m_0}{2}\left(\frac{x_j-x_{j-1}}{a_\ell}\right)^2\\
    &\qquad\qquad+\;\frac{m_0\mu^2}{2}x_j^2+\frac{\lambda}{4}(x_j-\eta)^4\Bigg\}.
    \end{aligned}
\end{equation*}
To construct the action $\tilde{S}_\ell$ defined in Eq. \eqref{eqn:pi_tilde}, observe that 
\begin{equation}
\begin{aligned}
\pi_\ell(\vec{x}) &= \pi^{\mathrm{even}}_\ell(x_0,x_2,\dots,x_{d_\ell-2})\times\\
&\qquad\times\;\;\prod_{j=0}^{d_{\ell-1}-1} \pi^{\mathrm{odd}}_\ell(x_{2j+1}\vert x_{2j},x_{2j+2})
\end{aligned}\label{eqn:pi_odd_even}
\end{equation}
where $\pi^{\mathrm{even}}_\ell$ is the marginal distribution of the even points
\begin{equation*}
\pi^{\mathrm{even}}_\ell(x_0,x_2,\dots,x_{d_\ell-2}) = \int_{\mathcal{D}} \dots \int_{\mathcal{D}}\pi_\ell(\vec{x}) dx_1\;dx_3\dots dx_{d_\ell-1}
\end{equation*}
and
\begin{equation}
\pi_\ell^{\mathrm{odd}}(x_{2j+1}\vert x_{2j},x_{2j+2}) = \mathcal{Z}^{-1}_{\ell,j} \exp\left[-W_\ell(x_{2j+1}\vert x_{2j},x_{2j+2})\right].\label{eqn:pi_odd}
\end{equation}
Here $W_\ell$ is defined for arbitrary values $x_-$, $x_+$ as
\begin{equation*}
\begin{aligned}
W_\ell(x\vert x_-,x_+) &= \frac{m_0}{a_\ell}\left(x^2-(x_-+x_+)x\right)\\
&\qquad+\;\; a_\ell\left(\frac{m_0\mu^2}{2}x^2 + \frac{\lambda}{4}\left(x-\eta\right)^4\right).
\end{aligned}
\end{equation*}
and $\mathcal{Z}_{\ell,j}=\mathcal{Z}_{\ell,j}(x_{2j},x_{2j+2})$ is a normalisation constant which depends on $x_{2j}, x_{2j+2}$. The distribution in Eq. \eqref{eqn:pi_odd} can be approximated by a Gaussian by writing
\begin{equation*}
W_\ell(x\vert x_-,x_+) \approx G_\ell(x\vert x_-,x_+)
\end{equation*}
with
\begin{equation*}
G_\ell(x\vert x_-,x_+) = \frac{m_0}{a_\ell}\sigma_\ell(x_-,x_+) \left(x-\zeta_\ell(x_-,x_+)\right)^2,
\end{equation*}
where $\zeta_\ell=\zeta_\ell(x_-,x_+)$ is the minimum of $W_\ell(x\vert x_-,x_+)$ and satisfies the non-linear equation
\begin{equation}
\left(1+\frac{1}{2}a_\ell^2\mu^2\right)\zeta_\ell + a_\ell^2\frac{\lambda}{2m_0}(\zeta_\ell-\eta)^3 = \frac{x_-+x_+}{2}.\label{eqn:zeta_doublewell}
\end{equation}
In the code $\zeta_\ell$ is found by a small number of fixed point iterations of Eq. \eqref{eqn:zeta_doublewell}, using $\overline{x}=(x_-+x_+)/2$ as a starting guess.
Further, $2m_0\sigma_\ell/a_\ell$ is the curvature of the function $W_\ell$, evaluated at the point $\overline{x}\approx \zeta_\ell(x_-,x_+)$, i.e.
\begin{equation*}
\begin{aligned}
\frac{2m_0}{a_\ell}\sigma_\ell(x_-,x_+) &= \frac{\partial^2 W_\ell}{\partial x^2}(\overline{x}|x_-,x_+)\\
&= \frac{2m_0}{a_\ell}\left(1+\frac{1}{2}a_\ell^2\left(\mu^2 +\frac{3\lambda}{m_0} (\overline{x}-\eta)^2\right)\right)\\
&\approx \frac{\partial^2 W_\ell}{\partial x^2}(\zeta_\ell(x_-,x_+)|x_-,x_+).
\end{aligned}
\end{equation*}
Now write $\vec{x}=[\tilde{\vec{x}},\vec{x}']$ as in Eq. \eqref{eqn:x_decomposition}. Given $\zeta_\ell(x'_{j},x'_{j+1})$ and $\sigma_\ell(x'_{j},x'_{j+1})$ for all $j=0,1,\dots,d_{\ell-1}-1$, we can then construct
\begin{equation*}
\tilde{S}_\ell([\tilde{\vec{x}},\vec{x}']) = \sum_{j=0}^{d_{\ell-1}-1} G_\ell(\tilde{x}_j\vert x'_j,x'_{j+1}).
\end{equation*}
The resulting probability density $\tilde{\pi}_\ell(\cdot\vert \vec{x}')$ defined in Eq. \eqref{eqn:pi_tilde} is a multivariate normal distribution with diagonal covariance matrix, which can be easily sampled; the normalisation constant in Eq. \eqref{eqn:pi_tilde} is
\begin{equation*}
\tilde{\mathcal{Z}}_{\ell}(\vec{x}') = \sqrt{\left(\frac{4\pi m_0}{a_\ell}\right)^{d_{\ell-1}} \prod_{j=0}^{d_{\ell-1}-1} \sigma_\ell(x'_{j},x'_{j+1})}.
\end{equation*}
%%%%%%%%%%%%%%%%%%%%%%%%%%%%%%%%%%%%%%%%%%%%%%%%%%%%%
\subsection{Topological oscillator}
%%%%%%%%%%%%%%%%%%%%%%%%%%%%%%%%%%%%%%%%%%%%%%%%%%%%%
The second model system is the topological oscillator, described for example in \cite{Ammon2016}. This is an interesting problem since it has a topological quantum number which can only take on integer values. The Lagrangian is
\begin{equation}
\mathcal{L}(x,t) = \frac{I_0}{2} \left(\frac{dx}{dt}\right)^2\label{sec:action_rotor_cont}
\end{equation}
where now crucially $x\in[-\pi,\pi)$, i.e. the particle is confined to a finite interval.  The Lagrangian in Eq. \eqref{sec:action_rotor_cont} can be obtained from the action of a free particle with mass $m_0$ confined to a circle with radius $R$,
\begin{equation*}
\mathcal{L}(y,z,t) = \frac{m_0}{2}\left(\left(\frac{dy}{dt}\right)^2+\left(\frac{dz}{dt}\right)^2\right),
\end{equation*}
with $(y,z)\in\mathbb{R}^2$, $y^2+z^2=R^2$ by setting $y(t)=R\cos(x(t))$, $z(t)=R\sin(x(t))$ and $I_0=R^2m_0$. The form of the discretised action chosen here is
\begin{equation*}
S(\vec{x}) = \frac{I_0}{a}\sum_{j=0}^{d-1}\left(1-\cos(x_{j}-x_{j-1})\right).
\end{equation*}
As above we used periodic boundary conditions $x_d=x_0$. Note that
\begin{equation*}
\frac{1-\cos(x_{j}-x_{j-1})}{a^2} = \frac{1}{2} \left(\frac{dx}{dt}\right)^2+\mathcal{O}\left(a^2\right).
\end{equation*}
For a given path $x(t)$ the topological charge $q(x)$ of the system describes the number of complete revolutions during the time period $T$. Mathematically it is defined as
\begin{equation*}
q(x(t)) = \frac{1}{2\pi}\int_0^T \frac{dx(t)}{dt}\;dt\in\mathbb{Z}.
\end{equation*}
For the discretised system this becomes
\begin{equation*}
q(\vec{x}) = \frac{1}{2\pi}\sum_{j=0}^{d-1}\Big\{ (x_j-x_{j-1})\mod [-\pi,\pi)\Big\}\in\mathbb{Z}.
\end{equation*}
Following the notation in \cite{Ammon2016}, for any $x\in \mathbb{R}$ the quantity $z=x\mod[-\pi,\pi)$ is defined as $z=x+2\pi k$ with $k\in \mathbb{Z}$ such that $-\pi\le z< \pi$.
The observable we consider is the topological susceptibility
\begin{equation}
Q(\vec{x}) = \chi_t(\vec{x}) = \frac{q^2(\vec{x})}{T}.\label{eqn:qoi_rotor}
\end{equation}
Defining $\xi:=T/I_0$ and $z:=a/I_0$, a tedious but straightforward calculation shows that the expectation value of $\chi_t$ for finite $a$, $T$ is given by
\ifpreprint % PREPRINT
\begin{strip}
\else % PREPRINT
  \begin{widetext}
\fi % PREPRINT
\begin{equation}
\mathbb{E}[\chi_t] = \frac{1}{4\pi^2 I_0}\left(1-\xi\widehat{\Sigma}_2(\xi)+\left[\frac{1}{2}-\xi\widehat{\Sigma}_2(\xi)+\frac{1}{4}\xi^2\left(\widehat{\Sigma}_4(\xi)-\widehat{\Sigma}_2(\xi)^2\right)\right]z\right)+\mathcal{O}(z^2)\;\xrightarrow{a\rightarrow 0}\;\frac{1-\xi\widehat{\Sigma}_2(\xi)}{4\pi^2 I_0}\;\xrightarrow{T\rightarrow\infty}\;\frac{1}{4\pi^2 I_0},\label{eqn:chi_t_theory}
\end{equation}
\ifpreprint % PREPRINT
\end{strip}
\else % PREPRINT
\end{widetext}
\fi % PREPRINT
where for any $p\in\mathbb{N}$, $\xi>0$ the function $\widehat{\Sigma}_p$ is defined as
\begin{xalignat}{2}
\Sigma_p(\xi) &:= \sum_{m\in\mathbb{Z}}m^p \exp\left[-\frac{1}{2}\xi m^2\right],&
\widehat{\Sigma}_p(\xi) &:= \frac{\Sigma_p(\xi)}{\Sigma_0(\xi)}.\label{eqn:Sigma_hat}
\end{xalignat}
Eq. \eqref{eqn:chi_t_theory} allows the calculation of the constant $\Delta_0$ in the Taylor expansion $\mathbb{E}[\chi_t]=\mathbb{E}[\chi_t(a=0)] + \Delta_0 a+\mathcal{O}(a^2)$ of the topological susceptibility. In other words, we can work out the bias for a given lattice spacing. This will also allow us to balance the discretisation- and statistical- errors in the MLMC estimator if we choose $\epsdisc=\epsstat$. In the continuum limit ($a\rightarrow 0$) the variance of $\chi_t$ can be shown to be
\begin{equation}
\operatorname{Var}[\chi_t] = \mathbb{E}\left[(\chi_t-\mathbb{E}[\chi_t])^2\right]
    = \frac{R(4\pi^2/\xi)}{8\pi^4I_0^2} \xrightarrow{T\rightarrow \infty} \frac{1}{8\pi^4I_0^2}\label{eqn:variance_rotor_exact}
\end{equation}
with the function $R$ defined by
\begin{equation*}
R(\zeta):=\frac{1}{2}\zeta^2\left(\widehat{\Sigma}_4(\zeta)-\widehat{\Sigma}_2(\zeta)^2\right).
\end{equation*}
%%%%%%%%%%%%%%%%%%%%%%%%%%%%%%%%%%%%%%%%%%%%%%%%%%%%%
\subsubsection{Coarse level action}\label{sec:coarse_level_action_rotor}
%%%%%%%%%%%%%%%%%%%%%%%%%%%%%%%%%%%%%%%%%%%%%%%%%%%%%
For the topological oscillator the coarse level action is
\begin{equation*}
S_\ell(\vec{x}) = \frac{I_{0}^{(\ell)}}{a_\ell}\sum_{j=0}^{d_\ell-1}\left(1-\cos(x_{j}-x_{j-1})\right),
\end{equation*}
where the moment of inertia $I^{(\ell)}_{0}$ is level dependent. In the simplest case one could simply set $I^{(\ell)}_{0}=I_0$ for all $\ell=0,1,\dots,L-1$. However, as will be shown below, performance can be improved significantly by using a perturbative matching procedure to construct $I^{(\ell)}_{0}$ on the coarser levels. To obtain $\tilde{S}_\ell$, rewrite $\pi_\ell$ as in Eqs. \eqref{eqn:pi_odd_even} and \eqref{eqn:pi_odd}, where now
\begin{equation*}
W_\ell(x\vert x_-,x_+) = \overline{W}_\ell(x\vert x_-,x_+) + 2-\frac{1}{2}\sigma_{\ell}(x_-,x_+)
\end{equation*}
with
\begin{equation*}
\begin{aligned}
  \overline{W}_\ell(x\vert x_-,x_+) &= \frac{I_{0}^{(\ell)}}{a_\ell}\sigma_\ell(x_-,x_+)\sin^2\left(\frac{x-\zeta_\ell(x_-,x_+)}{2}\right)\\
\sigma_\ell(x_-,x_+)&=4\left\vert\cos\left(\frac{x_+-x_-}{2}\right)\right\vert,\\
\tan\zeta_\ell(x_-,x_+) &= \frac{\sin(x_+)+\sin(x_-)}{\cos(x_+)+\cos(x_-)}.
\end{aligned}
\end{equation*}
Again write $\vec{x}=[\tilde{\vec{x}},\vec{x}']$ as in Eq. \eqref{eqn:x_decomposition}, and given $\zeta_\ell(x'_{j},x'_{j+1})$ and $\sigma_\ell(x'_{j},x'_{j+1})$ for all $j=0,1,\dots,d_{\ell-1}-1$ construct
\begin{equation*}
\tilde{S}_\ell([\tilde{\vec{x}},\vec{x}']) = \sum_{j=0}^{d_{\ell-1}-1} \overline{W}_\ell(\tilde{x}_j\vert x'_j,x'_{j+1}).
\end{equation*}
The normalisation constant in Eq. \eqref{eqn:pi_tilde} is
\begin{equation*}
\begin{aligned}
\mathcal{Z}_\ell(\vec{x}') &= (2\pi)^{d_{\ell-1}} \exp\left[-\frac{I_0^{(\ell)}}{2a_\ell}\sum_{j=0}^{d_{\ell-1}-1}\sigma_\ell(x'_j,x'_{j+1})\right]\\
&\quad\times\;\;\prod_{j=0}^{d_{\ell-1}-1}B_0\left(\frac{I_0^{(\ell)}}{2a_\ell}\sigma_\ell(x'_j,x'_{j+1})\right)
\end{aligned}
\end{equation*}
where $B_0$ is the zero-order modified Bessel function of the first kind. The resulting probability density $\tilde{\pi}_\ell$ is the product of one-dimensional densities of the form
\begin{equation}
\begin{aligned}
p_{\sigma,\delta x}(x) &= \mathcal{Z}_\sigma^{-1} \exp\left[-2\sigma \sin^2\left(\frac{x-\delta x}{2}\right)\right]\quad\text{with}\\
\mathcal{Z}_\sigma &= 2\pi e^{-\sigma}B_0(\sigma),
\label{eqn:expsin2_distribution}
\end{aligned}
\end{equation}
which can be easily sampled for arbitrary values of $\sigma$ and $\delta x$. In our code we find that rejection sampling with a suitable Gaussian envelope (as described in Appendix \ref{sec:rejection_sampling}) gives good results.
%%%%%%%%%%%%%%%%%%%%%%%%%%%%%%%%%%%%%%%%%%%%%%%%%%%%%
\subsubsection{Coarse level matching}\label{sec:coarse_level_matching}
%%%%%%%%%%%%%%%%%%%%%%%%%%%%%%%%%%%%%%%%%%%%%%%%%%%%%
Ideally, the coarse level actions should be obtained by recursively integrating out the modes that can be represented on a given lattice, but not on the next coarser one. In other words, $S_{\ell-1}$ is an effective action obtained from $S_\ell$. While for an arbitrary action this can not be done exactly, an approximate effective action can be constructed by a perturbative renormalisation group transformation or through (approximate) matching. Here we follow the latter procedure for the topological oscillator to adjust the moment of inertia $I_{0}^{(\ell)}$ on the coarser levels, starting from the physical value $I_0=I_0^{(L-1)}$ on the finest lattice. Let $\chi_t(a,I_0,T)$ be the topological susceptibility calculate for a given $I_0$, $T$ and lattice spacing $a$, and recall that we can compute $\chi_t(a,I_0,T)$ up to corrections of $\mathcal{O}((a/I_0)^2)$. We now require that
\begin{equation*}
\chi_t(a_{\ell-1},I_0^{(\ell-1)},T) = \chi_t(a_\ell,I_0^{(\ell)},T) + \mathcal{O}((a_\ell/I_0^{(\ell)})^2)
\end{equation*}
for all $\ell=1,\dots,L-1$. Using Eq. \eqref{eqn:chi_t_theory} this gives
\begin{equation*}
  I_0^{(\ell-1)} = \left(1+\frac{a_\ell}{I_0^{(\ell)}}\cdot \delta_I\left(T/I_0^{(\ell)}\right) \right)I_0^{(\ell)} + \mathcal{O}((a_\ell/I_0^{(\ell)})^2).
\end{equation*}
with
\begin{equation*}
  \delta_I(\xi) = \frac{1}{2}\cdot \frac{1-2\xi\widehat{\Sigma}_2(\xi)+\frac{1}{2}\xi^2\left(\widehat{\Sigma}_4(\xi)-\widehat{\Sigma}_2(\xi)^2\right)}{1-2\xi\widehat{\Sigma}_2(\xi)+\xi^2\left(\widehat{\Sigma}_4(\xi)-\widehat{\Sigma}_2(\xi)^2\right)}
\end{equation*}
and $\widehat{\Sigma}_p$ as defined in Eq. \eqref{eqn:Sigma_hat}. As the following numerical results show, computing $I_\ell^{(0)}$ with this approximate coarse level matching procedure significantly improves performance both for the hierarchical sampler in Alg. \ref{alg:hierarchical_sampler} and the MLMC method in Alg. \ref{alg:MLMC}.
%%%%%%%%%%%%%%%%%%%%%%%%%%%%%%%%%%%%%%%%%%%%%%%%%%%%%
\section{Results}\label{sec:results}
%%%%%%%%%%%%%%%%%%%%%%%%%%%%%%%%%%%%%%%%%%%%%%%%%%%%%
We now quantify the performance gains of the numerical algorithms described above. All results were generated with a C++ code developed by the authors which is freely available at\begin{center}\url{https://bitbucket.org/em459/mlmcpathintegral/}\end{center} The reported runtimes were obtained by running a sequential version of the code (which was compiled with version 18.5.274 of the Intel C compiler) on a single core of an Intel E5-2650 v2 (2.60 GHz) CPU.

For all numerical results we set $T=4$; as remarked above we do not consider finite-volume errors here, i.e. we assume that the exact value is the expectation value of the observable in the limit $a\rightarrow 0$ at a given $T$. As can be seen from Eq. \eqref{eqn:chi_t_theory}, finite-volume errors are exponentially suppressed for the topological oscillator\footnote{To see this, note that for $T\gg I_0$ the leading order term in the sum $\hat{\Sigma}_2(T/I_0)$ defined in Eq. \eqref{eqn:Sigma_hat} is $2e^{-T/(2I_0)}$.}. For the double-well potential the mass is set to $m_0=1.0$ whereas the moment of inertia for the topological oscillator is $I_0=0.25$.
%%%%%%%%%%%%%%%%%%%%%%%%%%%%%%%%%%%%%%%%%%%%%%%%%%%%%
\subsection{Autocorrelations}\label{sec:results_autocorrelations}
%%%%%%%%%%%%%%%%%%%%%%%%%%%%%%%%%%%%%%%%%%%%%%%%%%%%%
To quantify the significant reduction of autocorrelations which is achieved by hierarchical sampling, we measure the integrated autocorrelation time $\tau_{\mathrm{int}}$ for the single level Metropolis-Hastings algorithm (Alg. \ref{alg:standardMC}) if either a simple HMC algorithm or the hierarchical delayed acceptance sampler in Alg. \ref{alg:hierarchical_sampler} is used. We refer to the first method as ``StMC'' from now on, whereas the latter is denoted as ``HSMC''. In the latter case the number of levels is chosen such that the coarsest level is fixed and always has $d_0=16$ points for the double-well potential and $d_0=32$ for the topological oscillator (corresponding to lattice spacings of $a_0=0.25$ and $a_0=0.125$ respectively). A HMC sampler is used to generate proposals on the coarsest level. In all cases (i.e. either on the fine level for the StMC method or on the coarsest level for HSMC) 100 HMC steps are carried out and the size of the HMC timestep is tuned such that the acceptance probability of the HMC sampler is close to $80\%$. We implemented a simple HMC method based on a symplectic leapfrog integrator.
The integrated autocorrelation time defined in Eq. \eqref{eqn:tauint_definition} is estimated by measuring the QoI for $N=10^5$ samples and computing
\begin{equation*}
  \begin{aligned}
   \widehat{\tau}_{\mathrm{int}} &= 1+2\sum_{s=1}^{W} \frac{\widehat{\rho}(s)}{\widehat{\rho}(0)}\approx\tau_{\mathrm{int}}\qquad\text{with}\\
   \widehat{\rho}(s) &= \frac{1}{N-s}\sum_{j=1}^{N-s} Q^{(j)} Q^{(j+s)}\\
   &\approx\mathbb{E}[Q(\vec{x}^{(t_{\text{meas}})})Q(\vec{x}^{(t_{\text{meas}}+s)})],
    \end{aligned}
  \end{equation*}
where $t_{\text{meas}}$ is defined as in Eq. \eqref{eqn:tauint_definition}. As described in \cite{Wolff2004} the size of the window $W$ is chosen such that systematic and statistical errors on $\widehat{\tau}_{\mathrm{int}}$ are balanced.
% ====== plot tau_{int} double well ======
\begin{figure}
\begin{center}
\includegraphics[width=\linewidth]{\figdir/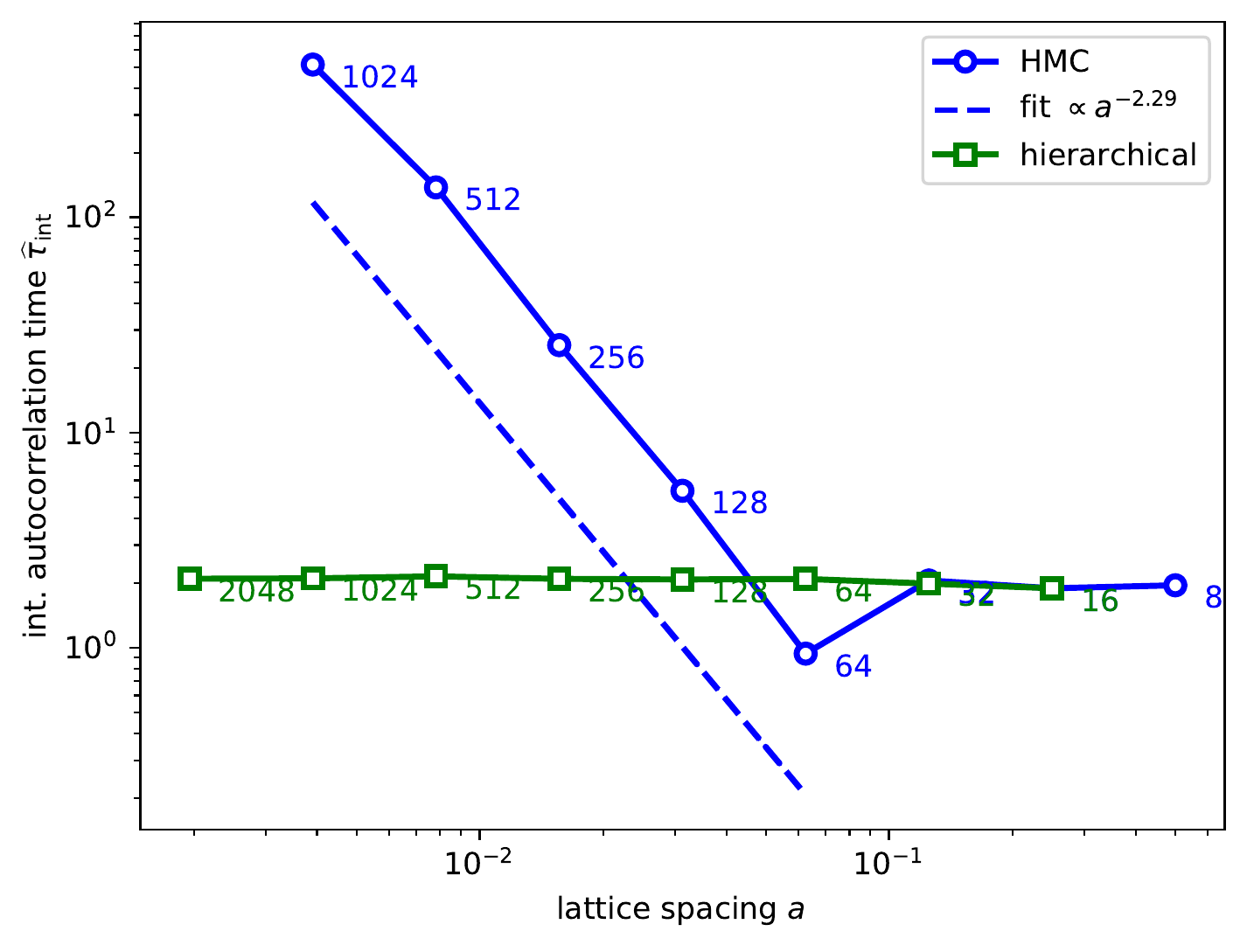}
\end{center}
\caption{Integrated autocorrelation time for double well potential. Results are shown both for a standard HMC and the hierarchical sampler.}
\label{fig:tau_int_doublewell}
\end{figure}
Fig. \ref{fig:tau_int_doublewell} shows the integrated autocorrelation time of the quantity of interest defined in Eq. \eqref{eqn:QoI_doublewell} for the double well potential. As can be seen from this plot, $\tau_{\mathrm{int}}$ increases in proportion to $a^{-z}$ with $z\approx 2.29$ for small lattice spacings, whereas it is completely flat for the hierarchical sampler.

% ====== plot tau_{int} rotor ======
\begin{figure}
\begin{center}
\includegraphics[width=\linewidth]{\figdir/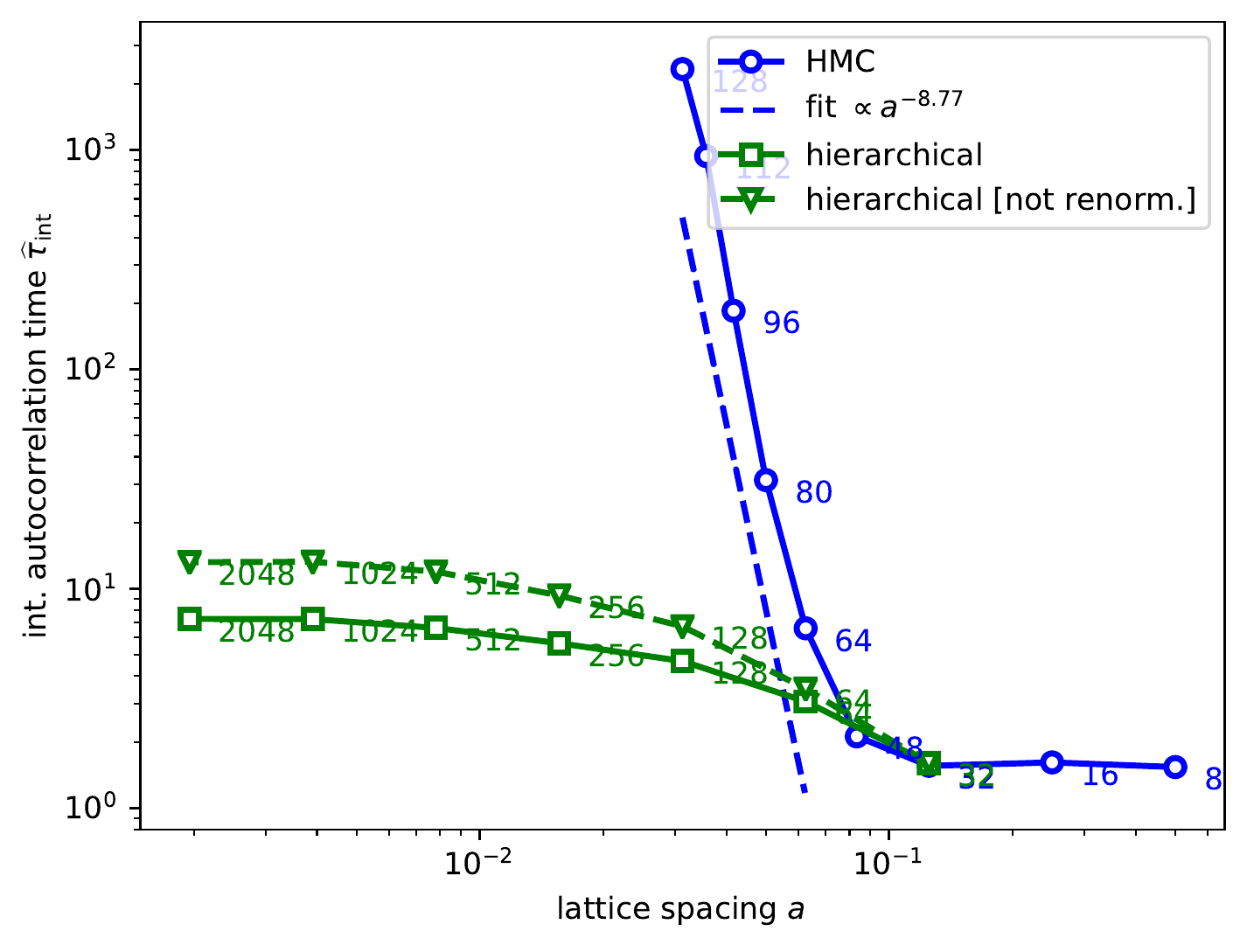}
\end{center}
\caption{Integrated autocorrelation time for topological oscillator. Results are shown both for a standard HMC and the hierarchical sampler.}
\label{fig:tau_int_rotor}
\end{figure}
For the topological oscillator the observable is the topological susceptibility defined in Eq. \eqref{eqn:qoi_rotor}. Here two different setups are considered for the hierarchical sampler: in the first setup the value of $I_0^{(\ell)}$ on the coarse levels is adjusted with the perturbative matching procedure described in Section \ref{sec:coarse_level_matching}. For comparison we also consider the case where $I_0^{(\ell)}=I_0=0.25$ is kept fixed on all levels, we refer to this as the ``not renormalised'' setup in the plots. As Fig. \ref{fig:tau_int_rotor} shows, for the topological susceptibility the integrated autocorrelation time increases very rapidly with approximately $\tau_{\mathrm{int}}\propto a^{-z}$, $z=8.77$ for small lattice spacings if a standard HMC sampler is used. In fact, the measured $\widehat{\tau}_{\mathrm{int}}$ is larger than 1000 for lattice spacings smaller than $0.03$, and the single level method becomes practically unusable if $a$ is reduced further. This is consistent with the results shown in \cite[Fig. 1]{Ammon2016} and can be attributed to freezing of the integer-valued topological charge $q$: for small lattice spacings tunnelling between sectors with different values of $q$ becomes increasingly unlikely. If the hierarchical sampler is used, this problem is dramatically reduced: $\tau_{\mathrm{int}}$ is around 10 and grows only weakly for small lattice spacings. Perturbative matching reduces $\tau_{\mathrm{int}}$ by a factor of approximately two.

The slow growth of the integrated autocorrelation for the hierarchical sampler is related to the acceptance probability of Alg. \ref{alg:hierarchical_sampler}. Recall that a proposal on the finest level is only accepted (i.e. $\vec{x}_{L-1}^{(t+1)}\ne \vec{x}_{L-1}^{(t)}$) if all coarse level proposals have been accepted. In other words, the overall acceptance probability $p_{\mathrm{acc}}=\mathbb{P}(\vec{x}^{(t+1)}_{L-1}\ne \vec{x}^{(t)}_{L-1})$ is the probability of accepting the proposal generated with HMC on the coarsest level (this probability is tuned to around $80\%$), times the probabilities of accepting the proposals generated with the two-level step in Alg. \ref{alg:twolevelMC} on all levels $\ell=L-1,L-2,\dots,1$.

% ====== plot acceptance rate (hierarchical sampler) ======
\begin{figure}
\begin{center}
\includegraphics[width=\linewidth]{\figdir/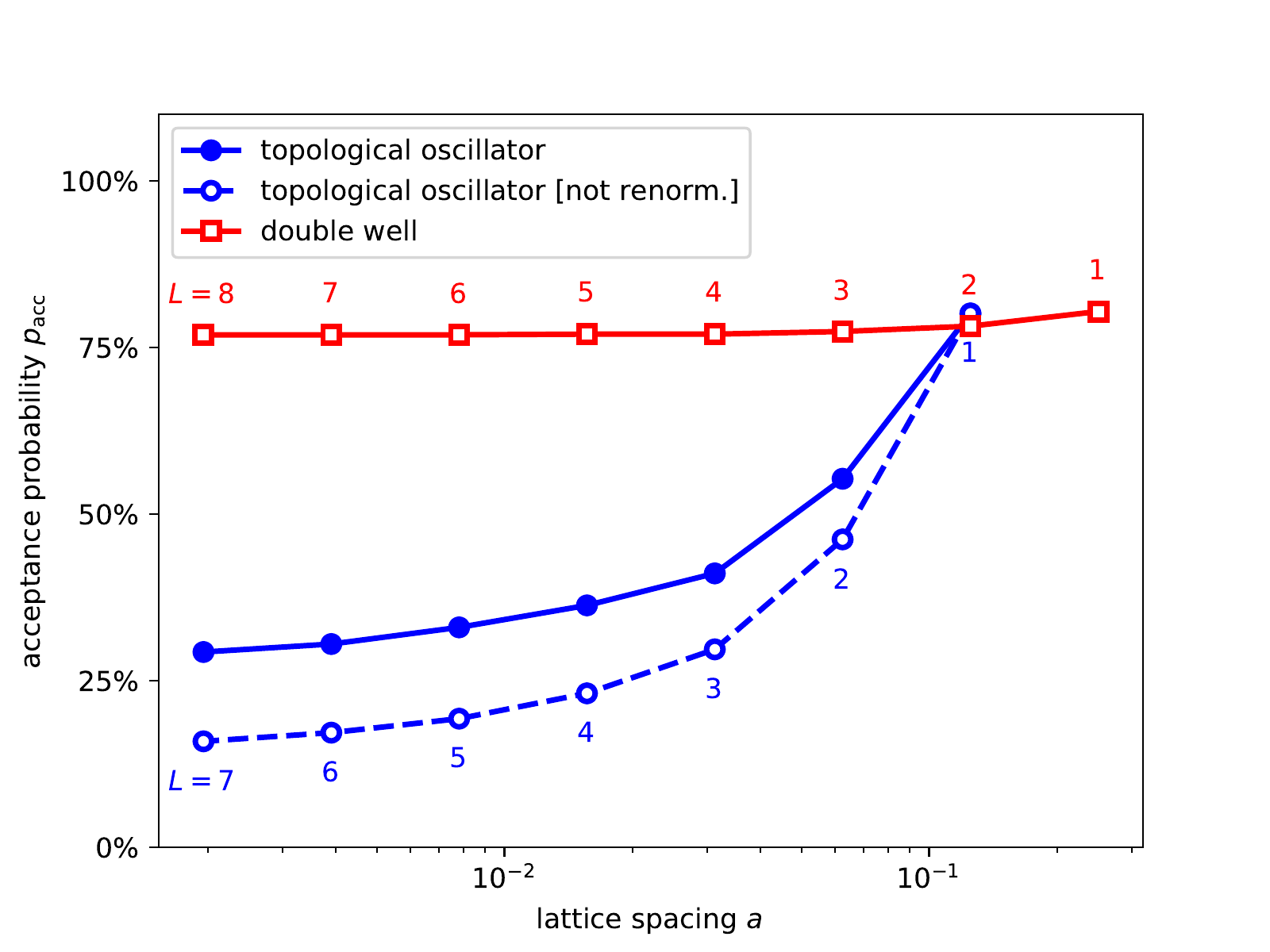}
\end{center}
\caption{Acceptance probability $p_{\mathrm{acc}}=\mathbb{P}(\vec{x}^{(t+1)}_{L-1}\ne \vec{x}^{(t)}_{L-1})$ of standard Monte Carlo with hierarchical sampling (HSMC). Results are shown both for the double well potential and the topological oscillator action.}
\label{fig:p_acceptance}
\end{figure}
Fig. \ref{fig:p_acceptance} shows this overall acceptance probability $p_{\mathrm{acc}}$ as the number of levels $L$ increases for both the double-well potential and the topological oscillator. For the double-well potential the overall acceptance rate does not drop below $75\%$ which implies that the acceptance probability of an individual two-level Metropolis-Hastings step approaches $100\%$ on the finer levels. For the topological oscillator a similar behaviour can be observed, although the curve flattens slower and the total acceptance rate approaches a smaller value for small lattice spacings. As expected, the acceptance probability is higher for the renormalised action. This is not surprising since in the two-level Metropolis-Hastings step the coarse level proposal is a better approximation of the even modes on the next-finer level. Although this explains the smaller absolute value of the autocorrelation time in Fig. \ref{fig:tau_int_rotor}, measurements of the runtime (see Tab. \ref{tab:speedup_matching_HSMC} below) show that using the renormalised action for the HSMC method has a smaller impact on the overall runtime since the average cost per sample grows as the acceptance probability increases. This can be seen immediately from Alg. \ref{alg:twolevelMC}: if the proposal is already rejected on one of the coarser levels, it is no longer necessary to carry out the more expensive two-level Metropolis Hastings steps on the finer levels.
%%%%%%%%%%%%%%%%%%%%%%%%%%%%%%%%%%%%%%%%%%%%%%%%%%%%%
\subsection{Discretisation error and variance decay}\label{sec:results_discretisationerror}
\begin{figure}
\begin{center}
\includegraphics[width=\linewidth]{\figdir/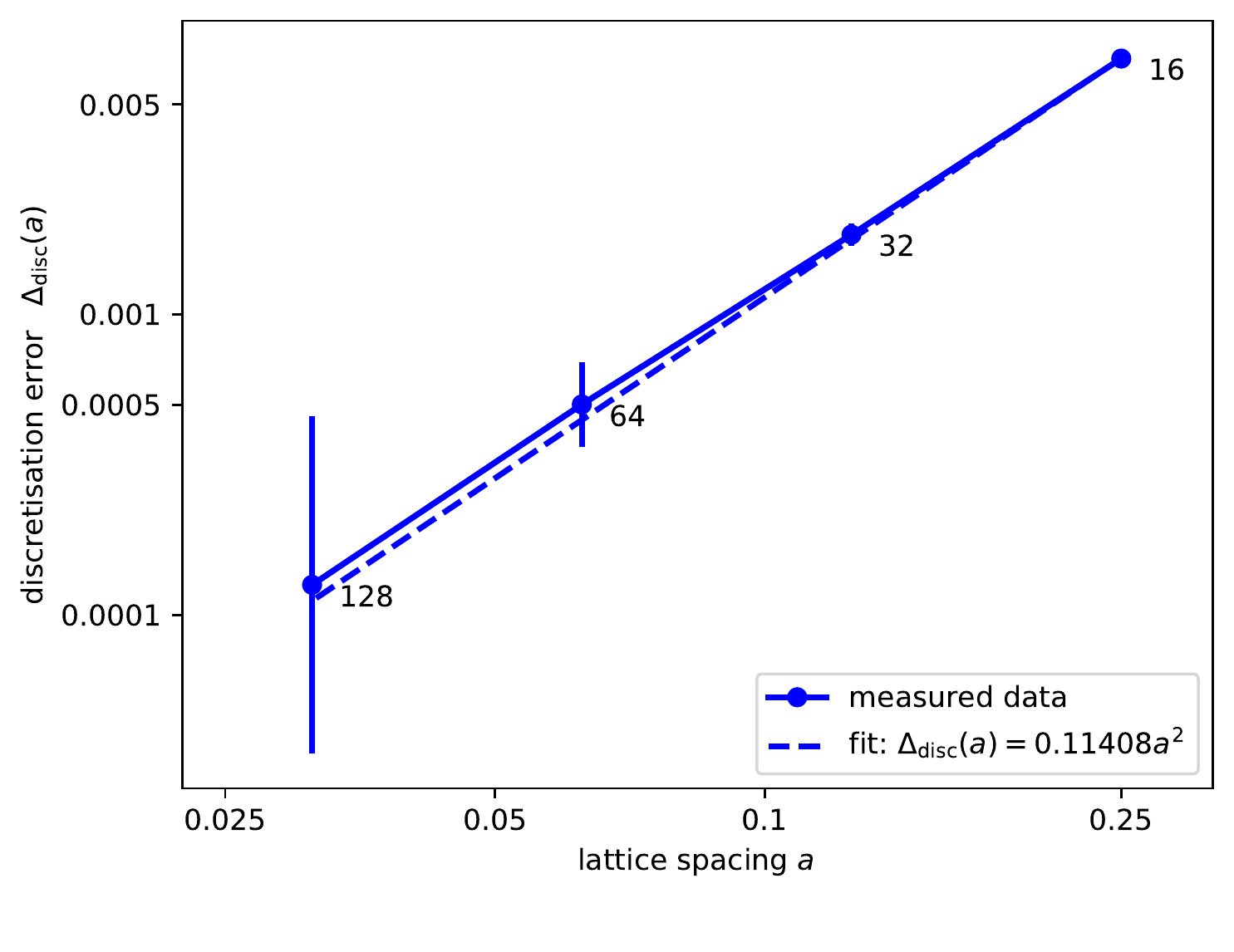}
\end{center}
\caption{Discretisation error $\Delta_{\mathrm{disc}}(a)$ as a function of the lattice spacing $a$ for the double well potential. The fit takes the form $\Delta_0 a^2$. Statistical errors are shown as vertical bars, and the data points are labelled with the number of dimensions $d$ for each lattice spacing.}
\label{fig:bias_doublewell}
\end{figure}

%%%%%%%%%%%%%%%%%%%%%%%%%%%%%%%%%%%%%%%%%%%%%%%%%%%%%
To quantify the discretisation error $\Delta_{\mathrm{disc}}(a)$ as a function of the lattice spacing, we derive an asymptotic bound on $\Delta_{\mathrm{disc}}(a)$. For this assume that
\begin{equation*}
  \Delta_{\mathrm{disc}}(a) = \mathbb{E}[Q(a)]-\mathbb{E}[Q(a=0)] = \Delta_0 a^\alpha+\mathcal{O}(a^{\alpha+1}).
\end{equation*}
For the double-well potential the parameters $\Delta_0$ and $\alpha$ are obtained by calculating $\widehat{Q}(a)$ with $N=4\cdot 10^8$ samples (using the hierarchical method in Alg. \ref{alg:hierarchical_sampler}) for a range of lattice spacings $a=1/32, 1/16,1/8, 1/4$. As shown in Fig. \ref{fig:bias_doublewell}, the measured data is consistent with $\alpha=2$. The coefficient $\Delta_0$ is estimated by approximating $\mathbb{E}[Q(a=0)]$ by $\widehat{Q}(a_{\mathrm{fine}})$ with $a_{\mathrm{fine}}=1/512$ and fitting a function of the form $\log\Delta_0 + 2\log a$ to $\log(\widehat{Q}(a)-\widehat{Q}(a_{\mathrm{fine}}))$ to obtain $\Delta_0=0.11408$. Based on this result we use the relationship $\epsdisc =\Delta_0 a^2$ to relate the lattice spacing to the tolerance on the discretisation error in the following.
% ====== plot bias double well ======
For the topological oscillator the asymptotic form of the discretisation error can be deduced from  Eq. \eqref{eqn:chi_t_theory}, which implies that at leading order the error is linear in the lattice spacing ($\alpha=1$). For our choice of numerical values we find that $\Delta_0=0.21567$.

\begin{figure}
\begin{center}
\includegraphics[width=\linewidth]{\figdir/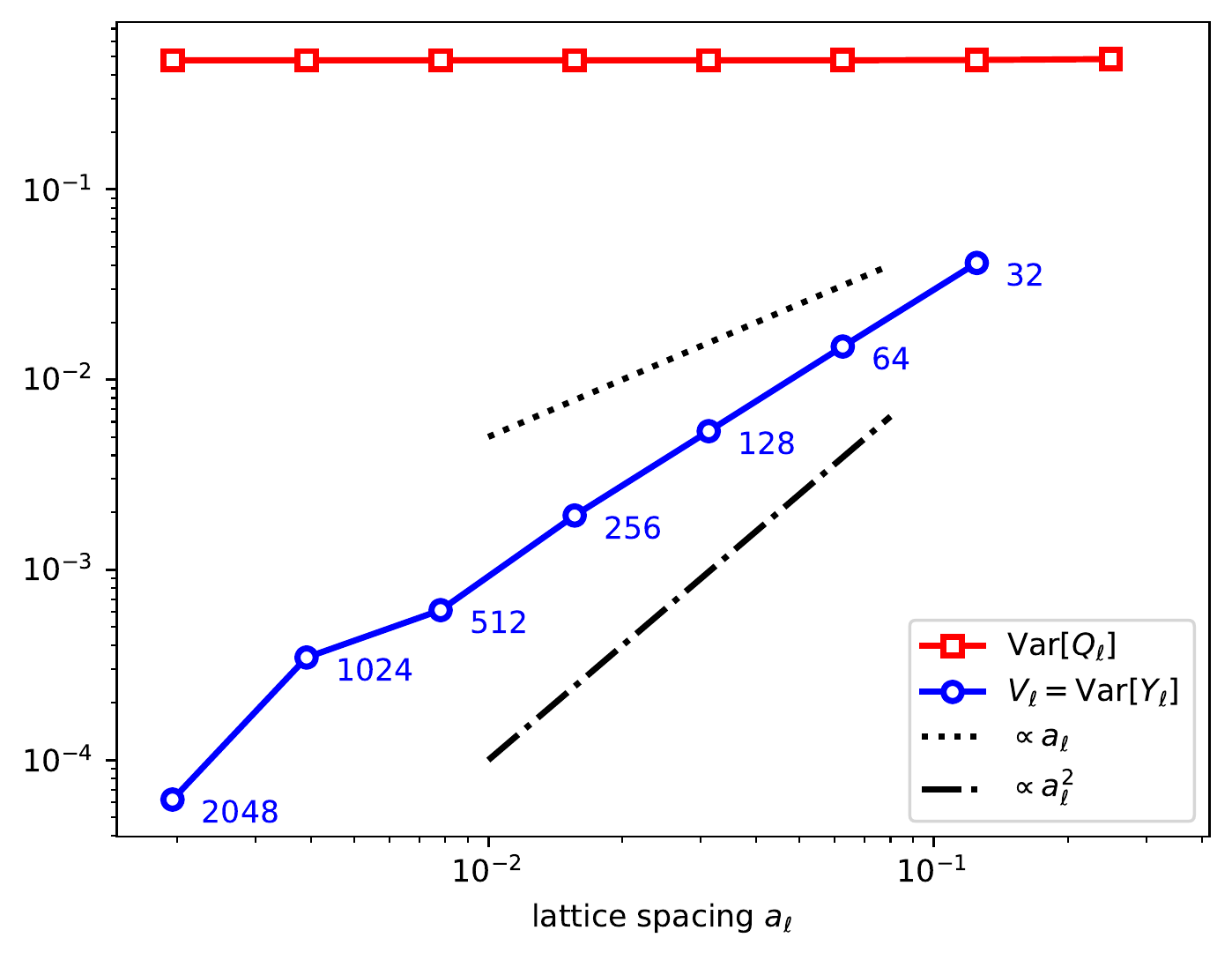}
\end{center}
\caption{Variance of difference estimators $Y_\ell$ and the quantity of interest $Q_\ell$ for the double well potential. The lattice spacing on level $\ell$ is $a_\ell=2^{L-1-\ell}a$. The data points are labelled with the number of dimensions $d_\ell$ for each lattice spacing.}
\label{fig:variance_doublewell}
\end{figure}

For the performance of the multilevel Monte Carlo method the behaviour of the variance $V_\ell$ of the difference of the quantity of interest between subsequent levels is important. Recall in particular that the computational complexity of the MLMC algorithm given in Eq. \eqref{eqn:MLMC_complexity} depends on value of $\beta$ which bounds $V_{\ell}/V_{\ell-1}\le 2^{-\beta}$. Fig. \ref{fig:variance_doublewell} shows $V_\ell$ for the double well potential as well as the variance of the quantity of interest itself. As can be seen from this plot, $\beta$ is larger than $1$ but smaller than $2$, and hence (since $\alpha=2$, as discussed above) we expect the computational complexity of MLMC to be $\mathcal{O}(\epsstat^{-2}+\epsdisc^{-1/2})$. This assumes that the sub-sampling rates and integrated autocorrelation times can be bounded, which appears plausible given the results shown in Figs. \ref{fig:tau_int_doublewell} and \ref{fig:tau_int_rotor}.
% ====== plot variance rotor ======
% ====== plot variance rotor ======
\begin{figure}
\begin{center}
\includegraphics[width=\linewidth]{\figdir/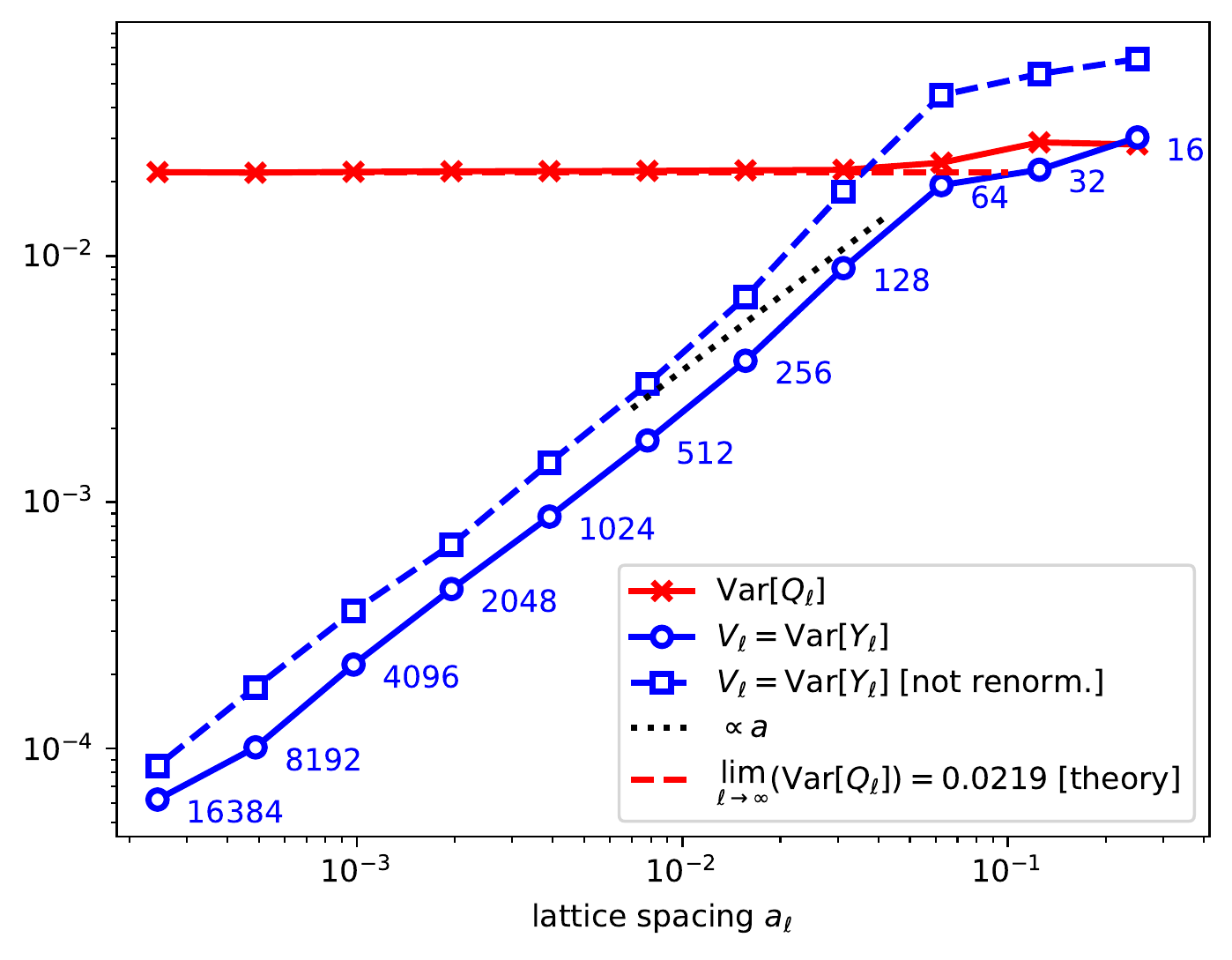}
\end{center}
\caption{Variance of difference estimators $Y_\ell$ and the quantity of interest $Q_\ell$ for the topological oscillator. The lattice spacing on level $\ell$ is $a_\ell=2^{L-1-\ell}a$. The continuum limit as given in Eq. \eqref{eqn:variance_rotor_exact} is shown as a red dashed line. The data points are labelled with the number of dimensions $d_\ell$ for each lattice spacing.}
\label{fig:variance_rotor}
\end{figure}
The variance decay for the topological oscillator is shown in Fig. \ref{fig:variance_rotor}, both for the perturbatively renormalised action and the un-renormalised action. Renormalising the action reduces the absolute value of $V_\ell$.  In both cases it is safe to assume that $\beta \ge 1$, and hence we expect the computational complexity to be no worse than $\mathcal{O}(\epsstat^{-2}|\log\epsdisc|^2+\epsdisc^{-1})$, provided the sub-sampling rates and integrated autocorrelation times can be bounded as $a\rightarrow 0$.
%%%%%%%%%%%%%%%%%%%%%%%%%%%%%%%%%%%%%%%%%%%%%%%%%%%%%
\subsection{Total runtime}
%%%%%%%%%%%%%%%%%%%%%%%%%%%%%%%%%%%%%%%%%%%%%%%%%%%%%
Finally, we compare the total runtime for three different setups:
\begin{description}
\item[StMC] The standard single level Monte Carlo method in Alg. \ref{alg:standardMC} with a HMC sampler.
\item[HSMC] The standard Monte Carlo method in Alg. \ref{alg:standardMC} with the hierarchical delayed-acceptance sampler written down in Alg. \ref{alg:hierarchical_sampler}.
  \item[MLMC] The multilevel Monte Carlo method in Alg. \ref{alg:MLMC}
  \end{description}
  The configuration of StMC and HSMC is described in Section \ref{sec:results_autocorrelations}. For the multilevel method the coarsest level has $d_0=16$ points for the double-well potential and $d_0=32$ points for the topological oscillator. The sub-sampling rates $t_\ell$ in Alg. \ref{alg:MLMC} are set to $\lceil 2\widehat{\tau}_{\text{int},\ell}\rceil$ where $\widehat{\tau}_{\text{int},\ell}$ is the estimated integrated autocorrelation time of the quantity of interest on level $\ell$ obtained with the hierarchical sampler. We confirmed that this choice of sub-sampling rate is sufficient to generate approximately independent samples and that any additional bias in the final MLMC estimator due to imperfect sub-sampling is comparable to the discretisation error. In all cases we generated and discarded a sufficiently large number of samples before computing estimators to ensure that the Markov chains are equilibrated on all levels. The runtimes reported here do not include the time spent in this burn-in phase of the simulation.
% ====== plot runtime double well ======
\begin{figure}
\begin{center}
\includegraphics[width=\linewidth]{\figdir/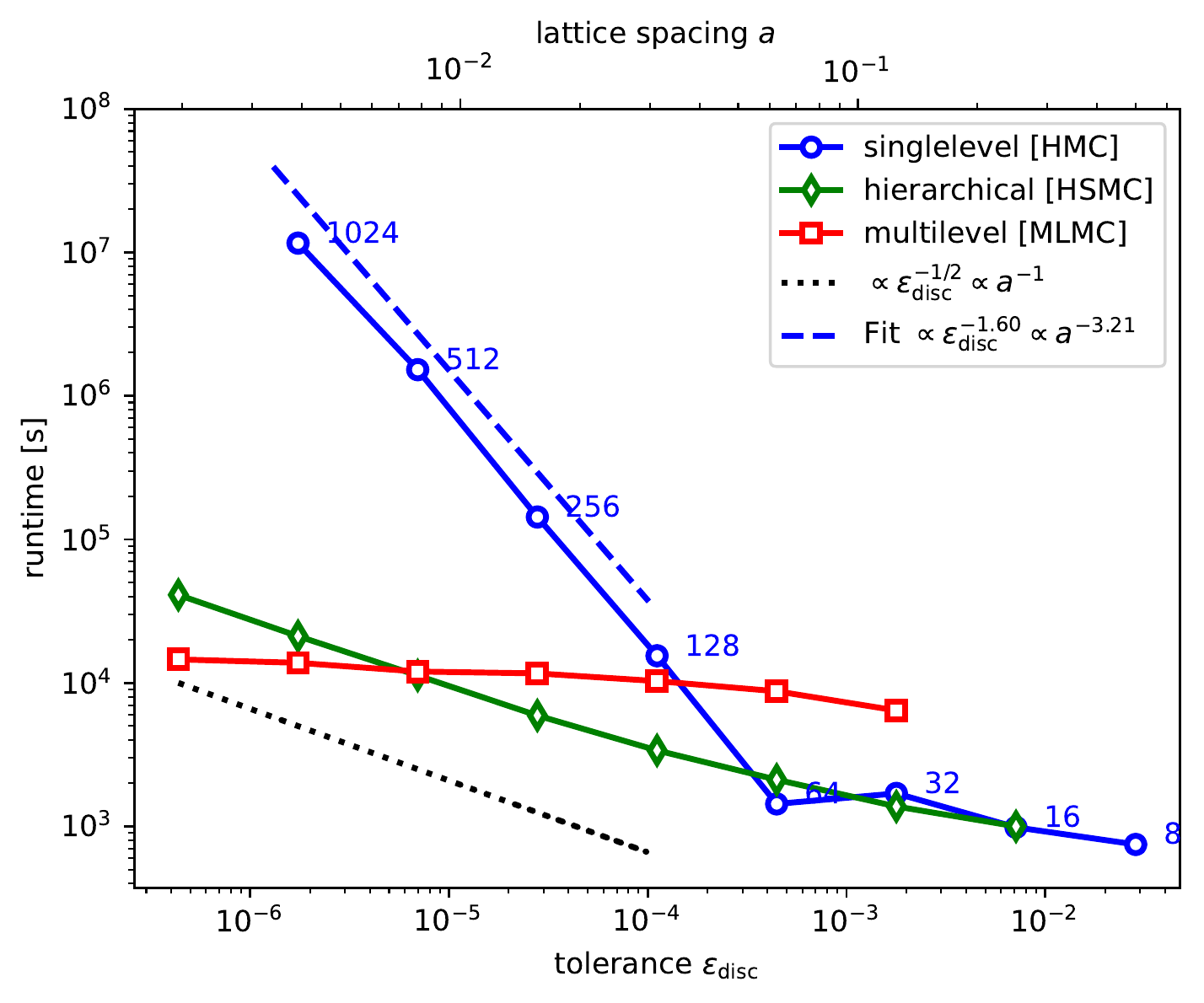}
\caption{Runtime of different Monte Carlo sampling algorithms for the double well potential with a fixed tolerance $\epsstat=10^{-4}$ on the statistical error. Results are shown in seconds and as a function of the tolerance $\epsdisc$. The data points are labelled with the number of dimensions $d$ for each lattice spacing.}
\label{fig:runtime_doublewell}
\end{center}
\end{figure}
% ====== plot runtime rotor ======
\begin{figure}
\begin{center}
\includegraphics[width=\linewidth]{\figdir/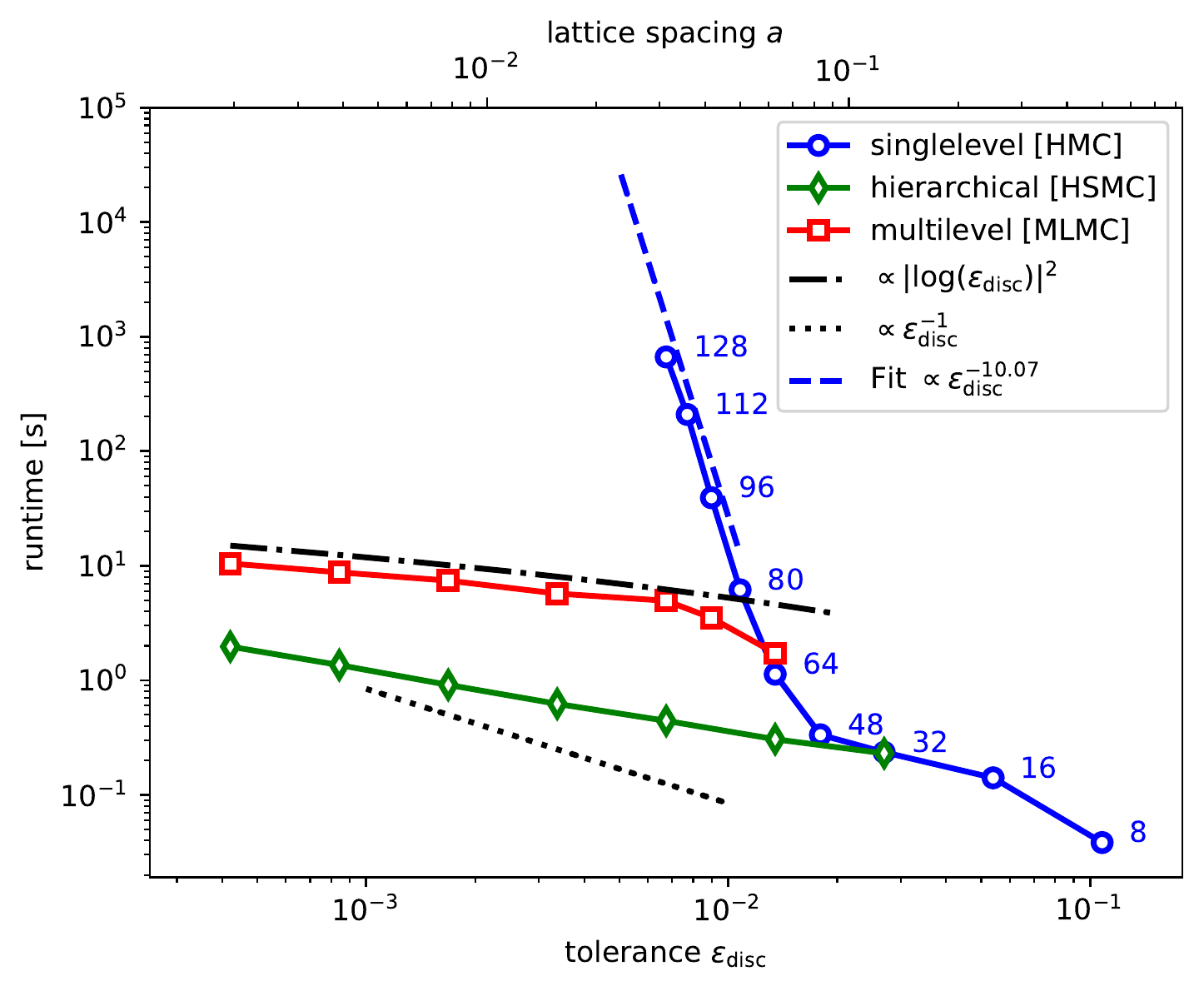}
\caption{Runtime of different Monte Carlo sampling algorithms for the topological oscillator with a fixed tolerance $\epsstat=10^{-2}$ on the statistical error. Results are shown in seconds and as a function of the tolerance $\epsdisc$. The data points are labelled with the number of dimensions $d$ for each lattice spacing.}
\label{fig:runtime_rotor}
\end{center}
\end{figure}
The tolerance on the statistical error is set to a fixed value of $\epsstat=10^{-4}$ for the double well potential and $\epsstat=10^{-2}$ for the topological oscillator, where the difference in size accounts for the fact that the discretisation error decreases much more rapidly for the double well problem. Figs. \ref{fig:runtime_doublewell} and \ref{fig:runtime_rotor} show the total runtime for those values of $\epsstat$ and different lattice spacings $a$, corresponding to different values of $\epsdisc$: as discussed in Section \ref{sec:results_discretisationerror}, for both considered problems we bound the discretisation error by $\epsdisc = \Delta_0 a^\alpha$ with $\alpha=2$, $\Delta_0=0.11408$ for the double-well potential and $\alpha=1$, $\Delta_0=0.21567$ for the topological oscillator. The times reported for HSMC and MLMC in Figs. \ref{fig:runtime_doublewell} and \ref{fig:runtime_rotor} were obtained with the renormalised coarse level action. As can be seen from those figures, the runtime grows rapidly with $\epsdisc^{-(1+z)/\alpha}$ for the StMC method, which is proportional to a high power $a^{-1-z}$ of the inverse lattice spacing since the discretisation error is first order in all cases. Here a factor $a^{-1}$ arises since the cost of generating a path is $\mathcal{O}(a^{-1})$ and the remaining power $a^{-z}$ can be explained by the growth in $\tau_{\mathrm{int}}$ discussed in Section \ref{sec:results_autocorrelations}. As the results in Figs. \ref{fig:runtime_doublewell} and \ref{fig:runtime_rotor} show, by taming autocorrelations the HSMC method reduces the growth of computational cost. In fact, for the lattice spacings considered here the cost grows slower than predicted by the theoretical $\mathcal{O}(\epsdisc^{-1})$ complexity bound for the topological oscillator. This is a pre-asymptotic effect and the reason for it is twofold: firstly, the (fixed) cost of the expensive coarse level HMC sampler still contributes significantly to the overall cost of the hierarchical sampler which is not yet dominated by the evaluation of the action on the finer levels. Secondly, as can be seen from the initial drop of the total acceptance probability in Fig. \ref{fig:p_acceptance}, the probability of accepting a proposed sample in the two-level Metropolis-Hastings step on a given level is smaller than $1$ on the coarser levels, before is approaches $1$ on the finer levels. As a consequence, in a significant proportion of cases generating a hierarchical sample does not require the evaluation of the fine-level action since the proposal is already rejected on a coarser level.

MLMC reduces the asymptotic rate of growth further, and for the double-well potential MLMC is significantly faster than HSMC for the smallest tolerance $\epsdisc$ considered here.
% ====== table speedup MLMC over StMC ======
\ifpreprint % PREPRINT
\begin{table*}
\else % PREPRINT
\begin{table}
\fi % PREPRINT
\begin{center}
\begin{tabular}{rcccrrr}
\hline
& $d$ & $a$ & $\epsdisc$ & $t_{\text{StMC}}$ & $t_{\text{MLMC}}$ & speedup\\ 
\hline\hline
  \multirow{6}{*}{doublewell $\left\{\begin{matrix}\\[5ex]\end{matrix}\right.$}
&$32$ & $0.1250$ & $1.78\cdot10^{-3}$ & $0.17$ & $0.64$ & $0.3\times$\\
&$64$ & $0.0625$ & $4.46\cdot10^{-4}$ & $0.14$ & $0.87$ & $0.2\times$\\
&$128$ & $0.0312$ & $1.11\cdot10^{-4}$ & $1.54$ & $1.03$ & $1.5\times$\\
&$256$ & $0.0156$ & $2.79\cdot10^{-5}$ & $14.32$ & $1.16$ & $12.3\times$\\
&$512$ & $0.0078$ & $6.96\cdot10^{-6}$ & $152.21$ & $1.20$ & $126.9\times$\\
&$1024$ & $0.0039$ & $1.74\cdot10^{-6}$ & $1160.02$ & $1.38$ & $840.9\times$\\
  \hline
  \multirow{3}{*}{$\begin{matrix}\text{topological}\\\text{oscillator}\end{matrix}$ $\left\{\begin{matrix}\\[5ex]\end{matrix}\right.$}
&$64$ & $0.0625$ & $1.35\cdot10^{-2}$ & $1.13$ & $1.72$ & $0.7\times$\\
&$96$ & $0.0417$ & $8.99\cdot10^{-3}$ & $39.26$ & $3.51$ & $11.2\times$\\
&$128$ & $0.0312$ & $6.74\cdot10^{-3}$ & $665.32$ & $4.95$ & $134.4\times$\\
  \hline
\end{tabular}
\caption{Comparison of runtime for standard, single-level Monte Carlo
  (StMC) and multilevel Monte Carlo (MLMC). All times for the double well potential were obtained with $\epsstat=10^{-4}$ and are given in units of $10^4$ seconds. For the topological oscillator a value of $\epsstat=10^{-2}$ was used and times are given in seconds.}
\label{tab:speedup_StMC_MLMC}
\end{center}
\ifpreprint % PREPRINT
\end{table*}
\else % PREPRINT
\end{table}
\fi % PREPRINT
% ====== table speedup MLMC over HSMC ======
\ifpreprint % PREPRINT
\begin{table*}
\else % PREPRINT
\begin{table}
\fi % PREPRINT
\begin{center}
\begin{tabular}{rcccrrr}
\hline
& $d$ & $a$ & $\epsdisc$ & $t_{\text{HSMC}}$ & $t_{\text{MLMC}}$ & speedup\\ 
\hline\hline
  \multirow{7}{*}{doublewell $\left\{\begin{matrix}\\[9ex]\end{matrix}\right.$}
&$32$ & $0.1250$ & $1.78\cdot10^{-3}$ & $0.14$ & $0.64$ & $0.2\times$\\
&$64$ & $0.0625$ & $4.46\cdot10^{-4}$ & $0.21$ & $0.87$ & $0.2\times$\\
&$128$ & $0.0312$ & $1.11\cdot10^{-4}$ & $0.34$ & $1.03$ & $0.3\times$\\
&$256$ & $0.0156$ & $2.79\cdot10^{-5}$ & $0.59$ & $1.16$ & $0.5\times$\\
&$512$ & $0.0078$ & $6.96\cdot10^{-6}$ & $1.13$ & $1.20$ & $0.9\times$\\
&$1024$ & $0.0039$ & $1.74\cdot10^{-6}$ & $2.12$ & $1.38$ & $1.5\times$\\
&$2048$ & $0.0020$ & $4.35\cdot10^{-7}$ & $4.10$ & $1.46$ & $2.8\times$\\
\hline
  \multirow{5}{*}{$\begin{matrix}\text{topological}\\\text{oscillator}\end{matrix}$ $\left\{\begin{matrix}\\[7ex]\end{matrix}\right.$}
&$64$ & $0.0625$ & $1.35\cdot10^{-2}$ & $0.31$ & $1.72$ & $0.2\times$\\
&$128$ & $0.0312$ & $6.74\cdot10^{-3}$ & $0.44$ & $4.95$ & $0.1\times$\\
&$256$ & $0.0156$ & $3.37\cdot10^{-3}$ & $0.62$ & $5.72$ & $0.1\times$\\
&$512$ & $0.0078$ & $1.68\cdot10^{-3}$ & $0.91$ & $7.42$ & $0.1\times$\\
&$1024$ & $0.0039$ & $8.42\cdot10^{-4}$ & $1.36$ & $8.76$ & $0.2\times$\\
&$2048$ & $0.0020$ & $4.21\cdot10^{-4}$ & $1.97$ & $10.47$ & $0.2\times$\\
\hline
\end{tabular}
\caption{Comparison of runtime for Monte Carlo with hierarchical sampling (HSMC) and multilevel Monte Carlo (MLMC). All times for the double well potential were obtained with $\epsstat=10^{-4}$ and are given in units of $10^4$ seconds. For the topological oscillator a value of $\epsstat=10^{-2}$ was used and times are given in seconds.}
\label{tab:speedup_HSMC_MLMC}
\end{center}
\ifpreprint % PREPRINT
\end{table*}
\else % PREPRINT
\end{table}
\fi % PREPRINT
Tab. \ref{tab:speedup_StMC_MLMC} summarises the speedup of MLMC over StMC for both problems. The relative gain of MLMC over HSMC is shown in Tab. \ref{tab:speedup_HSMC_MLMC}. Although the gap between the runtime of the two methods also reduces for the topological oscillator, for the tolerances considered here HSMC is still faster than MLMC.

While here we kept the tolerance $\epsstat$ fixed, in Appendix \ref{sec:results_epsilon} we also show the runtime as a function of the tolerance $\epsilon$ on the total root mean square error, i.e. for $\epsdisc=\epsstat=\epsilon/\sqrt{2}$.
%%%%%%%%%%%%%%%%%%%%%%%%%%%%%%%%%%%%%%%%%%%%%%%%%%%%%
\subsubsection{Breakdown of MLMC cost}
\begin{figure}
\begin{center}
\includegraphics[width=\linewidth]{\figdir/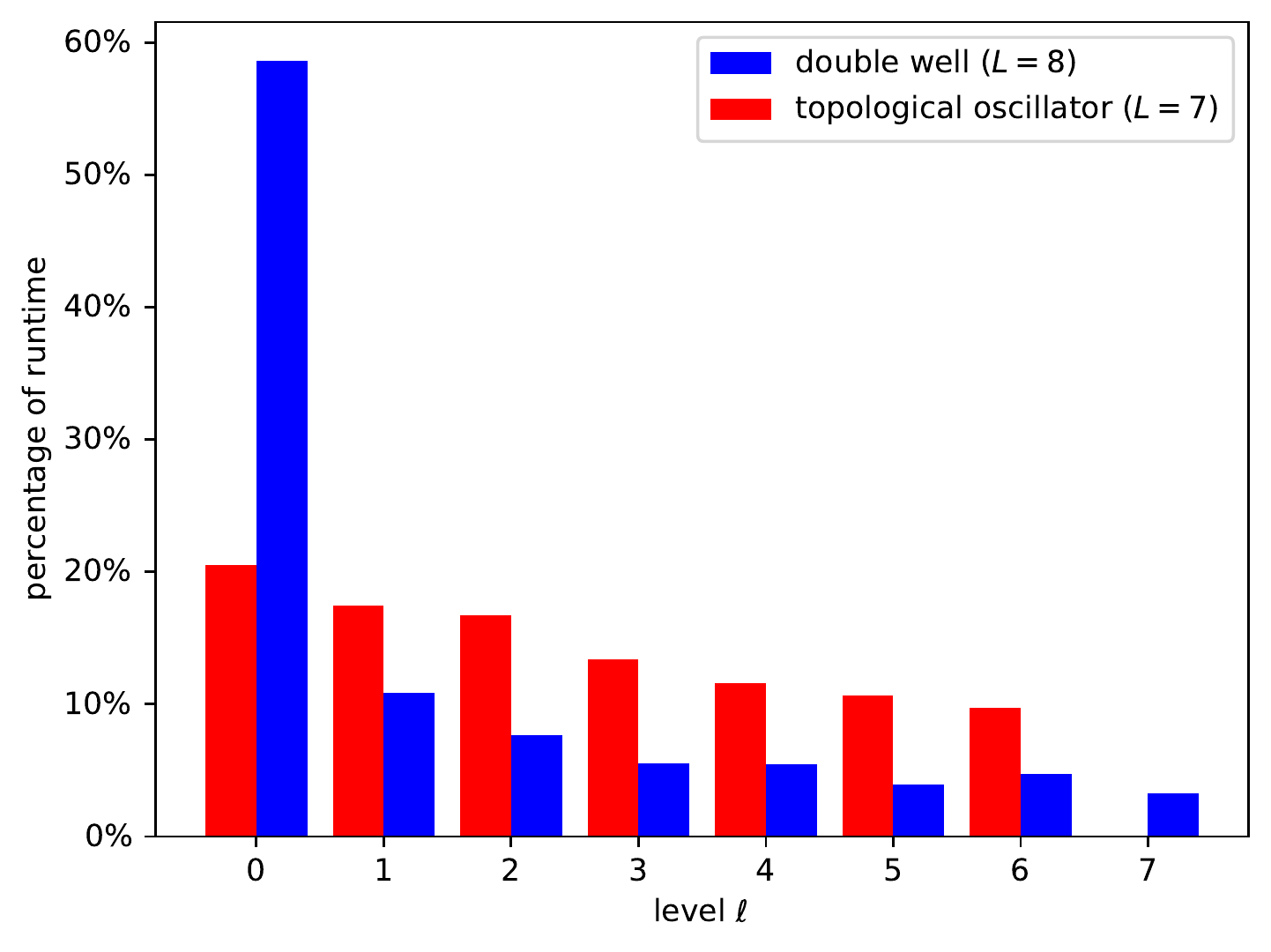}
\end{center}
\caption{Estimated breakdown of cost per level for MLMC. Results are shown for the finest lattice spacing with $d=2048$ for the double well potential and $d=1024$ for the topological oscillator.}
\label{fig:cost_breakdown}
\end{figure}

%%%%%%%%%%%%%%%%%%%%%%%%%%%%%%%%%%%%%%%%%%%%%%%%%%%%%
For the multilevel method it is instructive to break down the total computational cost into the time spent on the individual levels of the lattice hierarchy. To estimate the fraction of the runtime spent on level $\ell$ we computed
\begin{equation*}
  \frac{N_\ell^{\mathrm{eff}}\mathcal{C}_\ell^{\mathrm{eff}}}{\sum_{\ell=0}^{L-1}N_\ell^{\mathrm{eff}}\mathcal{C}_\ell^{\mathrm{eff}}},
\end{equation*}
which is plotted in Fig. \ref{fig:cost_breakdown}. 
% ====== plot cost breakdown ======
As can be seen from this plot, for the double well potential more than half of the time is spent on the coarsest level of the lattice hierarchy. This can be explained by the fact that, as Fig. \ref{fig:variance_doublewell} shows, the variance of difference estimators decreases by a factor between 2 and 4 between subsequent levels. The cost is more evenly distributed between levels for the topological oscillator problem since in this case the variance decays with a near-linear rate (see Fig. \ref{fig:variance_rotor}).
%%%%%%%%%%%%%%%%%%%%%%%%%%%%%%%%%%%%%%%%%%%%%%%%%%%%%
\subsubsection{Gains from coarse level matching}
%%%%%%%%%%%%%%%%%%%%%%%%%%%%%%%%%%%%%%%%%%%%%%%%%%%%%
Finally, we quantify the gains from coarse-level matching for the topological oscillator. For this the HSMC and MLMC runs were repeated without coarse level matching, i.e. with $I_0^{(\ell)}=I_0=0.25$ for all $\ell=0,1,\dots,L-1$. Tab. \ref{tab:speedup_matching_HSMC} shows that this results in a relatively modest reduction of the runtime for the HSMC sampler. As already discussed at the end of Section \ref{sec:results_autocorrelations}, this can be explained by the fact that renormalising the coarse level action leads to a reduction of the integrated autocorrelation time, but this effect is largely compensated by the increased cost per sample.
% ====== table speedup HSMC over non-renormalised HSMC ======
\begin{table}
\begin{center}
\begin{tabular}{rccrrr}
\hline
$d$ & $a$ & $\epsdisc$ & $t^{(0)}_{\text{HSMC}}$ & $t_{\text{HSMC}}$ & speedup\\ 
  \hline\hline
$64$ & $0.0625$ & $1.35\cdot10^{-2}$ & $0.36$ & $0.31$ & $1.2\times$\\
$128$ & $0.0312$ & $6.74\cdot10^{-3}$ & $0.64$ & $0.44$ & $1.4\times$\\
$256$ & $0.0156$ & $3.37\cdot10^{-3}$ & $0.96$ & $0.62$ & $1.5\times$\\
$512$ & $0.0078$ & $1.68\cdot10^{-3}$ & $1.44$ & $0.91$ & $1.6\times$\\
$1024$ & $0.0039$ & $8.42\cdot10^{-4}$ & $1.90$ & $1.36$ & $1.4\times$\\
$2048$ & $0.0020$ & $4.21\cdot10^{-4}$ & $2.57$ & $1.97$ & $1.3\times$\\  
\hline
\end{tabular}
\caption{Comparison of Monte Carlo with hierarchical sampling (HSMC) without and with coarse level mass matching, denoted by  $t^{(0)}_{\text{HSMC}}$ and $t_{\text{HSMC}}$ respectively. All times were obtained with $\epsstat=10^{-2}$ and are given in seconds.}
\label{tab:speedup_matching_HSMC}
\end{center}
\end{table}
As the corresponding speedups for MLMC in Tab. \ref{tab:speedup_matching_MLMC} show, the gain is significantly larger for the multilevel method, where coarse level matching more than halves the runtime. This is because for MLMC matching the actions has the additional effect of reducing the absolute value of the variance of the difference estimators, as can be seen from Fig. \ref{fig:variance_rotor}.
% ====== table speedup MLMC over non-renormalised MLMC ======
\begin{table}
\begin{center}
\begin{tabular}{rccrrr}
\hline
$d$ & $a$ & $\epsdisc$ & $t^{(0)}_{\text{MLMC}}$ & $t_{\text{MLMC}}$ & speedup\\ 
  \hline\hline
$64$ & $0.0625$ & $1.35\cdot10^{-2}$ & $4.07$ & $1.72$ & $2.4\times$\\
$128$ & $0.0312$ & $6.74\cdot10^{-3}$ & $10.68$ & $4.95$ & $2.2\times$\\
$256$ & $0.0156$ & $3.37\cdot10^{-3}$ & $16.47$ & $5.72$ & $2.9\times$\\
$512$ & $0.0078$ & $1.68\cdot10^{-3}$ & $19.27$ & $7.42$ & $2.6\times$\\
$1024$ & $0.0039$ & $8.42\cdot10^{-4}$ & $21.92$ & $8.76$ & $2.5\times$\\
$2048$ & $0.0020$ & $4.21\cdot10^{-4}$ & $25.64$ & $10.47$ & $2.5\times$\\
\hline
\end{tabular}
\caption{Comparison of Multilevel Monte Carlo (MLMC) runtime without and with coarse level mass  matching, denoted by  $t^{(0)}_{\text{MLMC}}$ and $t_{\text{MLMC}}$ respectively. All times were obtained with $\epsstat=10^{-2}$  and are given in seconds.}
\label{tab:speedup_matching_MLMC}
\end{center}
\end{table}
%%%%%%%%%%%%%%%%%%%%%%%%%%%%%%%%%%%%%%%%%%%%%%%%%%%%%
\subsubsection{Multilevel accelerated cluster algorithm}
\begin{figure}
\begin{center}
\includegraphics[width=\linewidth]{\figdir/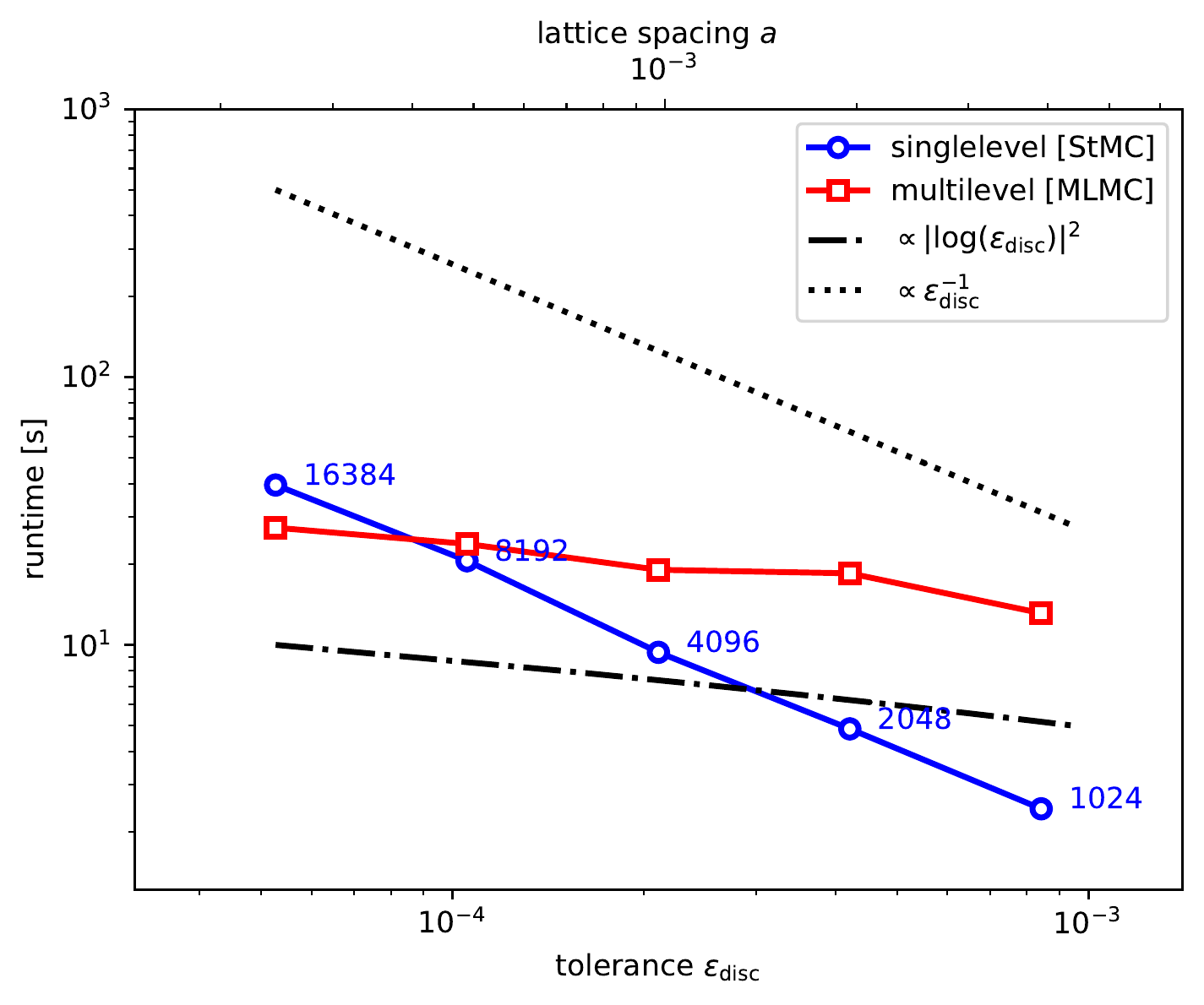}
\end{center}
\caption{Runtime of singlelevel (StMC) and multilevel (MLMC) cluster algorithm for the topological oscillator with a tolerance $\epsstat=10^{-3}$ on the statistical error. Results are shown in seconds and as a function of the tolerance $\epsdisc$. The data points are labelled with the number of dimensions $d$ for each lattice spacing.}
\label{fig:runtime_cluster_rotor}
\end{figure}

%%%%%%%%%%%%%%%%%%%%%%%%%%%%%%%%%%%%%%%%%%%%%%%%%%%%%
For the topological oscillator the cluster algorithm \cite{Wolff1989} can be used to generate Monte Carlo updates with extremely small autocorrelations for arbitrarily small lattice spacing. We implemented a variant of Alg. \ref{alg:MLMC} in which the new samples $\vec{x}_{0}^{(t+t_0)}$ (line 4) and the coarse level samples $\vec{z}^{(t+t_{\ell-1})}_{\ell-1}$ (line 7) are generated with the (single-update) cluster algorithm instead of the hierarchical sampler in Alg. \ref{alg:hierarchical_sampler}. Again the sub-sampling rates are set to $t_\ell=\lceil2\widehat{\tau}_{\text{int},\ell}\rceil$ and we find numerically that $\widehat{\tau}_{\text{int},\ell}\approx 3$ for all levels $\ell$ considered here. The number of unknowns on the coarsest level was fixed to $d_0=512$, while increasing to fine-level problem size from $d=1024$ to $d=16384$. The performance of this MLMC algorithm is compared to the standard, single-level cluster algorithm, again updating a single cluster in each Metropolis Hastings step. As the numerical results in Fig. \ref{fig:runtime_cluster_rotor} show, MLMC is around $40\%$ faster than the standalone cluster algorithm for the smallest lattice spacing considered here. The speedup of the MLMC accelerated cluster algorithm over the single-level cluster algorithm for all lattice spacings is shown in Tab. \ref{tab:speedup_cluster}. More importantly, the numerical experiments show that the runtime of MLMC increases roughly as $\mathcal{O}(|\log(\epsdisc)|^2)$ and thereby grows significantly slower than the runtime of the cluster algorithm, which shows the expected $\mathcal{O}(\epsdisc^{-1})$ growth.
Again we also show the corresponding results for varying $\epsstat=\epsdisc=\epsilon/\sqrt{2}$ in Fig. \ref{fig:runtime_cluster_rotor_epsilon} in Appendix \ref{sec:results_epsilon}.
% ====== plot runtime rotor [cluster sampler] ======
% ====== table speedup MLMC over StMC for cluster algorithm ======
\begin{table}
\begin{center}
\begin{tabular}{rccrrr}
\hline
$d$ & $a$ & $\epsdisc$ & $t_{\text{StMC}}$ & $t_{\text{MLMC}}$ & speedup\\ 
  \hline\hline
  $1024$ & $0.003906$ & $8.42\cdot10^{-4}$ & $2.44$ & $13.13$ & $0.2\times$\\
$2048$ & $0.001953$ & $4.21\cdot10^{-4}$ & $4.86$ & $18.50$ & $0.3\times$\\
$4096$ & $0.000977$ & $2.11\cdot10^{-4}$ & $9.38$ & $19.06$ & $0.5\times$\\
$8192$ & $0.000488$ & $1.05\cdot10^{-4}$ & $20.60$ & $23.81$ & $0.9\times$\\
  $16384$ & $0.000244$ & $5.27\cdot10^{-5}$ & $39.51$ & $27.38$ & $1.4\times$\\
\hline
\end{tabular}
\caption{Comparison of singlelevel (StMC) and multilevel (MLMC) cluster algorithm for the topological oscillator. All times were obtained with $\epsstat=10^{-3}$ and are given in seconds.}
\label{tab:speedup_cluster}
\end{center}
\end{table}

It should be stressed at this point, that while the cluster algorithm proved to be highly efficient for the topological oscillator, its applicability is highly problem dependent and can for example not be directly used for the double well potential problem considered in this work or many other problems in quantum field theory.
%%%%%%%%%%%%%%%%%%%%%%%%%%%%%%%%%%%%%%%%%%%%%%%%%%%%%
\section{Conclusion}\label{sec:conclusion}
%%%%%%%%%%%%%%%%%%%%%%%%%%%%%%%%%%%%%%%%%%%%%%%%%%%%%
In this paper we have described a hierarchical sampling algorithm and applied it for simulations in quantum mechanics. We demonstrated that this can overcome the rapid growth of autocorrelations as the continuum limit is approached. In particular, we considered the anharmonic oscillator with a non-symmetric double well potential and the quantum mechanical topological oscillator model described in \cite{Ammon2016}. Empirically we find that for both cases the integrated autocorrelation time does not show any signifcant increase towards the continuum limit when the lattice spacing approaches zero. This result is particularly significant for the susceptibility of a topological oscillator, which suffers from freezing of the topological charge if a single level method with a standard HMC sampler is used.

Combining this new hierarchical sampling technique with a multilevel Monte Carlo acceleration results in a dramatic reduction of the computational complexity and a significant reduction of the overall runtime. For the finest considered lattice spacings the additional speedup from MLMC (compared to hierarchical sampling for a non-symmetric double well potential or the cluster algorithm for a topological oscillator) is around $1.4\times$ to $2.8\times$. We find that the accurate construction of coarse level theories with an approximate matching procedure is important to achieve optimal performance.

In this paper we have concentrated on reducing the time spent in the sampling phase of the Markov Chain Monte Carlo simulation and did not include burn-in times in the reported runtimes. However, also burn-in can be accelerated with hierarchical sampling since the reduction in autocorrelation time allows chains to equilibrate much faster.

While here we have demonstrated the methods for quantum mechanical systems, the same techniques can be used in lattice field theory simulations. In fact, as explained in the introduction, we expect the speedup to be more significant in this case since the relative cost of computations on coarser levels is further suppressed. A crucial step will be to construct suitable coarse grained theories which could be achieved analytically in perturbation theory or by adopting the framework of Symanzik's effective theory, where the improvement coefficients can be computed perturbatively.  Since many physically interesting theories, and in particular lattice QCD, is asymptotically free, this is expected to work increasingly well as the continuum limit is approached. Of, course, finding non-perturbative methods to construct the course grained theory would be even better.

Recently, multilevel Monte Carlo has received significant attention in other areas, which led to further innovations. While here the method is described in the most natural setup, where coarse levels are constructed by increasing the lattice spacing, coarsening in other categories is also possible and potentially leads to further performance gainsby using the multiindex Monte Carlo \cite{Haji2016} technique. For example the complexity of the theory could be reduced on coarser levels or the physical volume of the lattices could be increased with $\ell$, thus aiming to approach the continuum limit $a\rightarrow 0$ and large volume limit $T\rightarrow \infty$ simultaneously. In lattice QCD one might increase the dynamical quark masses on the coarser levels, which simplifies the computation of the fermion determinant.

In summary, the success of the benchmark computations presented in this paper suggests that applying MLMC techniques to higher dimensional theories, e.g. to gauge theories, is indeed a promising approach which we plan to follow in the future.
%%%%%%%%%%%%%%%%%%%%%%%%%%%%%%%%%%%%%%%%%%%%%%%%%%%%% 
\section*{Acknowledgements}
%%%%%%%%%%%%%%%%%%%%%%%%%%%%%%%%%%%%%%%%%%%%%%%%%%%%%
This research made use of the Balena High Performance Computing (HPC) Service at the University of Bath. We would like to thank Stefan Sch\"{a}fer (DESY Zeuthen) for useful discussions and comments on this paper. We are grateful to Christopher Anders and Pan Kessel (BIFOLD \& TU Berlin) for pointing out a bug in our code which resulted in incorrect results in the original version of this paper.
\clearpage
\appendix
%%%%%%%%%%%%%%%%%%%%%%%%%%%%%%%%%%%%%%%%%%%%%%%%%%%%%
\section{Extension to higher dimensions}\label{sec:two_dimensional_lattice}
\begin{figure}
  \begin{center}
    \includegraphics[width=\linewidth]{\figdir/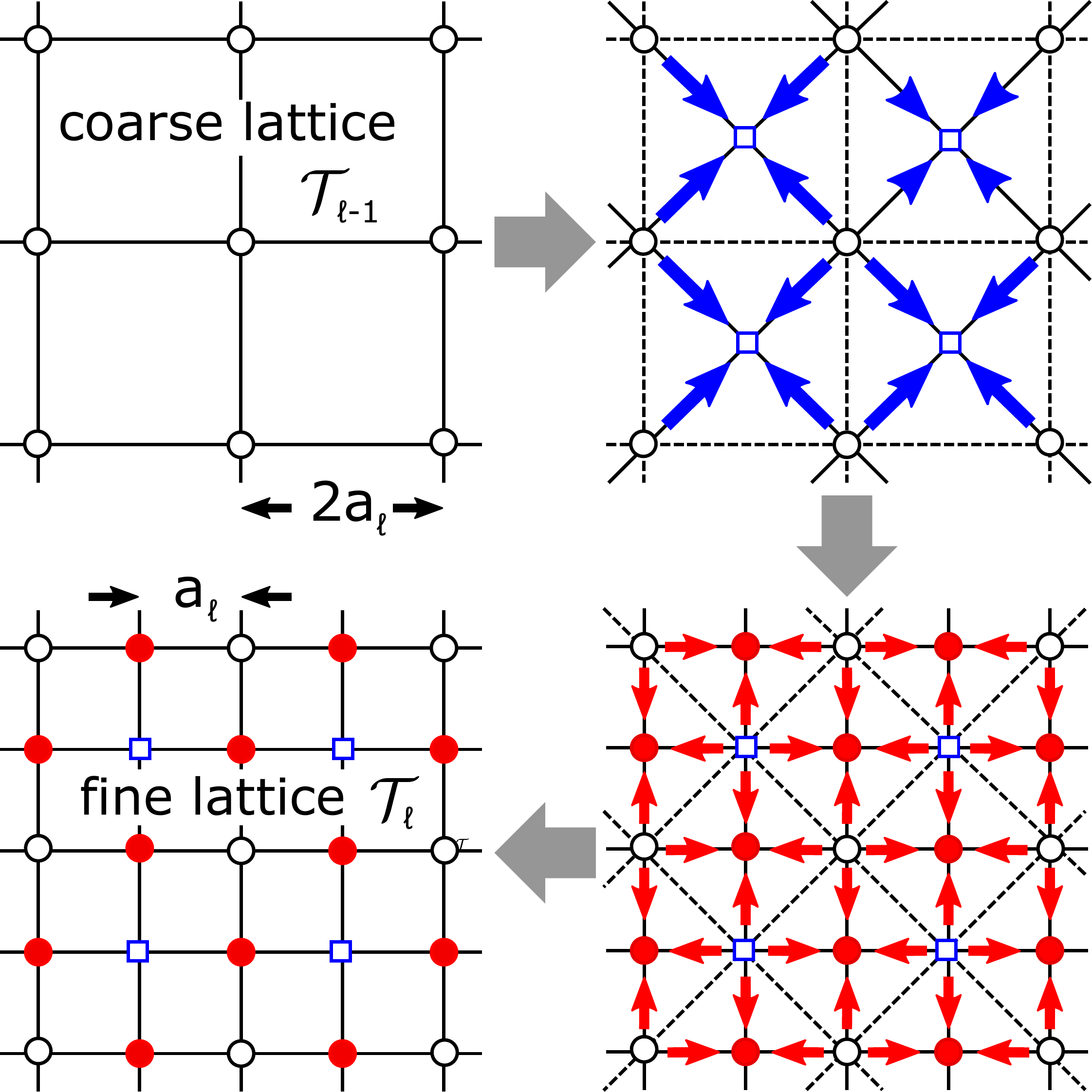}
    \caption{Fill-in of fine-level unknowns on a hierarchical two-dimensional lattice as required in lines 5 and 6 of Alg. \ref{alg:twolevelMC_2d}. Starting from the coarse level unknowns on a rotated lattice (upper left), first the unknowns in $\Omega_{\ell}^{(1)}$ associated with the empty blue squares are filled in using $\tilde{\pi}_\ell^{(1)}$ (upper right). Next, the distribution $\tilde{\pi}^{(2)}_\ell$ is used to fill in the unknowns in $\Omega_{\ell}^{(2)}$ associated with the solid red circles (lower right) to finally obtain the state on the fine lattice (lower left).}
    \label{fig:2d_fillin}
\end{center}
\end{figure}

%%%%%%%%%%%%%%%%%%%%%%%%%%%%%%%%%%%%%%%%%%%%%%%%%%%%%
To illustrate how the methods in this paper, and in particular the two-level Metropolis-Hastings step in Alg. \ref{alg:twolevelMC}, can be extended to higher dimensions, consider a discretised two-dimensional theory for which the degrees of freedom are located at the vertices of a uniform lattice with spacing $a$. If $\Omega$ is the state space and $S:\Omega\rightarrow \mathbb{R}$ is the lattice action, the probability density $\pi^*$ which is sampled with a Monte Carlo method is defined by
\begin{equation*}
  \pi^*(\vec{\Phi}) = \mathcal{Z}^{-1} e^{-S(\vec{\Phi})} \qquad\text{for all $\vec{\Phi}\in\Omega$}.
\end{equation*}
Starting from the finest lattice $\mathcal{T}$, construct a hierarchy of $L$ lattices $\mathcal{T}=\mathcal{T}_{L-1},\mathcal{T}_{L-2},\dots,\mathcal{T}_0$ by doubling the lattice spacing simultaneously in both dimensions; this is shown for two subsequent levels $\mathcal{T}_{\ell}$, $\mathcal{T}_{\ell-1}$ of the hierarchy in Fig. \ref{fig:2d_fillin} (left). On level $\ell$ the lattice spacing is written as $a_{\ell}=2^{\ell-L+1}a$ and we assume that there is an action $S_\ell:\Omega_\ell\rightarrow\mathbb{R}$ with associated probability density $\pi_\ell$ where
\begin{equation}
  \pi_\ell(\vec{\Phi}) = \mathcal{Z}_\ell^{-1} e^{-S_\ell(\vec{\Phi})} \qquad\text{for all $\vec{\Phi}\in\Omega_\ell$}.
  \label{eqn:pi_ell_2d}
\end{equation}
Again, the coarse level theories are naturally obtained as (approximate) effective theories of the original fine-level theory on level $L-1$ where $S_{L-1}=S$ and $\pi_{L-1} = \pi^*$.

While the coarse level actions are constructed by starting from the original lattice, the hierarchical sampler in Alg. \ref{alg:twolevelMC} constructs new fine level samples by generating a proposal on the coarsest lattice and successively adding fine-level modes. On each level $\ell$ this requires a mechanism for filling in the values of unknowns in the fine-level space $\Omega_\ell$ for a given state $\vec{\Phi}'\in\Omega_{\ell-1}$ in the coarse level space $\Omega_{\ell-1}$. We use two iterations of the construction described for the Ising model in \cite{Faas1986} to achieve this. As illustrated in Fig. 1(a) there, the key idea is to use a rotated lattice with a lattice spacing that is reduced by a factor $\sqrt{2}$. The values at the additional sites that are generated in each rotation are drawn from a distribution which depends only on the values at the already existing sites\footnote{Although sampling is particularly simple if the action only contains nearest-neighbour interactions (as for the Ising model in \cite{Faas1986}) so that the value at each new site can be drawn independently, this is not a necessary condition.}.

To explain this process in more detail, observe that on each level $\ell$ the state space $\Omega_\ell$ can be written as the direct sum of three spaces $\Omega_{\ell}^{(2)}$, $\Omega_{\ell}^{(1)}$ and $\Omega_{\ell-1}$ with \mbox{$\Omega_\ell:=\Omega_{\ell}^{(2)}\oplus \Omega_{\ell}^{(1)}\oplus\Omega_{\ell-1}$}, which should be compared to the decomposition in Eq. \eqref{eq:Omega_sum}. To see this and to define $\Omega_\ell^{(1)}$, $\Omega_\ell^{(2)}$, separate the unknowns $\vec{\Phi}\in\Omega_\ell$ into three different classes, depending on which topological entity of a coarse grid cell on level $\ell-1$ they are associated with, namely
\begin{enumerate}
\item coarse-level unknowns associated with coarse level vertices, collected in a vector $\vec{\Phi}'\in\Omega_{\ell-1}$ and shown as empty black circles in Fig. \ref{fig:2d_fillin},
\item fine-level unknowns associated with the interior of coarse level cells, collected in $\tilde{\vec{\Phi}}^{(1)}\in\Omega_{\ell}^{(1)}$ and shown as empty blue squares and
\item fine-level unknowns associated with edges of coarse level cells, collected in $\tilde{\vec{\Phi}}^{(2)}\in\Omega_{\ell}^{(2)}$ and shown as solid red circles in the figure.
\end{enumerate}
Given $\vec{\Phi}'\in\Omega_{\ell-1}$, $\tilde{\vec{\Phi}}^{(1)}\in\Omega_{\ell}^{(1)}$ and $\tilde{\vec{\Phi}}^{(2)}\in\Omega_{\ell}^{(2)}$ write $\overline{\vec{\Phi}}=[\tilde{\vec{\Phi}}^{(1)},\vec{\Phi}']\in\overline{\Omega}_{\ell}:=\Omega_{\ell}^{(1)}\oplus \Omega_{\ell-1}$ and
\begin{equation}
  \vec{\Phi}=[\tilde{\vec{\Phi}}^{(2)}, \overline{\vec{\Phi}}] =[\tilde{\vec{\Phi}}^{(2)}, [\tilde{\vec{\Phi}}^{(1)},\vec{\Phi}']]\in\Omega_{\ell},
  \label{eqn:brackets_2d}
\end{equation}
which should be compared to Eq. \eqref{eqn:x_decomposition} in the main text. Assume that there is a conditional probability density $\tilde{\pi}^{(1)}(\cdot|\vec{\Phi}')$ on the state space $\Omega_{\ell}^{(1)}$, given the values of the coarse level unknowns $\vec{\Phi}'\in\Omega_{\ell-1}$. Since the empty black circles and empty blue squares in the top right of Fig. \ref{fig:2d_fillin} define a rotated lattice with spacing $\sqrt{2}a_\ell$, the density $\tilde{\pi}^{(1)}$ can be constructed by writing down an action $\tilde{S}^{(1)}:\overline{\Omega}_\ell\rightarrow\mathbb{R}$ on this rotated lattice, namely
\begin{equation}
  \tilde{\pi}^{(1)}(\tilde{\vec{\Phi}}^{(1)}|\vec{\Phi}') = \left(\tilde{\mathcal{Z}}^{(1)}_{\ell}(\vec{\Phi}')\right)^{-1}\exp\left[-\tilde{S}_\ell^{(1)}([\tilde{\vec{\Phi}}^{(1)},\vec{\Phi}'])\right].
    \label{eqn:pi_ell_1_2d}
\end{equation}
This action could for example be obtained by a renormalisation group transformation on $S_\ell$, followed by some approximations that guarantee that it is possible to effectively generate states in $\Omega^{(1)}_\ell$ for a given $\vec{\Phi}'$. Similarly define a conditional probability density $\tilde{\pi}^{(2)}(\cdot|\overline{\vec{\Phi}})$ on $\Omega_{\ell}^{(2)}$, given the values of the unknowns $\overline{\vec{\Phi}}\in\overline{\Omega}_\ell$. Here $S^{(2)}:\Omega_\ell\rightarrow \mathbb{R}$ can be expressed as an approximation of $S_\ell$ with
\begin{equation}
  \begin{aligned}
    \tilde{\pi}^{(2)}(\tilde{\vec{\Phi}}^{(2)}|\overline{\vec{\Phi}}) &= \left(\tilde{\mathcal{Z}}^{(2)}_{\ell}(\overline{\vec{\Phi}})\right)^{-1}\exp\left[-\tilde{S}_\ell^{(2)}([\tilde{\vec{\Phi}}^{(2)},\overline{\vec{\Phi}}])\right]\\
    &=\left(\tilde{\mathcal{Z}}^{(2)}_{\ell}([\tilde{\vec{\Phi}}^{(1)},\vec{\Phi}'])\right)^{-1}\\
    &\quad\times\;\exp\left[-\tilde{S}_\ell^{(2)}([\tilde{\vec{\Phi}}^{(2)}, [\tilde{\vec{\Phi}}^{(1)},\vec{\Phi}']])\right].
  \end{aligned}
    \label{eqn:pi_ell_2_2d}
  \end{equation}
The exact choice of $\tilde{S}^{(1)}$ and $\tilde{S}^{(2)}$ influences the acceptance rate, but does not have any impact on the fine-level discretisation error. Alg. \ref{alg:hierarchical_sampler} for the Metropolis-Hastings step $\vec{\Phi}^{(t)}_\ell\rightarrow \vec{\Phi}^{(t)}_\ell$ can now be rewritten for a two-dimensional theory as shown in the following Alg. \ref{alg:twolevelMC_2d}.
  \begin{algorithm}[H]
\caption{Two-level Metropolis Hastings step for two-dimensional theories.\newline Input: level $\ell$, current sample $\vec{\Phi}_{\ell}^{(t)}\sim\pi_{\ell}$, proposal distribution $q_{\ell-1}$ \newline Output: new sample $\vec{\Phi}_{\ell}^{(t+1)}\sim\pi_{\ell}$}
\label{alg:twolevelMC_2d}
\begin{algorithmic}[1]
  \State{Let $\vec{\Phi}_\ell^{(t)}=[\tilde{\vec{\Phi}}_{\ell}^{(2,t)},[\tilde{\vec{\Phi}}_{\ell}^{(1,t)},\vec{\Phi}_{\ell-1}^{(t)}]]$ with $\vec{\Phi}_{\ell-1}^{(t)}\in\Omega_{\ell-1}$, $\vec{\Phi}_{\ell}^{(1,t)}\in\Omega_{\ell}^{(1)}$, $\vec{\Phi}_{\ell}^{(2,t)}\in\Omega_{\ell}^{(2)}$ as in Eq. \eqref{eqn:brackets_2d} and pick $\vec{\Psi}_{\ell-1}$ from $q_{\ell-1}(\cdot|\vec{\Phi}_{\ell-1}^{(t)})$.}
  \If{$\vec{\Phi}_{\ell-1}^{(t+1)}=\vec{\Phi}_{\ell-1}^{(t)}$ (the coarse level proposal was rejected)}
  \State{Set $\vec{\Phi}_\ell^{(t+1)}\mapsfrom \vec{\Phi}_\ell^{(t)}$}
  \Else
  \State {Pick $\tilde{\vec{\Psi}}^{(1)}_{\ell}$ from $\tilde{\pi}^{(1)}_{\ell}(\cdot\vert\vec{\Psi}_{\ell-1})$}
  \State{Pick $\tilde{\vec{\Psi}}^{(2)}_{\ell}$ from $\tilde{\pi}^{(2)}_{\ell}(\cdot\vert[\tilde{\vec{\Psi}}^{(1)},\vec{\Psi}_{\ell-1}])$}
  \State{Let $\vec{\Psi}_{\ell}=[\tilde{\vec{\Psi}}^{(2)}_{\ell},[\tilde{\vec{\Psi}}^{(1)}_{\ell},\vec{\Psi}_{\ell-1}]]$ and compute
\begin{equation*}
  \begin{aligned}
   \exp[-\Delta S_\ell] &= \rho_{\ell}\cdot \rho_{\ell}^{(2)}\cdot \rho_{\ell}^{(1)}\cdot \rho'_{\ell-1}\quad\text{with}\\
\rho_{\ell} &:= \frac{\pi_{\ell}(\vec{\Psi}_{\ell})}{\pi_{\ell}(\vec{\Phi}^{(t)}_{\ell})}\\
\rho_{\ell}^{(2)}&:=\frac{\tilde{\pi}^{(2)}_{\ell}(\tilde{\vec{\Phi}}_\ell^{(2,t)}\vert [\tilde{\vec{\Phi}}_\ell^{(1,t)},{\vec{\Phi}}^{(t)}_{\ell-1}])}{\tilde{\pi}^{(2)}_{\ell}(\tilde{\vec{\Psi}}^{(2)}_\ell\vert [\tilde{\vec{\Psi}}_\ell^{(1)},\vec{\Psi}_{\ell-1})]}\\
\rho_{\ell}^{(1)}&:=\frac{\tilde{\pi}^{(1)}_{\ell}(\tilde{\vec{\Phi}}_\ell^{(1,t)}\vert {\vec{\Phi}}^{(t)}_{\ell-1})}{\tilde{\pi}^{(1)}_{\ell}(\tilde{\vec{\Psi}}^{(1)}_\ell\vert \vec{\Psi}_{\ell-1})}\\
\rho'_{\ell-1}&:=\frac{\pi_{\ell-1}({\vec{\Phi}}^{(t)}_{\ell-1})}{\pi_{\ell-1}(\vec{\Psi}_{\ell-1})}
\end{aligned}
\end{equation*}
}
\State{Accept the proposal $\vec{\Psi}_\ell$ and set $\vec{\Phi}_\ell^{(t+1)}\mapsfrom \vec{\Psi}_\ell$ with}
\StateX{probability $\min\{1,\exp[-\Delta S_\ell]\}$; set $\vec{\Phi}_\ell^{(t+1)}\mapsfrom \vec{\Phi}^{(t)}_\ell$ if}
\StateX{the proposal is rejected.}
      \EndIf
    \end{algorithmic}
  \end{algorithm}
  An explicit expression for $\Delta S_\ell$ is readily obtained from Eqs. \eqref{eqn:pi_ell_2d}, \eqref{eqn:pi_ell_1_2d} and \eqref{eqn:pi_ell_2_2d}. The key difference between Alg. \ref{alg:twolevelMC} and Alg. \ref{alg:twolevelMC_2d} is that the fine level states from $\Omega_{\ell}^{(1)}$ and $\Omega_{\ell}^{(2)}$ are filled in in two steps and that the triple product of ratios in Alg. \ref{alg:twolevelMC} has been replaced by the product of the four ratios $\rho_\ell$, $\rho_{\ell}^{(2)}$, $\rho_{\ell}^{(1)}$ and $ \rho'_{\ell-1}$ in Alg. \ref{alg:twolevelMC_2d}. A similar construction is possible in higher dimensions where it is necessary to successively fill in the fine level unknowns which are not in the coarse level state space.
%%%%%%%%%%%%%%%%%%%%%%%%%%%%%%%%%%%%%%%%%%%%%%%%%%%%%
\section{Multilevel Monte Carlo cost analysis}\label{sec:MLMC_cost_analysis}
%%%%%%%%%%%%%%%%%%%%%%%%%%%%%%%%%%%%%%%%%%%%%%%%%%%%%
We make the reasonable assumption that the cost $\mathcal{C}_{\mathrm{coarse}}$ of generating a sample $\vec{x}_{0}^{(t+1)}$ with the standard Metropolis sampler on the coarsest level is proportional to the number of unknowns $d_0$, and does not increase as the number of levels increases (while keeping $a_0$ fixed). More specifically, we assume that this cost $\mathcal{C}_{\mathrm{coarse}}$ can be bounded by
\begin{equation*}
\mathcal{C}_{\mathrm{coarse}} \le A_0d_{0}=2^{-L+1}A_0d
\end{equation*}
for some constant $A_0$. Furthermore, given the coarse level sample $\vec{y}_{\ell-1}$, the cost of executing Alg. \ref{alg:twolevelMC} is proportional to $d_\ell$ and can be bounded by
\begin{equation*}
\mathcal{C}_{\ell}^{\mathrm{2-level}} \le B_0d_\ell=2^{\ell-L+1}B_0d
\end{equation*}
for some other constant $B_0$. A straightforward calculation shows that the cost of obtaining a new sample $\vec{x}_\ell^{(t+1)}$ with Alg. \ref{alg:hierarchical_sampler} can be bounded by
\begin{equation*}
\mathcal{C}_{\ell} \le (A_0 + B_0) d_\ell = 2^{\ell-L+1}(A_0 + B_0) d,
\end{equation*}
i.e. does not grow more than linearly with the number $d_\ell$ of unknowns on level $\ell$. Taking into account the sub-sampling rates $t_{\ell}$, the cost of obtaining an independent measurement of $Y_{\ell}^{(j)}$ on level $\ell$ in Alg. \ref{alg:MLMC} is therefore
\begin{equation}
\mathcal{C}_{\ell}^{\mathrm{eff}}=
\begin{cases}
\lceil\tau_{\mathrm{int},\ell}\rceil \left(\mathcal{C}_\ell^{\mathrm{2-level}}+t_{\ell-1}\mathcal{C}_{\ell-1}\right) & \text{for $\ell=1,\dots,L-1$}\\
\lceil\tau_{\mathrm{int},0}\rceil t_0\mathcal{C}_{\mathrm{coarse}} & \text{for $\ell=0$}
\end{cases}
\label{eqn:cost_eff_expressions}
\end{equation}
where $\tau_{\mathrm{int},\ell}$ is the integrated autocorrelation time on level $\ell$.
In our code we measured $\mathcal{C}_{\mathrm{coarse}}$, $\mathcal{C}^{\mathrm{2-level}}_{\ell}$ and $\mathcal{C}_{\ell}$ during the setup phase of each run, and then used Eq. \eqref{eqn:cost_eff_expressions} to compute $\mathcal{C}_{\ell}^{\mathrm{eff}}$ required in Eq. \eqref{eqn:N_ell_eff}. The integrated autocorrelation time were updated on-the-fly in the multilevel Monte Carlo algorithm, generating additional samples if this increased the $N_\ell^{\mathrm{eff}}$ in Eq. \eqref{eqn:N_ell_eff}.

To quantify the cost of the multilevel Monte Carlo algorithm in Alg.~\ref{alg:MLMC}, further assume that the sub-sampling rates $t_\ell$ are bounded by some $t_{\max}\ge t_\ell$ for all $\ell=1,\dots,L-2$. By definition, this is also an upper bound on the integrated autocorrelation times on all levels, i.e. $\tau_{\mathrm{int},\ell}\le t_{\max}$ for $\ell=0,1,\dots,L-1$. Then, there exists a constant $\tilde{C}_0$ such that $\mathcal{C}_{\ell}^{\mathrm{eff}} \le \tilde{C}_0 d_\ell = 2^{\ell}\tilde{C}_0d_0$, in other words the cost for generating an independent measurement $Y_\ell^{(j)}$ on level $\ell$ does not grow at a faster rate than the number of unknowns $d_\ell=2^\ell d_0=2^{\ell-L+1}d$ on this particular level. More generally, to make the following derivation applicable for field theories in $D>1$ dimensions (where $D=1$ corresponds to quantum mechanics), we assume that there is a $C_0>0$ such that
\begin{equation*}
  \mathcal{C}_{\ell}^{\mathrm{eff}} \le 2^{D\ell}C_0\quad\text{for all $\ell=0,1,\dots,L-1$}.
\end{equation*}
We now show how the cost of the MLMC algorithm depends on the tolerances $\epsdisc$ and $\epsstat$ as $\epsdisc,\epsstat \rightarrow 0$. Using the definition of $N_{\ell}^{\mathrm{eff}}$ in Eq.~\eqref{eqn:N_ell_eff} and the fact that $\max\{A,B\}\le A+B$, the total cost of MLMC with a tolerance $\epsstat$ on the statistical error and a given number of levels $L$ can be bounded by
\begin{equation}
\mathcal{C}_{\mathrm{MLMC}} = \sum_{\ell=0}^{L-1} N_{\ell}^{\mathrm{eff}} \mathcal{C}_\ell^{\mathrm{eff}
} \le\epsstat^{-2} \sigma(L)^2+\tilde{\sigma}(L)
\label{eqn:MLMC_cost}
\end{equation}
where
\begin{xalignat*}{2}
  \sigma(L) &:= \sum_{\ell=0}^{L-1} \sqrt{V_\ell \mathcal{C}_\ell^{\mathrm{eff}}},&
  \tilde{\sigma}(L) &:= \sum_{\ell=0}^{L-1} \mathcal{C}_\ell^{\mathrm{eff}}.
\end{xalignat*}
Assuming that 
%the variance $V_\ell$ decays at a rate $2^{-\beta}$, i.e.
\begin{equation*}
V_\ell \le 2^{-\beta\ell} V_0,
\end{equation*}
a straightforward calculation shows that $\sigma(L)$ can be bounded as follows, depending on whether $\beta$ is larger, equal or smaller than $D$:
\begin{equation}
\begin{aligned}
\sigma(L) \le \kappa_0\sum_{\ell=0}^{L-1} 2^{\frac{D-\beta}{2}\ell}
\le\begin{cases}
\kappa_+ & \text{for $\beta>D$}\\
\kappa_0 L & \text{for $\beta=D$}\\
\kappa_- 2^{\frac{1-\beta}{2}L} & \text{for $\beta<D$}.
\end{cases}
\label{eqn:bound_sigma}
\end{aligned}
\end{equation}
with the constants
\begin{xalignat*}{2}
\kappa_0 &= \sqrt{C_0V_0}, &
\kappa_+ &= \frac{\kappa_0}{1-2^{\frac{D-\beta}{2}}} = -\kappa_-.
\end{xalignat*}
The sum $\tilde{\sigma}(L)$ is readily bounded by
\begin{equation}
  \tilde{\sigma}(L) \le C_0 \frac{2^{DL}}{2^D-1}
  \label{eqn:bound_sigma_tilde}
\end{equation}
To obtain a bound on the number of levels $L$, we further assume that the discretisation is of order $\alpha$, i.e. for a given lattice spacing $a$ the discretisation error $\Delta_{\mathrm{disc}}(a)$ can be bounded by
\begin{equation*}
\Delta_{\mathrm{disc}}(a) \le \tilde{\Delta}_0 a^\alpha
\end{equation*}
for some constants $\alpha\ge 1$, $\tilde{\Delta}_0$. If we set $a=2^{-L_{\max}+1} a_0$ with
\begin{equation*}
L_{\max} = 1+\left\lceil \log_2\left(a_0\tilde{\Delta}_0^{1/\alpha}\right) -\frac{1}{\alpha}\log_2 \epsdisc \right\rceil
\end{equation*}
the discretisation error will be smaller than $\epsdisc$. Hence, it is not necessary to use more than $L_{\max}$ levels, and $L$ in Eq.~\eqref{eqn:bound_sigma} can be bounded by
\begin{equation*}
L \le 2+\log_2\left(a_0\tilde{\Delta}_0^{1/\alpha}\right) -\frac{1}{\alpha}\log_2 \epsdisc.
\end{equation*}
Using this bound in Eqs.~\eqref{eqn:bound_sigma} and \eqref{eqn:bound_sigma_tilde} implies that the cost in Eq.~\eqref{eqn:MLMC_cost} has the following computational complexity as a function of $\epsdisc$ and $\epsstat$:
\begin{equation}
  \mathcal{C}_{\mathrm{MLMC}} = \begin{cases}
\mathcal{O}\left(\epsstat^{-2}+\epsdisc^{-D/\alpha}\right) & \text{for $\beta>D$}\\
\mathcal{O}\left(\epsstat^{-2}\vert\log \epsdisc\vert^2+\epsdisc^{-D/\alpha}\right) & \text{for $\beta=D$}\\
\mathcal{O}\left(\epsstat^{-2}\epsdisc^{-\frac{D-\beta}{\alpha}}+\epsdisc^{-D/\alpha}\right) & \text{for $\beta<D$}.
\end{cases}
\label{eqn:MLMC_complexity_appendix}
\end{equation}
For the quantum mechanical problems considered in this paper we have that $D=1$, which leads to the computational complexity in Eq. \eqref{eqn:MLMC_complexity}; Eq. \eqref{eqn:cost_MLMC} in the introduction is a special case of this for $\alpha=\beta=1$. In fact, as explained in \cite{Dodwell2015}, $\alpha=\beta$ holds more generally for the Markov Chain variant of the multilevel Monte Carlo algorithm. Hence, for quantum field theories in higher dimensions with $D>\alpha=\beta$ the third case in Eq. \eqref{eqn:MLMC_complexity_appendix} applies, which results in Eq. \eqref{eqn:cost_QFT} in the introduction. Finally, setting $\epsstat=\epsdisc=\epsilon/\sqrt{2}$ gives Eq. \eqref{eqn:cost_QFT_epsilon}.
%%%%%%%%%%%%%%%%%%%%%%%%%%%%%%%%%%%%%%%%%%%%%%%%%%%%%
\section{Memory requirements}\label{sec:memory_requirements}
%%%%%%%%%%%%%%%%%%%%%%%%%%%%%%%%%%%%%%%%%%%%%%%%%%%%%
To put the memory requirements of the algorithms described in this paper into context consider a $D$ dimensional quantum field theory and HMC sampling as an established reference method. In addition to the current state $\vec{x}^{(t)}$, both the proposal $\vec{y}$ and $\nu\ge 1$ temporary vectors have to be stored to implement the symplectic timestepping scheme in the enlarged phase space. For the simple leapfrog implementation used in this work $\nu=1$ since only one additional momentum vector is required. If there are $d$ lattice points in each direction this leads to a total storage requirement of $2+\nu$ state vectors of length $d^D$ or $\mathcal{M}_{\text{HMC}}=(2+\nu)d^D$ double precision variables in $D$ dimensions. Executing the two-level Metropolis Hastings step in Alg. \ref{alg:twolevelMC} on level $\ell$ of the hierarchy requires storage for $\vec{x}^{(t)}_{\ell}$ and the proposal $\vec{y}_{\ell}$, which are both vectors of length $d^D_\ell$. Depending on how the proposal on level $\ell-1$ is generated, this might require additional vectors of length $d^D_{\ell-1}$. For example, if the proposals $\vec{y}_{\ell-1}$ are drawn from $q_{\ell-1}(\cdot\vert \vec{x}_{\ell-1}^{(t)})$ with a single level Metropolis Hasting method and a HMC proposal distribution, one would require $\nu$ additional vectors. However, since unknowns on the finer levels are filled in recursively and existing entries of $\vec{x}_{\ell}^{(t)}$ are used to represent the current state on coarser levels, the hierarchical sampler in Alg. \ref{alg:hierarchical_sampler} only needs to store two vectors of length $d^D_\ell$ to represent $\vec{x}_\ell^{(t)}$ and the proposal $\vec{y}_{\ell}$ as well as $\nu$ vectors of length $d^D_0$ to account for the Metropolis Hastings step on the coarsest level with $\ell=0$. This leads to total storage requirements of $\mathcal{M}_{\text{HS}}(\ell)=2d^D_\ell + \nu d^D_0$ for Alg. \ref{alg:hierarchical_sampler}. In particular, on the finest level
\begin{equation}
  \mathcal{M}_{\text{HS}}(L-1) = \left(2 + 2^{-(L-1)D}\nu\right)d^D < \mathcal{M}_{\text{HMC}}.\label{eqn:mem_HS}
\end{equation}
To obtain the memory requirements of the MLMC method in Alg. \ref{alg:MLMC} note that on each level both the current state $\vec{x}_\ell^{(t)}$ and a proposal $\vec{y}_\ell$ have to be stored. In addition, the storage requirements of the hierarchical sampler in Alg. \ref{alg:hierarchical_sampler} have to be taken into account on all but the very finest level. Consequently for $L\ge 2$ levels the total amount of required memory is
\begin{equation}
  \begin{aligned}
    \mathcal{M}_{\text{MLMC}} &= 2\sum_{\ell=0}^{L-1} d_\ell^D+\sum_{\ell=0}^{L-2} \mathcal{M}_{\text{HS}}(\ell)\\
    &= \left(2+4\frac{1-2^{-(L-1)D}}{2^D-1}+\nu\frac{L-1}{2^{(L-1)D}}\right)d^D\\
      & < (6+\nu) d^D< 3\mathcal{M}_{\text{HMC}} .\label{eqn:mem_MLMC}
    \end{aligned}
  \end{equation}
  As Eqs. \eqref{eqn:mem_HS} and \eqref{eqn:mem_MLMC} show, the memory footprint of the hierarchical sampler in Alg. \ref{alg:hierarchical_sampler} is actually smaller than that of HMC, whereas the MLMC method in Alg. \ref{alg:MLMC} requires less than three times the amount of storage used by a standard HMC method for any dimension $D$. Limited storage usually restricts the size of systems that can be simulated for higher dimensions ($D\ge3$). As the second line of Eq. \eqref{eqn:mem_MLMC} shows, for those higher dimensional problems the additional memory overhead of MLMC (compared to HMC) is actually less than $30\%$.
%%%%%%%%%%%%%%%%%%%%%%%%%%%%%%%%%%%%%%%%%%%%%%%%%%%%%
\section{Rejection sampling}\label{sec:rejection_sampling}
%%%%%%%%%%%%%%%%%%%%%%%%%%%%%%%%%%%%%%%%%%%%%%%%%%%%%
To draw samples from the distribution $p_{\sigma,\delta x}$ defined in Eq. ~\eqref{eqn:expsin2_distribution} we use rejection sampling with a Gaussian envelope, as described in the following algorithm:
\begin{algorithm}[H]
\caption{Rejection sampling for distribution $p_{\sigma,\delta x}$ defined in Eq. \eqref{eqn:expsin2_distribution}}
\label{alg:rejection_sampling}
\begin{algorithmic}[1]
  \Loop
  \State{Draw sample $x$ from Gaussian distribution $g_\sigma$}
  \StateX{with $g_\sigma(x)=\sqrt{\frac{2\sigma}{\pi^3}}\exp\left[-\frac{2\sigma}{\pi^2}x^2\right]$.}
  \If{$-\pi \le x \le \pi$}
  \State{Draw uniformly distributed random $u\in[0,1)$.}
  \If{$u \le \exp\left[-2\sigma\left(\sin^2\left(\frac{x}{2}\right)-\frac{x^2}{\pi^2}\right)\right]$}
  \State{\Return{$x+\delta x$}}
  \EndIf
  \EndIf
  \EndLoop
\end{algorithmic}
\end{algorithm}
%%%%%%%%%%%%%%%%%%%%%%%%%%%%%%%%%%%%%%%%%%%%%%%%%%%%%
\section{Fixed tolerance on the total error}\label{sec:results_epsilon}
\begin{figure}
\begin{center}
\includegraphics[width=\linewidth]{\figdir/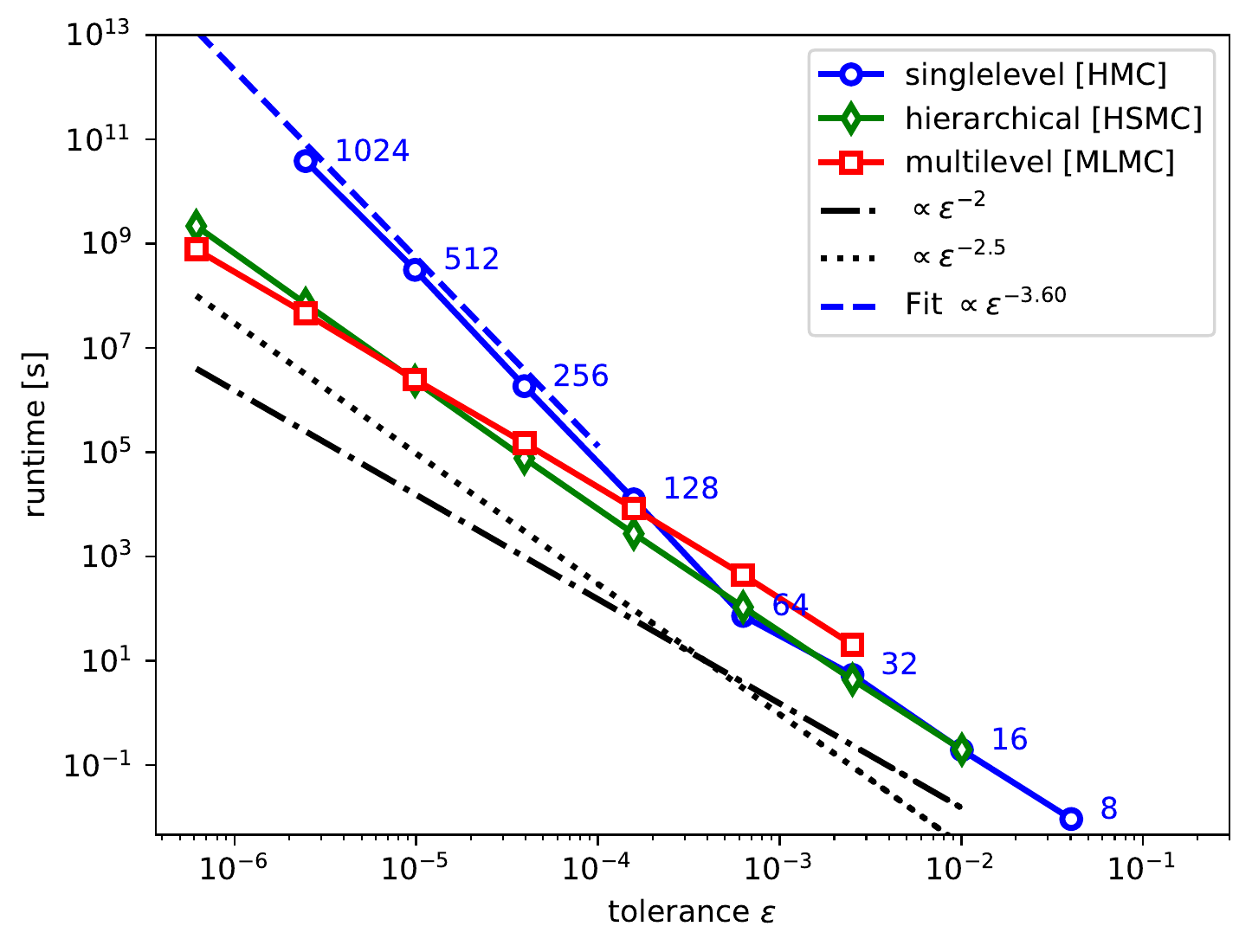}
\end{center}
\caption{Estimated runtime of different Monte Carlo sampling algorithms for the double well potential. Results are shown in seconds and as a function of the tolerance $\epsilon$ on the total error. The data points are labelled with the number of dimensions $d$ for each lattice spacing.}
\label{fig:runtime_doublewell_epsilon}
\end{figure}

\begin{figure}
\begin{center}
\includegraphics[width=\linewidth]{\figdir/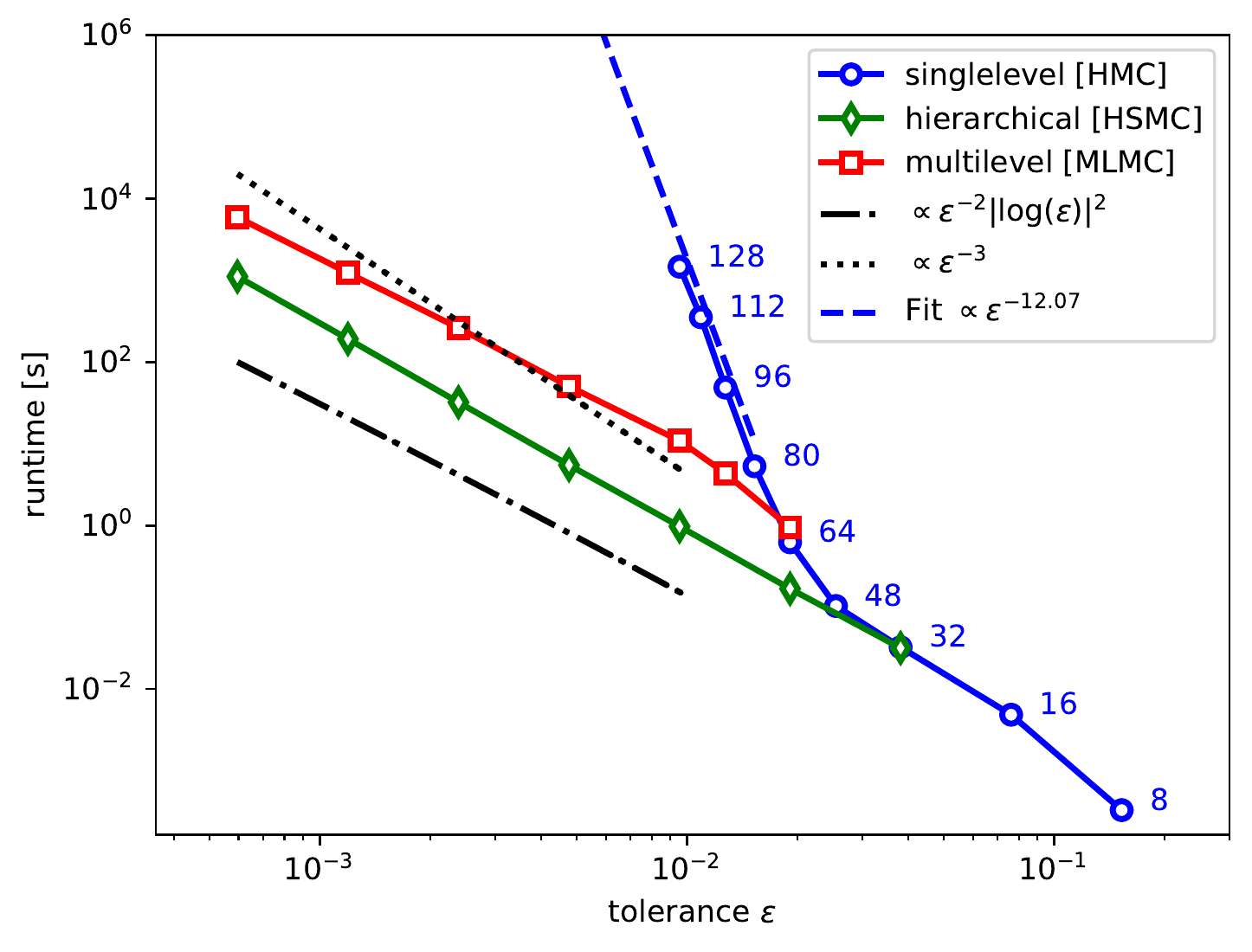}
\end{center}
\caption{Runtime of different Monte Carlo sampling algorithms for the topological oscillator. Results are shown in seconds and as a function of the tolerance $\epsilon$ on the total error. The data points are labelled with the number of dimensions $d$ for each lattice spacing.}
\label{fig:runtime_rotor_epsilon}
\end{figure}

\begin{figure}
\begin{center}
\includegraphics[width=\linewidth]{\figdir/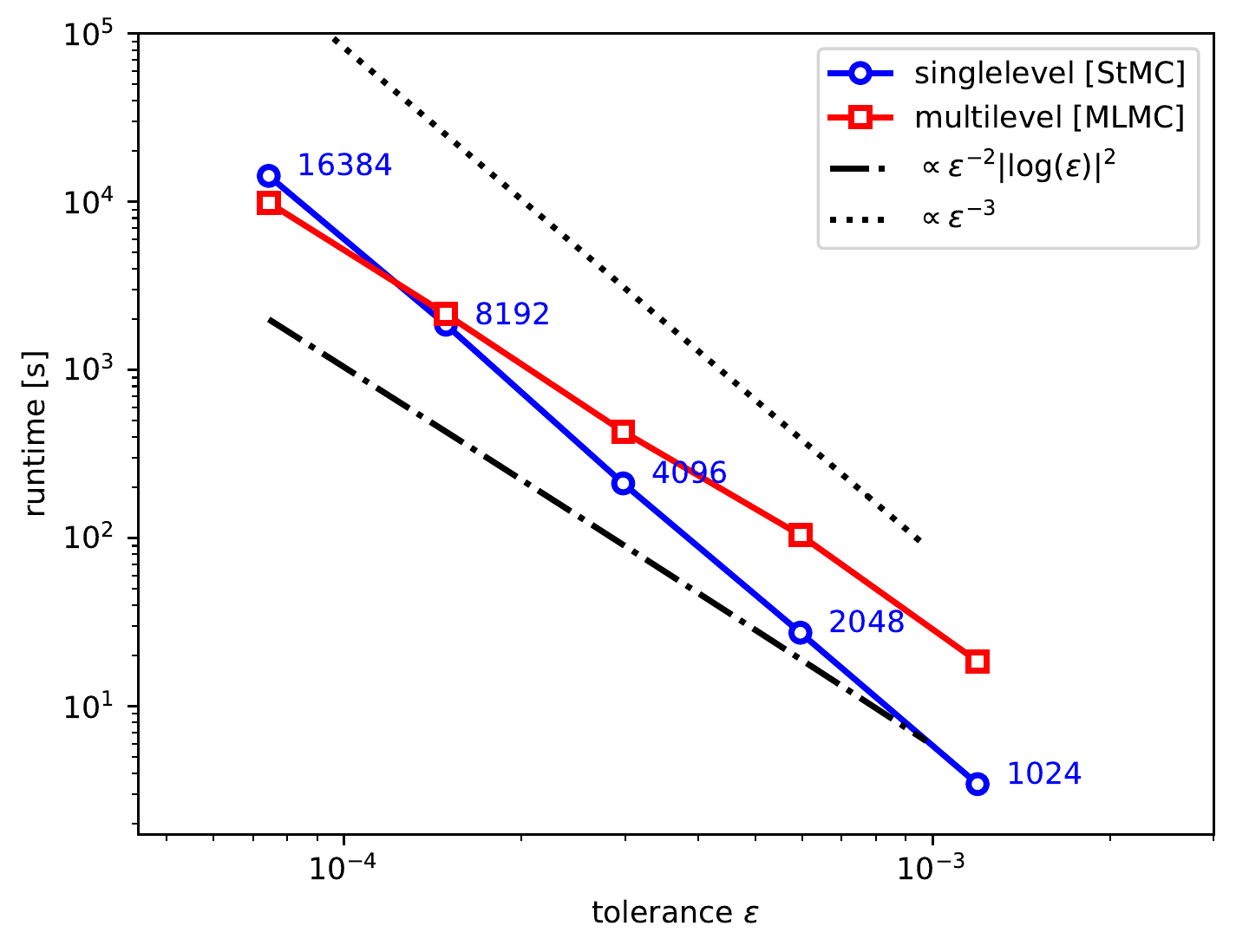}
\end{center}
\caption{Runtime of singlelevel (StMC) and multilevel (MLMC) cluster algorithm for the topological oscillator. Results are shown in seconds and as a function of the tolerance $\epsilon$ on the total error. The data points are labelled with the number of dimensions $d$ for each lattice spacing.}
\label{fig:runtime_cluster_rotor_epsilon}
\end{figure}

%%%%%%%%%%%%%%%%%%%%%%%%%%%%%%%%%%%%%%%%%%%%%%%%%%%%%
While for the results presented in the main text we fixed $\epsstat$ and varied the tolerance on the discretisation error, in the following we also show the (estimated) runtime as a function of the tolerance $\epsilon$ on the total root mean square error. For this we set $\epsstat=\epsdisc=\epsilon/\sqrt{2}$ as is common in the multilevel Monte Carlo literature. Figs. \ref{fig:runtime_doublewell_epsilon} and \ref{fig:runtime_rotor_epsilon} show the runtime of the single-level HMC method, the hierarchical sampler and the multilevel method as a function of the tolerance $\epsilon$ on the total error; they should be compared to Figs. \ref{fig:runtime_doublewell} and \ref{fig:runtime_rotor}. Finally, Fig. \ref{fig:runtime_cluster_rotor_epsilon} shows the runtime of the standard cluster-sampler and the multilevel-accelerated variant of the method; the corresponding plot in the main text is Fig. \ref{fig:runtime_cluster_rotor}.
% ====== plot runtime double well ======
% ====== plot runtime rotor ======
% ====== plot runtime rotor [cluster sampler] ======
\clearpage
\bibliographystyle{unsrt}
% \bibliography{paper}

\end{document}